\documentclass[a4paper,11pt]{article}
\pdfoutput=1
\usepackage{jheppub} 

\usepackage{graphicx}
\usepackage[english]{babel}
\usepackage{hyperref}
\usepackage{amsmath}
\usepackage{tabularx}
\usepackage{rotating}         
\usepackage{xspace}           
\usepackage{eurosym}

\allowdisplaybreaks[2]

\graphicspath{{figures/}}

\newcommand{\revised}[1]{#1}
\newcommand{\rev}[1]{#1}

\RequirePackage{color}
\definecolor{RED}{rgb}{1,0,0}
\definecolor{BLUE}{rgb}{0,0,1}
\definecolor{darkblue}{rgb}{0.24,0.45,0.70}
\definecolor{darkgrey}{rgb}{0.5,0.5,0.5}
\definecolor{darkred}{rgb}{0.55,0.0,0.0}
\definecolor{grey}{rgb}{0.75,0.75,0.75}
\definecolor{ForestGreen}{rgb}{0.0, 0.27, 0.13}

\newcommand{\NASixtyOne}{NA61\slash SHINE\xspace}

\newcommand{\gev}{\operatorname{GeV}}

\newcommand{\fm}{\operatorname{fm}}

\newcommand{\jpsi}{J\mskip -2mu/\mskip -0.5mu\Psi}

\newcommand{\gsim}{\raisebox{-4pt}{%
    $\,\stackrel{\textstyle >}{\sim}\,$}}

\newcommand{\DY}{Drell--Yan\xspace}

\begin{document}

\begin{flushleft}
\hfill \textbf{CERN-PBC-REPORT-2018-008} \\[-3em]
\includegraphics[width=5.5em]{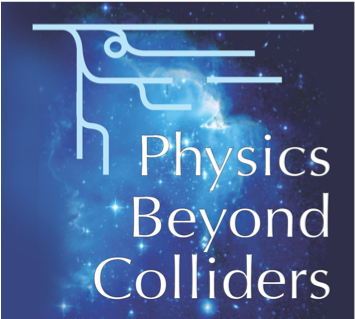}
\end{flushleft}

\begin{center}
{\Large \bf Physics Beyond Colliders \\[0.3em]
QCD Working Group Report} \\[2em]
A.~Dainese$^{1}$,
M.~Diehl$^{2, *}$,
P. Di Nezza$^{3}$,
J.~Friedrich$^{4}$,
M.~Ga{\'z}dzicki$^{5,6}$
G.~Graziani$^{7}$,
C.~Hadjidakis$^{8}$,
J.~J{\"a}ckel$^{9}$,
J.~P.~Lansberg$^{8}$,
A.~Magnon$^{10}$,
G.~Mallot$^{10}$,
F.~Martinez~Vidal$^{11}$,
L.~M.~Massacrier$^{8}$,
L.~Nemenov$^{12}$,
N.~Neri$^{13}$,
J.~M.~Pawlowski$^{9, *}$,
S.~M.~Pu{\l}awski$^{14}$,
J.~Schacher$^{15}$,
G.~Schnell$^{16, *}$,
A.~Stocchi$^{17}$,
G.~L.~Usai$^{18}$,
C.~Vall{\'e}e$^{19}$,
G.~Venanzoni$^{20}$ \\
\vspace{2em}
\parbox{0.8\textwidth}{\textbf{Abstract:} This report summarises the main findings of the QCD Working Group in the CERN Physics Beyond Colliders Study.}
\end{center}

\vfill

\parbox{0.92\textwidth}{\small
$^{1}$ INFN, Sezione di Padova, Padova, Italy\\
$^{2}$ Deutsches Elektronen-Synchroton DESY, 20603 Hamburg, Germany\\
$^{3}$ INFN, Laboratori Nazionali di Frascati, Frascati, Italy \\
$^{4}$ Technische Universit{\"a}t M{\"u}nchen, Physik Dept., 85748 Garching, Germany\\
$^{5}$ University of Frankfurt, Frankfurt, Germany\\
$^{6}$ Jan Kochanowski University in Kielce, Poland\\
$^{7}$ INFN, Sezione di Firenze, Firenze, Italy\\
$^{8}$ IPNO, CNRS-IN2P3, Univ.\ Paris-Sud, Universit{\'e} Paris-Saclay, 91406 Orsay Cedex, France\\
$^{9}$ Institut f\"ur Theoretische Physik, Universit\"at Heidelberg, Philosophenweg 16, 69120 Heidelberg, Germany\\
$^{10}$ CERN, 1211 Geneva 23, Switzerland\\
$^{11}$ IFIC, Universitat de Val{\`e}ncia-CSIC, Valencia, Spain\\
$^{12}$ JINR Dubna, Russia\\
$^{13}$ INFN, Sezione di Milano and Universit\`a di Milano, Milan, Italy\\
$^{14}$ University of Silesia, Katowice, Poland\\
$^{15}$ Albert Einstein Center for Fundamental Physics, Laboratory of High Energy Physics, Bern, Switzerland\\
$^{16}$ Department of Theoretical Physics, University of the Basque Country UPV/ EHU, 48080 Bilbao, Spain and IKERBASQUE, Basque Foundation for Science, 48013 Bilbao, Spain\\
$^{17}$ LAL, CNRS-IN2P3, Univ. Paris-Sud, Universit{\'e} Paris-Saclay, 91440 Orsay Cedex, France\\
$^{18}$ Dipartimento di Fisica dell'Universit{\`a} and Sezione INFN, Cagliari, Italy\\
$^{19}$ CPPM, CNRS-IN2P3 and Aix-Marseille University, Marseille, France \\
$^{20}$ INFN, Sezione di Pisa, Pisa, Italy
}

\vfill

${}^{*}${\small Working group conveners and editors of this summary report}

\newpage

\tableofcontents

\newpage



\section{Overview}
\label{sec:motivation}

Quantum chromodynamics (QCD) is a well-established part of the
Standard Model, but at the same time remains an area of active
research, with many aspects we would like to understand better.  Among
these are for instance the mechanism of confinement, the quark-gluon
structure of hadrons, and the nature of the QCD phase diagram.
Furthermore, an increasingly precise description of QCD processes is
required in several other fields.  Perhaps most prominent among these
is the description of $pp$ collisions at LHC, but QCD cross sections
are also required as an input for describing air showers induced by
cosmic rays and for neutrino oscillation experiments.

The CERN Physics Beyond Colliders (PBC) study, summarised in
\cite{PBC-summmary}, included a physics working group dedicated to
QCD, in which a wide range of proposals was discussed. These proposals
concern future measurements at the CERN SPS as well as fixed-target
installations at the LHC. Before presenting them, let us briefly
mention other major facilities worldwide at which the investigation of
QCD dynamics is a major goal.

\paragraph{Facilities in operation:}
\begin{itemize}
\item All four \textbf{LHC} experiments have a vigorous programme of
  measurements related to QCD.  Prominent examples are the
  determination of parton densities, the dynamics of QCD radiation
  in hadronic jets, and heavy ion physics.
\item The experiments PHENIX and STAR at the \textbf{RHIC} collider at
  BNL pursue both measurements in heavy-ion physics and, using the
  only polarised high-energy proton beams available so far, in proton
  spin physics.  Significant upgrades are foreseen to turn the PHENIX
  detector into sPHENIX, which is foreseen to start operation in the
  beginning of the 2020s.
\item Jefferson Lab has successfully completed the upgrade of its
  electron beam facility (called \textbf{JLab 12} in the following).
  It now operates fixed-target experiments with polarised electron
  beams at 11 GeV in Halls A, B, C, and with a 12 GeV photon beam in
  Hall D.  Whilst the focus of Hall D is on spectroscopy, the other
  three halls have a varied programme of investigations of the
  structure of the nucleon and of light nuclei.
\item The \textbf{J-PARC} complex in Japan operates a wide range of
  experiments based on its high-intensity primary proton beam, which
  currently has a maximum energy of 30 GeV.  In the hadron hall, a
  variety of secondary beams are used for experiments in hadron and
  nuclear physics.  An extension of this hall is under discussion, as
  well as a heavy ion programme with
  $\sqrt{s_\textrm{NN}}= 1.9 - 6.2$\, GeV (foreseen 2025).
\end{itemize}
\paragraph{Facilities under construction:}
\begin{itemize}
\item The \textbf{NICA} facility at JINR (Dubna) will pursue a broad
  range of measurements in heavy ion physics, hadron structure and
  spin physics, both in collider and in fixed target mode. The
  collider mode works at collision energies of
  $\sqrt{s_\textrm{NN}}= 4 - 11$\, GeV. Commissioning is foreseen for
  2020.
\item The \textbf{FAIR} facility at GSI will pursue measurements
  in heavy ion physics at the \textbf{CBM} experiment, with \textbf{SIS
    100} at collision energies of $\sqrt{s_\textrm{NN}}= 2.7 - 4.9$\,
  GeV. The experiment will investigate the phase structure at high
  densities, chiral symmetry and the equation of state at neutron star
  densities. Commissioning is planned for 2025. Parts of the detector will
  be used in the RHIC beam energy scan (BES-II) in 2019/2020.

  The \textbf{PANDA} experiment will use an anti-proton beam from SIS
  100 with momenta 1.5 -15\, GeV$/c$ for experiments on hadron
  spectroscopy, in medium effects of hadrons, nucleon structure and
  hypernuclei. Commissioning is planned for 2025.
\end{itemize}
\paragraph{Proposed facilities:}
\begin{itemize}
\item The Electron-Ion Collider \textbf{(EIC)} is a facility planned
  in the US that would collide polarised electrons with polarised
  protons or with light to heavy ions in the range $\sqrt{s} = 20$ to
  $100 \gev$ in its initial stage.  With its high luminosity
  ($10^{33}$ to $10^{34}$ cm$^{-2}$ s$^{-1}$ for $e p$) it is designed
  for in-depth studies of proton and nuclear structure in the regime
  dominated by gluons and sea quarks \cite{Accardi:2012qut}.  It has
  been identified as ``the highest priority for new facility
  construction following the completion of FRIB'' in the 2015
  Long-Range Plan for U.S. nuclear science, and both BNL and JLab are
  actively working towards its realisation.  Provided a successful
  outcome of various stages of reviews and of appropriate funding, an
  EIC could start operation around 2030 \cite{Tribble:EICUG2018}.
\item The \textbf{LHeC} would collide the LHC proton or ion beam with
  $60 \gev$ electrons, aiming at luminosities in the range $10^{33}$
  to $10^{34}$ cm$^{-2}$ s$^{-1}$ for $e p$.  Its primary physics
  motivation is the search of physics beyond the Standard Model, but
  -- similarly to the LHC -- it would also permit detailed studies of
  QCD at high energy and/or high resolution scale, with precise
  determinations of PDFs and of $\alpha_s$ being among the highlights.
  From a technical point of view, an LHeC could start operation in LHC
  run 5 \cite{Bordry:2018gri}.
\item In the context of planning future hadron colliders at the energy frontier,
the possibilities of electron-proton collisions are also being discussed, for
example for an $eh$ option at an FCC.  Such facilities would extend the
possibilities described for the LHeC to yet higher scales.
\end{itemize}
This list is not meant to be comprehensive; measurements at other facilities
that are relevant in the context of PBC proposals will be mentioned in the
appropriate sections.


\subsection{Proposals and physics topics}

The following proposals were discussed in the PBC QCD working group and will be
presented in this summary:
\begin{description}
\item[LHC-FT gas] A number of studies were performed for using one of
  the LHC beams in fixed-target mode. A viable solution is a gaseous
  target internal to the LHC ring. Corresponding studies were
  presented \revised{by members of ALICE}, one by LHCb with a focus on
  the improvement of the already existing SMOG programme, one by
  members of LHCb on the possibilities to install a polarised gas
  target for the investigation of proton spin phenomena (labelled
  ``LHCSpin'' in the following), and a study by the AFTER@LHC group.
  A comprehensive review by the latter has recently appeared
  in~\cite{Hadjidakis:2018ifr}. \revised{Studies for similar
    measurement campaigns, but with solid targets and beam extraction
    using bent crystals are ongoing
    \cite{Hadjidakis:2018ifr,Hadjidakis:2018-PBC}.}
\item[LHC-FT crystals] This proposal studies the possibility to use a
  setup with bent crystals at LHCb in order to measure the magnetic
  dipole moments of short-lived baryons such as the $\Lambda_c$
  \revised{\protect\cite{Burmistrov:2194564,Bagli:2017foe}}.  The
  measurement of their electric dipole moment was investigated in the
  same proposal and -- given its main physics motivation -- was
  discussed in the PBC BSM working group \cite{PBC-BSM}.  The prospect
  to measure the magnetic moment of the $\tau$ lepton in this setup
  was just recently discussed in \cite{Fomin:2018ybj}.
\item[COMPASS++] A diverse programme of QCD measurements,
  \revised{using major upgrades of the COMPASS detector} at the M2
  beamline of the SPS, is being proposed in the LoI
  \cite{Denisov:2018unj}.  For the purpose of this summary, this
  proposal is termed COMPASS++.  The first group of proposed
  measurements can be realised with the existing muon or hadron beams
  at M2, whereas a second group requires RF separated high-intensity
  kaon or antiproton beams.  The technical feasibility of the latter
  was studied in the PBC Conventional Beams working group
  \cite{PBC-convbeam}.
\item[MUonE]  This proposal aims at extracting the hadronic vacuum polarisation
at very small spacelike momentum transfers $t$ from elastic scattering of muons
on the shell electrons of a target.  Via a sum rule one could then evaluate the
contribution of hadronic vacuum polarisation to the anomalous magnetic moment
$(g-2)_{\mu}$ of the muon.  The location of this measurement would be the M2
beamline of the SPS.
\item[NA61++]  The NA61/SHINE collaboration has presented its plans for running
after LS2 as proposal addenda to the SPSC
\cite{Aduszkiewicz:2309890,Aduszkiewicz:2621751}.   The proposed measurements
range from charm hadron production in Pb Pb collisions for heavy ion physics to
nuclear fragmentation cross sections for cosmic ray physics and hadron
production in hadron-induced reactions for neutrino physics.
\item[NA60++]  Pursuing earlier studies by the NA60 experiment, this proposal
aims at probing the QCD phase transition in the production of low energy lepton
pairs in Pb Pb collisions.  In addition, comparison of $a_1$ and $\rho$ meson
production at this experiment would offer the possibility to investigate the
restoration of chiral symmetry.
\item[DIRAC++]  The DIRAC experiment at the CERN proton synchroton (PS) reported
the first observation of $\pi^+ K^-$ and $\pi^- K^+$ atoms.  With a similar
experiment at the SPS, reusing parts of the original DIRAC detector, such atoms
could be produced with significantly higher statistics, which would yield
precise information on the $\pi K$ scattering lengths.
\end{description}
Due to the beam intensities required by both the NA60++ and DIRAC++, they would
need to be installed in an underground hall.   As a consequence, the only
possible location for both experiments is the ECN3 cavern, where the NA62
experiment is currently installed and running.  A study in the PBC Conventional
Beams working group found that both experiments would fit together in that
cavern, but none of them could be installed in the presence of NA62
\cite{PBC-convbeam}.

Concerning the M2 beamline, different scenarios for concurrent running of
COMPASS++ and MUonE were discussed in the QCD working group.  A detailed study
of this issue is presented in section \ref{sec:compass-muone}.

\begin{figure}
\begin{center}
\includegraphics[width=0.98\textwidth]{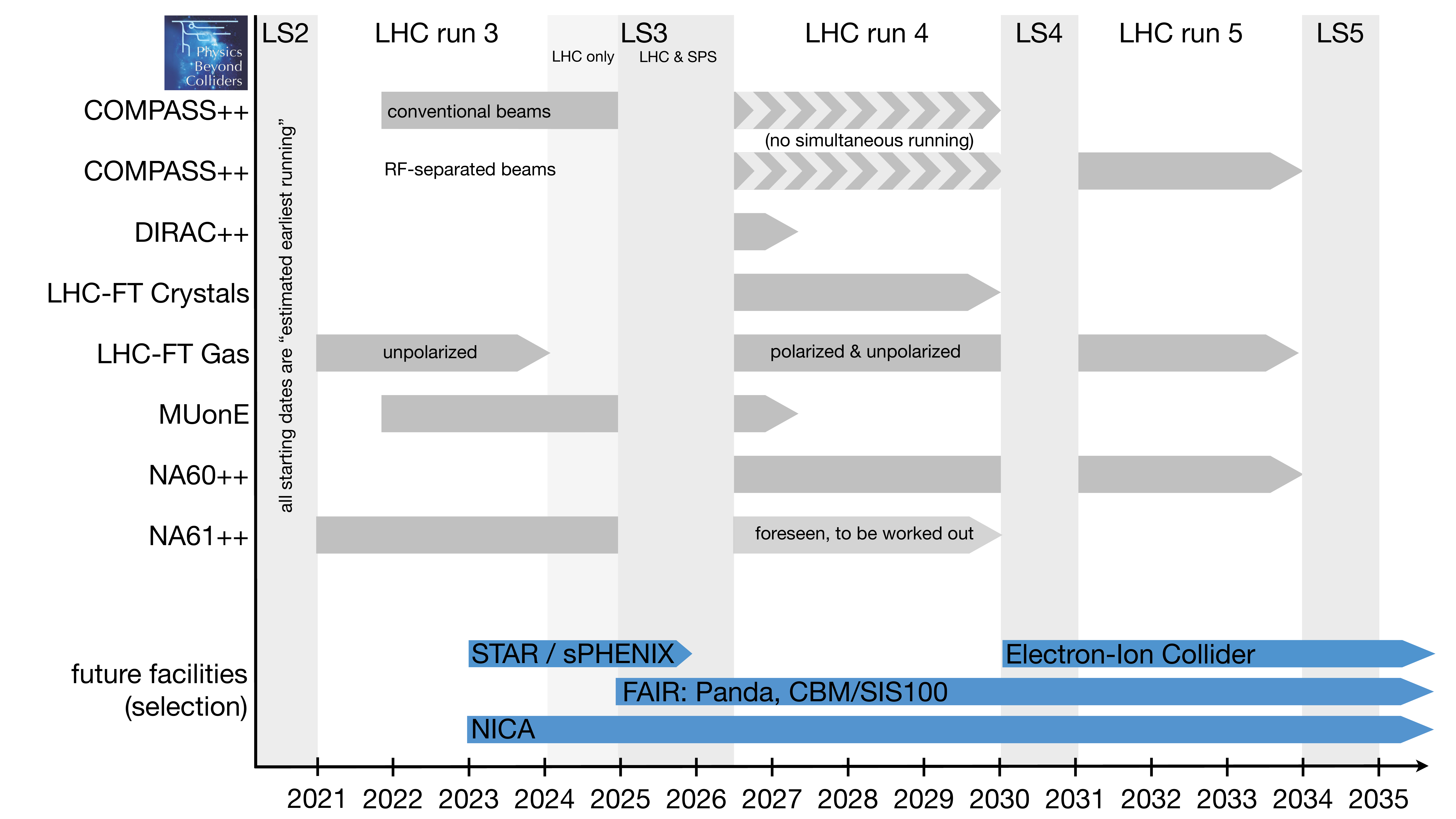}
\caption[Schematic overview of the proposals discussed in the working
group, together with their possible timelines.]{\label{fig:timelines}
  Schematic overview of the proposals discussed in the working group,
  together with their possible timelines.  The beginning of an arrow
  marks the estimated earliest possible running.  Possible timelines
  of \rev{selected other facilities} with a focus on QCD studies are
  shown for comparison.}
\end{center}
\end{figure}

A schematic overview of the different proposals is presented in
figure~\ref{fig:timelines}.  For comparison, possible timelines of other major
QCD facilities outside CERN are also given.  Not shown are the already running
experiments at Jefferson Lab, at BNL (RHIC) and at J-PARC, nor the different
options of high-energy colliders (LHeC, FCC, etc.) that are currently under
discussion.

We now briefly present the different physics topics addressed in the proposals
just listed.
\begin{description}
\item[Parton distributions.] Parton distribution functions (PDFs) are
  among the most prominent quantities that describe the structure of
  hadrons at the quark-gluon level.  Measurements with \textbf{LHC-FT
    gas} setups could significantly increase our knowledge of the
  unpolarised PDFs of the proton (which are a key ingredient for the
  description of LHC collider data) and PDFs of nuclei (which in
  particular describe the initial state of heavy-ion collisions).  The
  PDFs of pions and kaons are of special theory interest since these
  hadrons are the pseudo-Goldstone bosons of QCD.  Measurements
  proposed by \textbf{COMPASS++} could have a substantial impact on
  better determining their shape.  Several intriguing spin effects in
  QCD are quantified in parton distributions that depend on the
  transverse parton momentum and on the polarisation of the proton.
  Polarised Drell-Yan production as envisaged in the \textbf{LHC-FT}
  and the \textbf{COMPASS++} proposals is a prime reaction to
  determine these distributions and would be highly complementary to
  the spin physics programme of JLab 12 and the EIC.
\item[Heavy-ion physics.] The phase structure and physics of QCD at
  large densities $\mu_B/T\gtrsim 2-3$ is not uncovered yet. Neither
  experimental data from heavy ion collisions nor theoretical
  investigations have provided conclusive answers so far.  The
  proposed experiments \textbf{NA60++} and \textbf{NA61++} cover an
  important part of this density regime with
  $\sqrt{s_\textrm{NN}} \sim 5 - 17$\, GeV at the {CERN} {SPS}.  Major
  open questions concern the existence and location of a critical end
  point in the QCD phase diagram, the details of chiral symmetry
  restoration, as well as the onset of deconfinement. In this context
  the {SPS} experiments offer unique opportunities, in particular with
  the open charm measurements at \textbf{NA61++} and the precise
  determination of the fireball initial temperature at
  \textbf{NA60++}. The energy range covered by the \textbf{LHC-FT}
  experiments is unique and connects the energy range covered by the
  LHC collider experiments with the lower energy range covered by
  {SPS}, {RHIC}, {HADES} ({GSI}) and the planned experiments at {NICA}
  and {CBM}.

\item[Elastic muon scattering.]  Measurements of elastic $\mu p$
  scattering by \textbf{COMPASS++} and of elastic $\mu e$ scattering
  by \textbf{MUonE} would provide valuable input to two outstanding
  questions in precision physics.

  (i) The extraction of the electromagnetic proton radius from either
  spectroscopy (on ordinary or muonic hydrogen) or elastic $e p$
  scattering leads to a bewildering spread of results that are
  inconsistent within their quoted uncertainties.  Elastic $\mu p$
  scattering at SPS energies would give independent experimental
  input, with systematic uncertainties that are quite different from
  those in the $e p$ channel.

  (ii) The discrepancy between the theoretical and experimental values
  of $(g-2)_\mu$ is a significant problem of the Standard Model.  The
  contribution $a_\mu^{\text{HVP}}$ from hadronic vacuum polarisation
  is one of the two dominant sources of theoretical uncertainty on
  $a_\mu = (g-2)_\mu /2$.  The MUonE proposal aims at extracting
  $a_{\mu}^{\text{HVP}}$ with an accuracy comparable to the one of the
  currently most precise determination, which is based on data for
  $e^+ e^-$ annihilation into hadrons (supplemented in part by
  hadronic $\tau$ decays).
\item[Low-energy QCD.]  The behaviour of QCD in the low-energy limit
  is largely dominated by the dynamics of chiral symmetry breaking.
  This is described by an effective field theory, chiral perturbation
  theory ($\chi$PT), which beyond its role in QCD also serves as a
  blueprint for many extensions of the Standard Model.  $\chi$PT in
  the flavour SU(2) sector of $u$ and $d$ quarks has become precision
  physics, whilst the inclusion of $s$ quarks to flavour SU(3) remains
  more challenging.  Reasons for this are the larger value of the $s$
  quark mass, which results in a larger expansion parameter of the
  theory, and the higher number of low-energy constants that need to
  be determined in order to make the theory predictive at loop level.
  Information on the $\pi K$ scattering lengths from \textbf{DIRAC++}
  or from the electromagnetic kaon polarisabilities from
  \textbf{COMPASS++} would be highly valuable, as both quantities can
  be computed in three-flavour $\chi$PT.

  The spectrum of mesons carrying strangeness with masses beyond
  $1 \gev$ is known with much less precision than the one of
  non-strange mesons.  \textbf{COMPASS++} proposes a comprehensive
  measurement campaign of the strange meson spectrum with RF separated
  kaon beams.  The data analysis would significantly benefit from the
  partial wave analysis methods developed by the COMPASS collaboration
  during its spectroscopy programme with non-strange mesons
  \cite{Adolph:2015tqa}.

  Hadrons made from both heavy and light quarks play a particular role
  in QCD, both for practical reasons (such as the study of the CKM
  mechanism in $b$ or $c$ quark decays) and because the heavy quark
  mass opens up the possibility to employ modern field theory methods
  based on an expansion around the heavy quark limit.  The
  \textbf{LHC-FT crystal} proposal discussed here aims at measuring
  the magnetic moments of baryons containing a $c$ or a $b$ quark.
\item[Measurements for cosmic ray and for neutrino physics.]
  Measurements of hadronic cross sections that are relevant to the
  propagation of cosmic rays are proposed by \textbf{LHC-FT},
  \textbf{COMPASS++} and \textbf{NA61++}.  An example is the
  production cross section for antiprotons or antideuterons.  Both
  LHCb and NA61/SHINE have already published measurements in this area
  \cite{Aaij:2018svt,Aduszkiewicz:2017sei}.  \textbf{NA61++} also
  proposes to further pursue measurements of hadroproduction cross
  section that can reduce the uncertainties on neutrino fluxes in
  neutrino oscillation experiments.
\end{description}
A schematic overview of the different physics topics and the proposals that
address them is shown in Table \ref{tab:qcd-matrix}.

\begin{sidewaystable}[p]
\begin{tabular}{|l|cccc|c|c|c|c|c|c|}
  \hline
  & \multicolumn{4}{c|}{LHC FT gas} & LHC FT & COMPASS\footnotesize{++} &
                                                                          MUonE & NA61\footnotesize{++} &
                                                                                                          NA60\footnotesize{++} & DIRAC\footnotesize{++} \\
  & \scriptsize{ALICE} & \scriptsize{LHCb} & \rev{\scriptsize{LHCSpin}}  & \rev{\scriptsize{AFTER@LHC}} & crystals & & & & & \\
  \hline
  proton PDFs & $\times$ & $\times$ & & $\times$ & & & & & & \\
  nuclear PDFs & $\times$ & $\times$ & & $\times$ & & $\times$ & & & & \\
  spin physics & $\times$ & & $\times$ & $\times$ & & $\times$ & & & & \\
  meson PDFs & & & & & & $\times$ & & & & \\
  \hline
  heavy ion physics & $\times$ & & & $\times$ & & & & $\times$ & $\times$ & \\
  \hline
  elast.~$\mu$ scattering & & & & & & $\times$ & $\times$ & & & \\
  \hline
  chiral dynamics & & & & & & $\times$ & & & & $\times$ \\
  magnet.~moments & & & & & $\times$ & & & & & \\
  spectroscopy & & & & & & $\times$ & & & & \\
  \hline
  measurements for & & & & & & & & & & \\
  cosmic rays and & $\times$ & $\times$ & & $\times$ & & $\times$ & & $\times$ & &
  \\
  neutrino physics & & & & & & & & & & \\
  \hline
\end{tabular}
\caption{\label{tab:qcd-matrix} Schematic overview of the physics topics
addressed by \rev{the studies presented} in the QCD working group.}
\end{sidewaystable}

\clearpage

\section{Proposals}
\label{sec:proposals}

In the present section, we discuss the different proposals one by one,
with the exception of the proposed measurements for cosmic ray or
neutrino physics, which will be reviewed together in
section~\ref{sec:cosmics}.


\subsection{LHC Fixed Target}
\label{sec:LHC-FT}

A fixed-target program using the proton and ion beams of the LHC
offers highest ever collision energies in the fixed-target mode.  It
allows for a novel physics program for heavy-ion, hadron, spin, and
astroparticle physics with existing (or new set-ups), while at the
same time complying with constraints of a parasitic running along with
the LHC collider program.  The clear advantages of such program can be
summarised as~\cite{Hadjidakis:2018ifr}.
\begin{itemize}
\item large luminosities, comparable to those of the LHC and well above those of experiments at similar energies
\item wide kinematic coverage with access to backward rapidities in
  the centre-of-mass
\item a variety of target nuclei including polarised ones
\end{itemize}

The kinematical conditions put such a program in a unique position,
with little direct competition in the nearby future.  On the other
hand the program complements various ongoing and planned programs, and
provides important input to studies at the LHC and astroparticle
physics.  In particular, following highlights can be addressed:
\begin{itemize}
\item high-$x$ unpolarized and polarised quark and gluon distributions
  of the proton, difficult to access at other facilities;
\item nuclear parton distributions;
\item detailed studies related to heavy-ion physics;
\item hadron production (e.g., \(\bar{p}\)) as input to interpretation of astroparticle physics data.
\end{itemize}
The first two points will be further discussed below, while heavy-ion physics will be addressed in section~\ref{sec:HIC-Summary},
and the cross-links to astroparticle physics will be presented in section~\ref{sec:cosmics}, due to larger overlap with other proposals.

\subsubsection{Physics motivation}

\begin{figure}
\centering
\includegraphics[width=0.75\textwidth,angle=0]{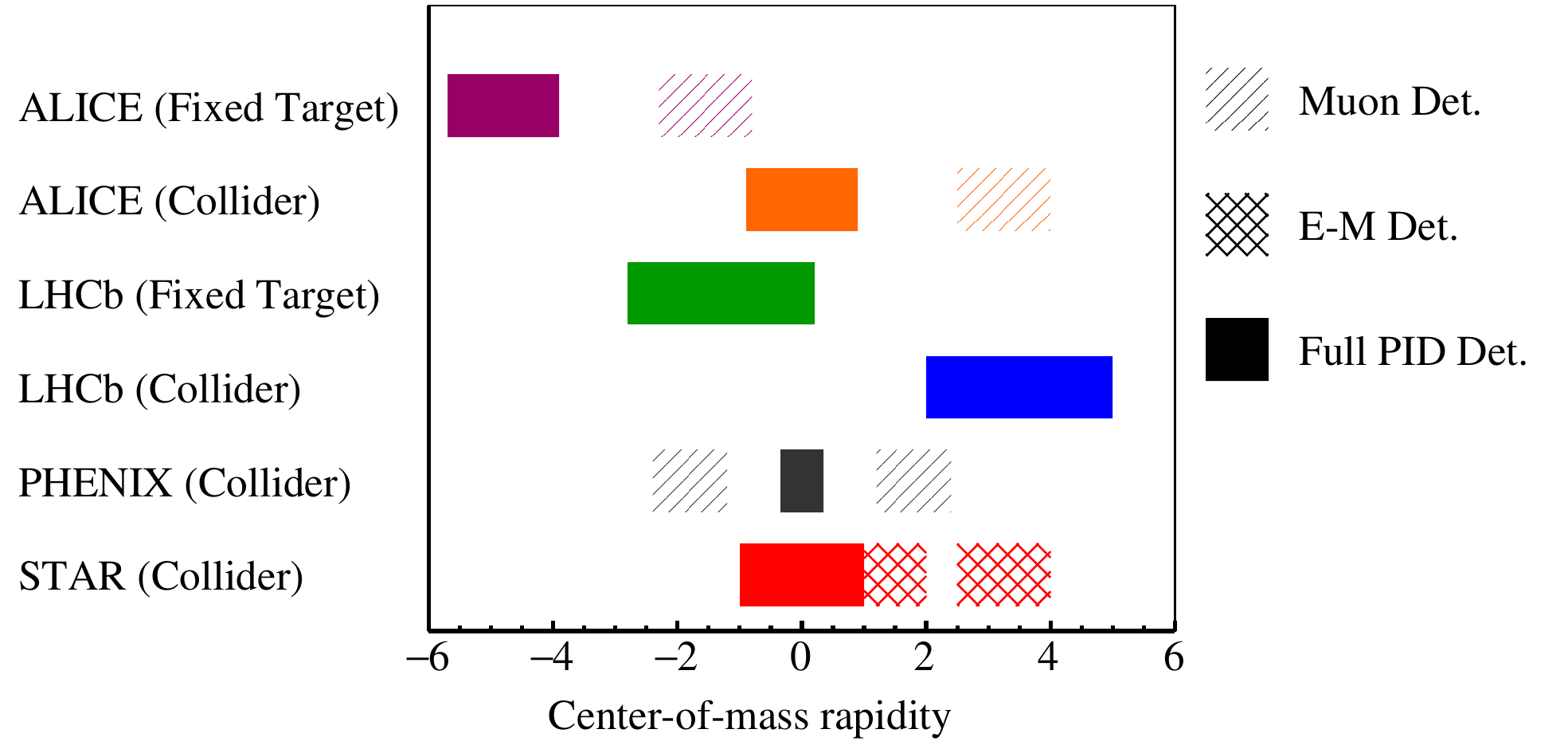}
\caption[Comparison of the kinematic coverage of the ALICE and LHCb
detectors at the LHC and the STAR and PHENIX detectors at
RHIC.]{Comparison of the kinematic coverage of the ALICE and LHCb
  detectors at the LHC and the STAR and PHENIX detectors at RHIC. For
  ALICE and LHCb, the acceptance is shown in the collider and the
  fixed-target modes with a target position at the nominal Interaction
  Point (IP) for a 7 TeV proton beam. The ``Full PID Det.'' label
  indicates detector with particle identification capabilities, ``E-M
  Det.'' an electromagnetic calorimeter, and ``Muon Det.'' a muon
  detector.  (Figure taken from~\cite{Hadjidakis:2018ifr}.)}
\label{fig:rapidity}
\end{figure}

\begin{figure}[!t]
\centering
\includegraphics[width=0.4\textwidth,angle=270]{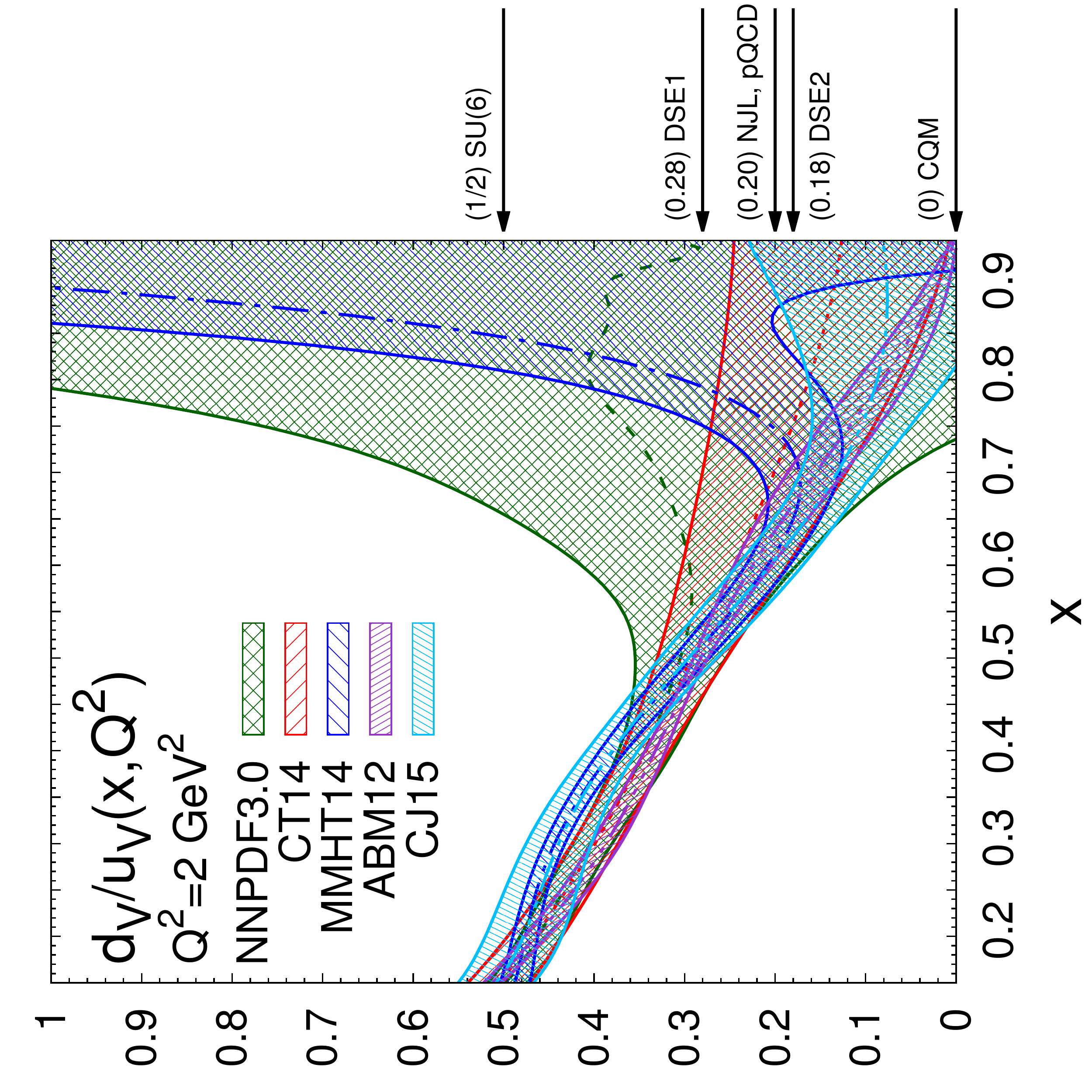}
\includegraphics[width=0.4\textwidth,angle=270]{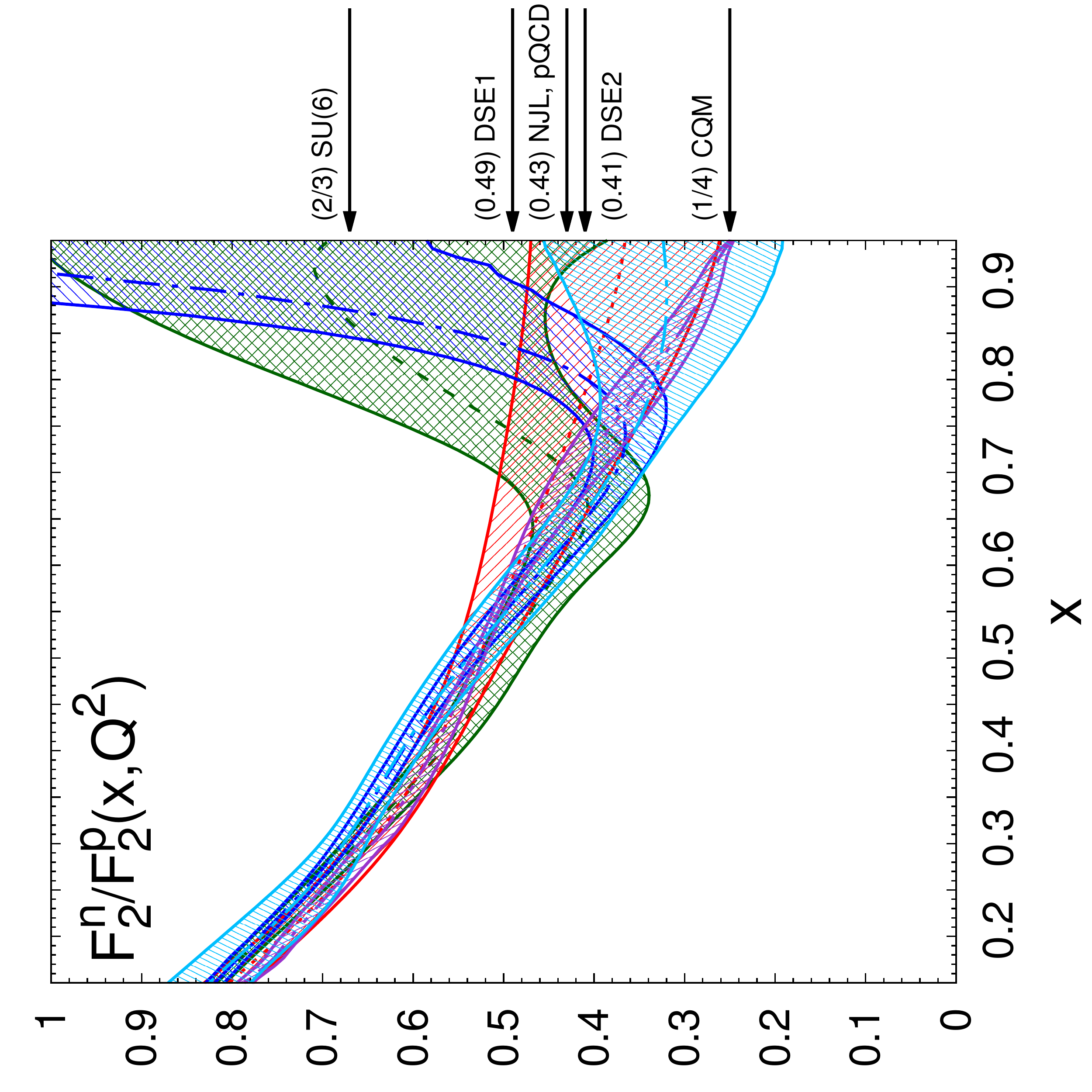}
\caption[The ratios of the valence down to up distributions and of
$F_2^n/F_2^p$ for various PDF sets, compared to predictions from
different nonperturbative models.]{The ratios of the valence down to
  up distributions (left) and of $F_2^n/F_2^p$ (right) at $Q^2=2$
  GeV$^2$ for various PDF sets. Indicated as well are the high-\(x\)
  predictions from different nonperturbative models (see
  Ref.~\cite{Ball:2016spl} for details).  (Figures taken
  from~\cite{Ball:2016spl}.)}
\label{fig:high-x-ratioplots}
\end{figure}

The driving physics processes for an LHC Fixed-Target (LHC-FT) program
are Drell--Yan and heavy-quark (including \(b\bar{b}\) mesons)
production, complemented by particle production in nuclear collisions.
The large boost of the system in a fixed-target forward-acceptance
setup using the LHC beams (cf.~figure~\ref{fig:rapidity}) permits to
probe quark and gluon distributions at very high parton momentum
fractions, $x$, \revised{and intermediate hard scales, $Q^2$, where
  data with high statistics is unavailable so far and hard to access
  in the near future at any other facility.}
Parton distributions are on one side an indispensable tool for the
interpretation of proton-induced processes as, e.g., searches at the
LHC, and on the other hand interesting by themselves.  They directly
touch the foundation of our understanding of the proton in the
Standard Model.  While parton distributions at low momentum fractions
are dominated by QCD-radiative effects and scale evolution can be
reliably predicted using perturbative methods, the initial shape
including the large-\(x\) behaviour of those distributions provide
insights into the non-perturbative quark and gluon dynamics, at the
moment still out of reach of theoretical calculations. A wide range of
predictions exists for the behaviour of quark distributions based on,
e.g., counting rules~\cite{Brodsky:1973kr}, spin-flavor SU(6)
symmetry, dominance of scalar valence diquarks etc., resulting in
differing predictions for the down- to up-quark ratio in the limit of
\(x\to 1\), where the parton carries practically all the proton
momentum.  That behaviour in the large-\(x\) behaviour, in turn, has
direct impact on the \(F_{2}\) structure-function ratio for proton and
neutrons (see, e.g., Ref.~\cite{Melnitchouk:1995fc} and references
therein).  This is illustrated in figure~\ref{fig:high-x-ratioplots},
where the ratios of valence-quark distributions and of $F_2^n/F_2^p$
are plotted for various PDF sets and compared to a set of
non-perturbative QCD predictions.

\begin{figure}
\centering
\includegraphics[width=0.45\textwidth,angle=0]{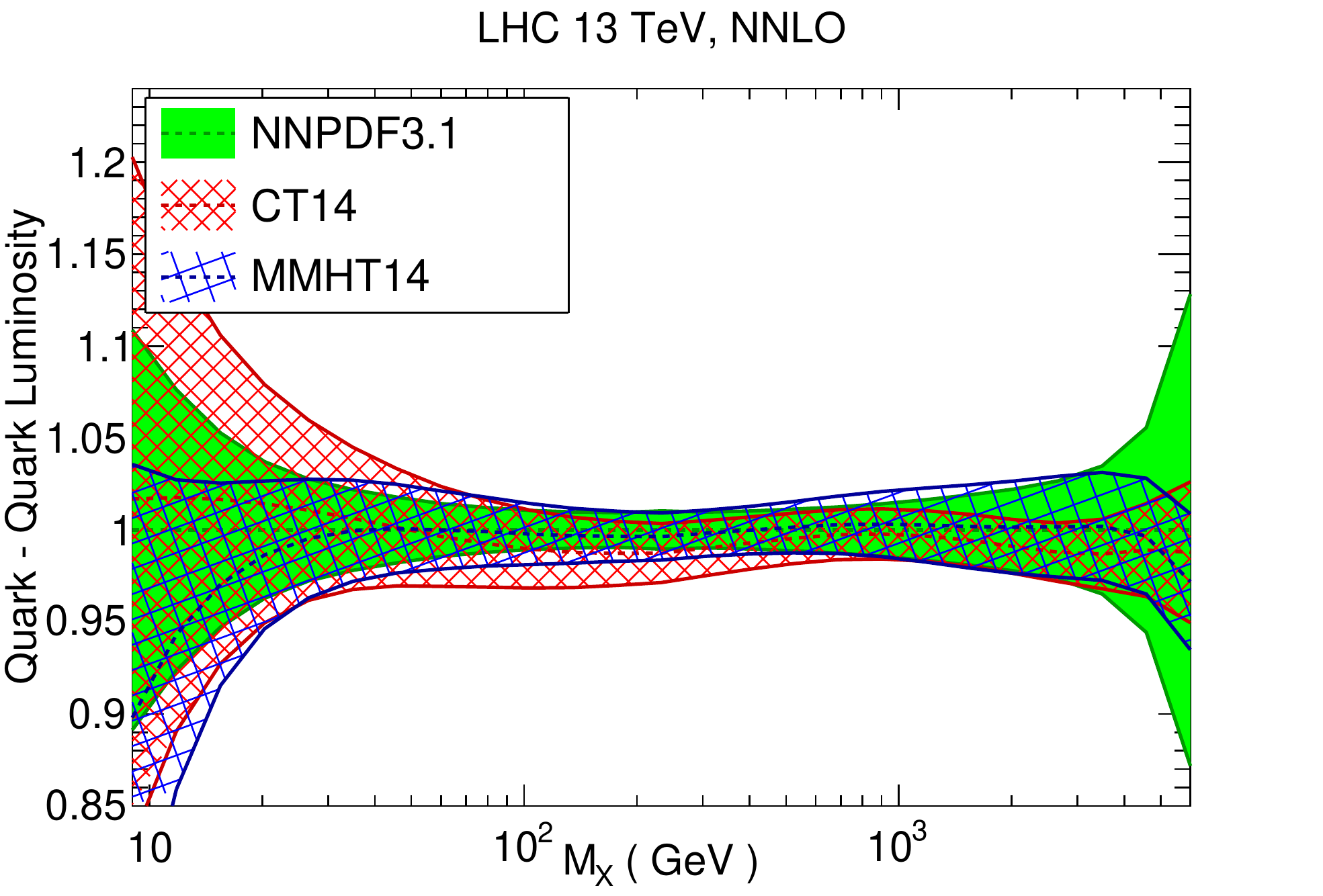}
\includegraphics[width=0.45\textwidth,angle=0]{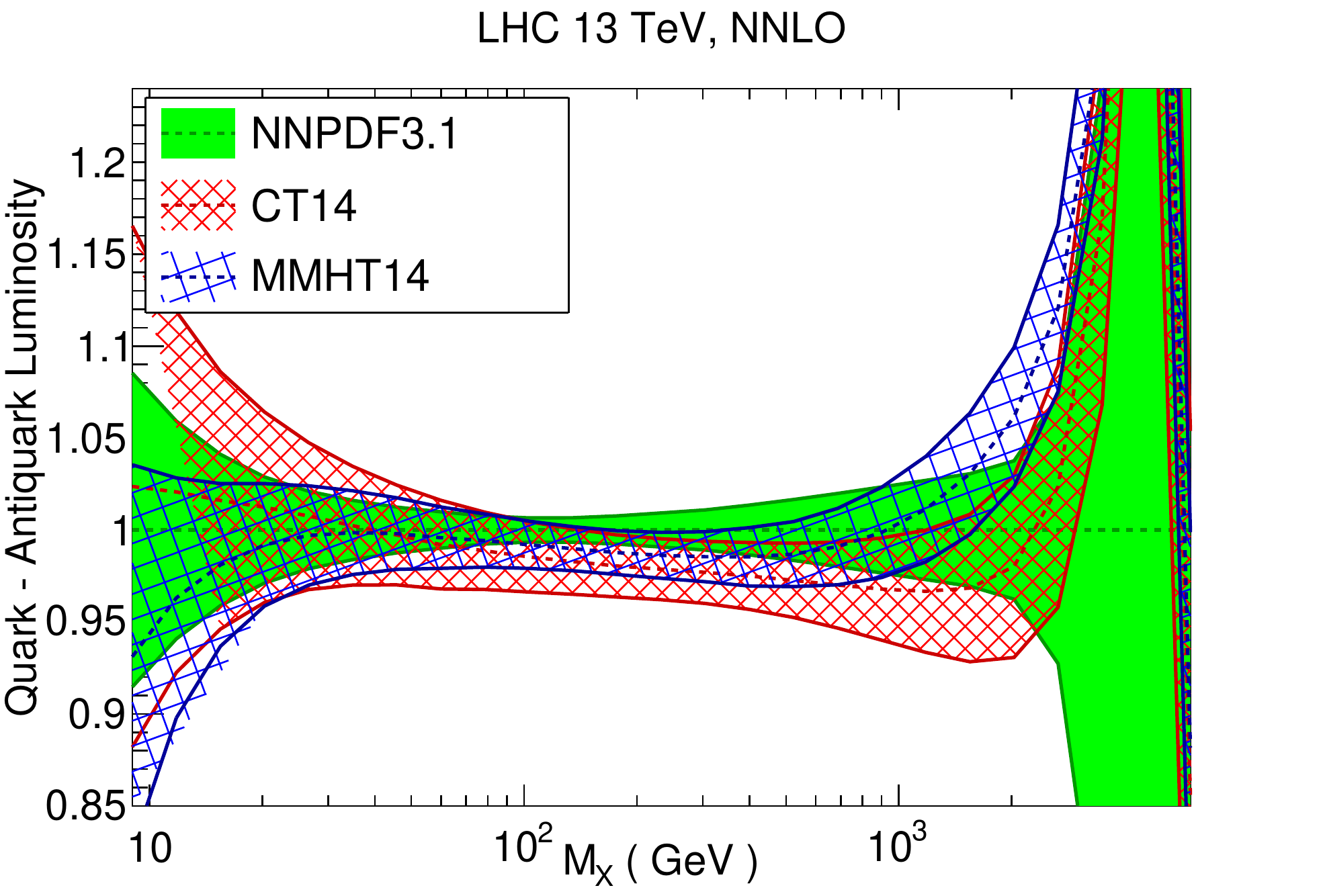}\\[.5cm]
\includegraphics[width=0.45\textwidth,angle=0]{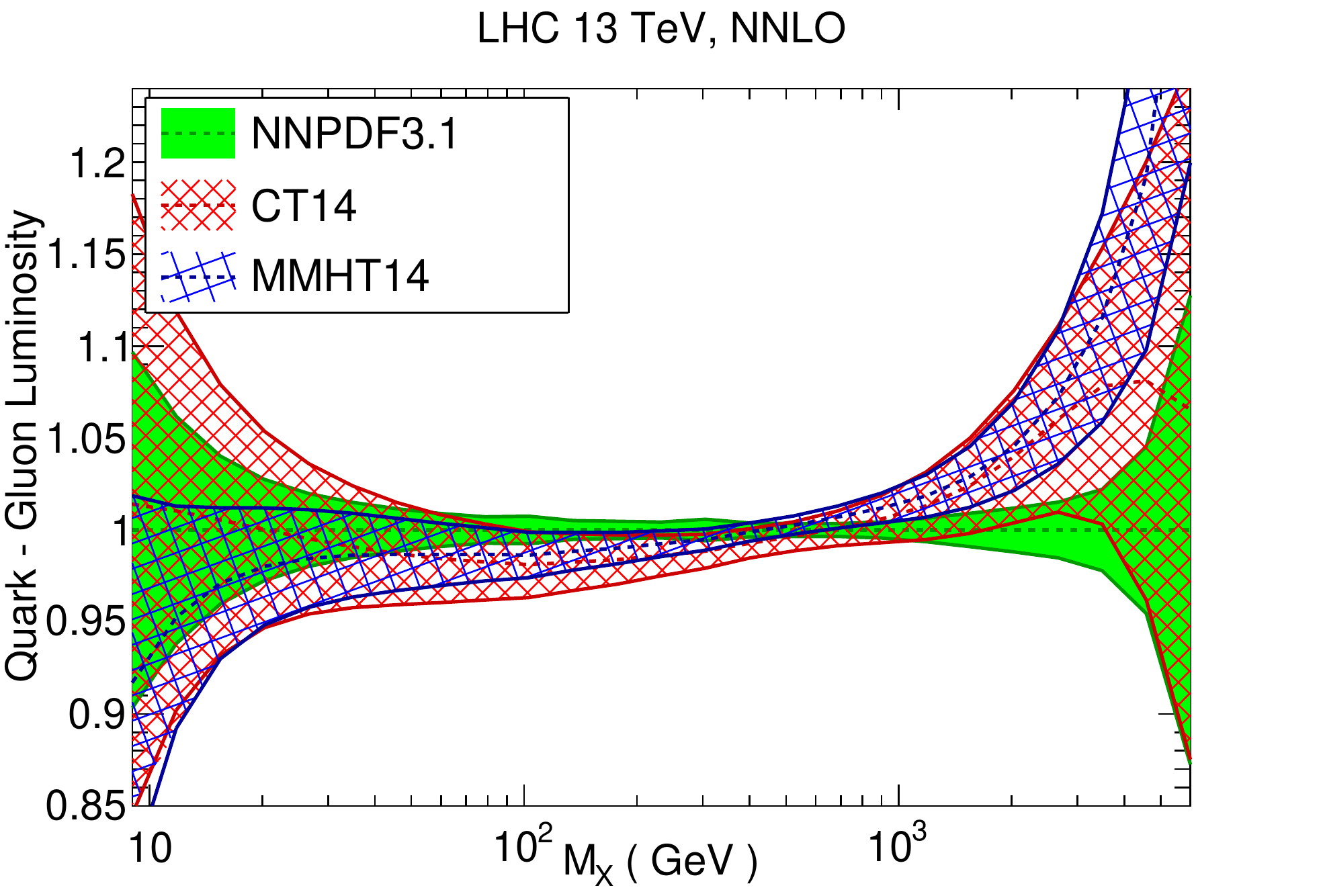}
\includegraphics[width=0.45\textwidth,angle=0]{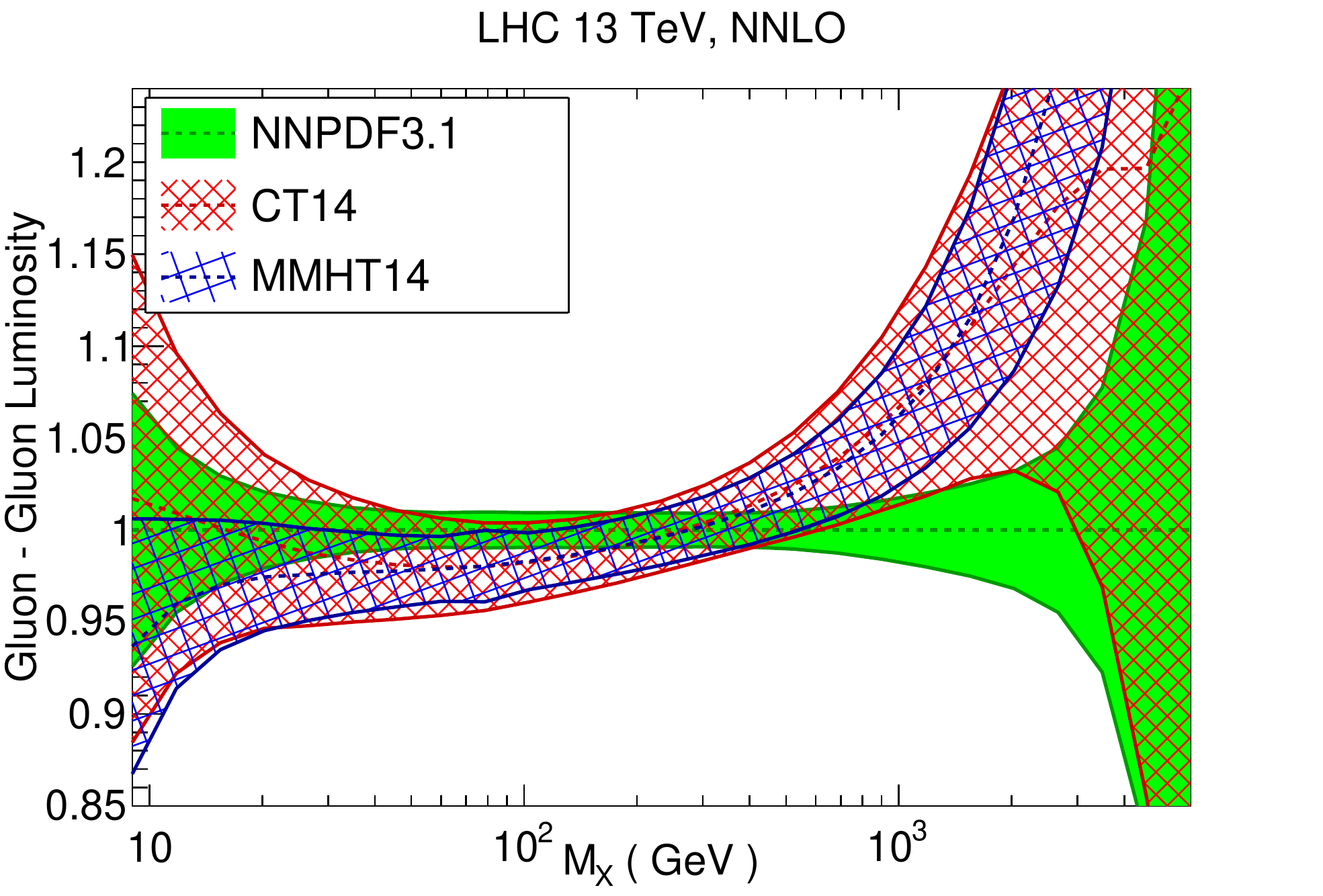}
\caption[Parton-parton luminosities for 13 TeV proton-proton
collisions for various PDF sets.]{Parton-parton luminosities for 13
  TeV proton-proton collisions for quark-quark (top left),
  quark-antiquark (top right), quark-gluon (bottom left), and
  gluon-gluon (bottom right) processes for various
  sets~\cite{Ball:2017nwa,Harland-Lang:2014zoa,Hou:2017khm} of parton
  distribution functions, as labelled.  (Figures taken
  from~\cite{Ball:2017nwa}.)}
\label{fig:parton-luminosities}
\end{figure}

It is apparent from figure~\ref{fig:high-x-ratioplots} that present
PDF fits cannot discriminate the various predictions and, much more,
that the ratios are hardly constrained starting from around \(x=0.6\).
A very similar picture emerges for the gluon
distribution~\cite{Dulat:2015mca,Ball:2017nwa}.  This has far-reaching
consequences, illustrated in figure~\ref{fig:parton-luminosities}.
High-mass particles at the LHC can only be probed reliably if the
high-\(x\) parton distributions are known well enough because they
determine the parton-parton luminosities and thus cross sections at
those masses.  In particular the gluon-gluon channel suffers from
precision already at masses above 1~TeV.

A similarly elusive parton distribution at high \(x\) is that of charm
quarks. In the 1980s it was already advocated that besides
perturbatively produced \(c\bar{c}\) pairs, the proton should possess
intrinsic charm~\cite{Brodsky:1980pb}, which should be probed at high
\(x\) to isolate it from the perturbative sea (for a recent review
see~Ref.~\cite{Brodsky:2015fna}).  In most PDF fits the charm-quark
PDFs are set to zero below an energy scale driven by the charm mass of
about 1.3~GeV. In contrast, the NNPDF collaboration finds that
including charm degrees of freedom the fits favour a non-vanishing
intrinsic large-\(x\) component, which carries about \(0.7\pm0.3\%\)
of the nucleon momentum at \(Q=1.65\)~GeV~\cite{Ball:2016neh}.

\begin{figure}
\centering
\includegraphics[width=0.49\textwidth,angle=0]{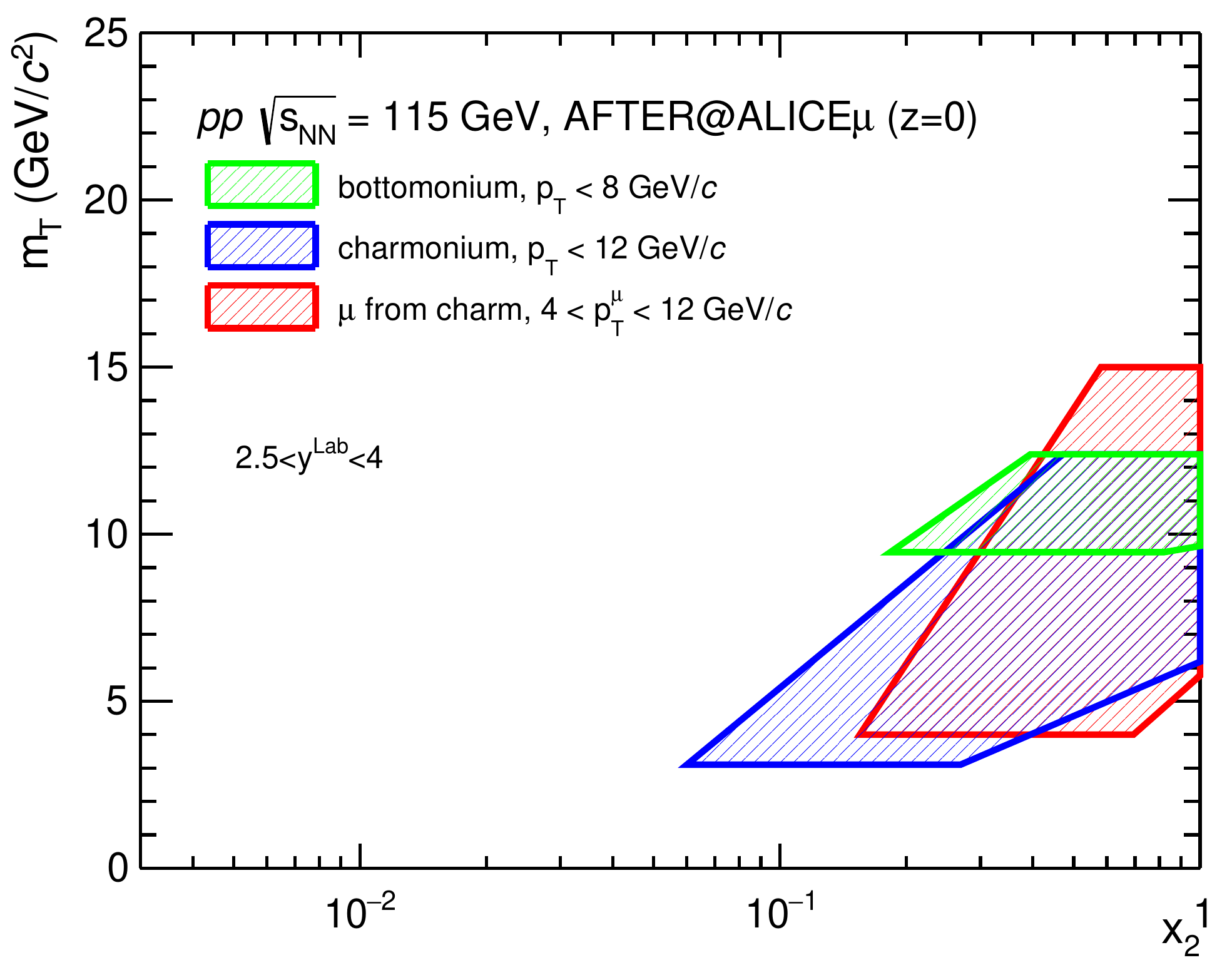}
\includegraphics[width=0.49\textwidth,angle=0]{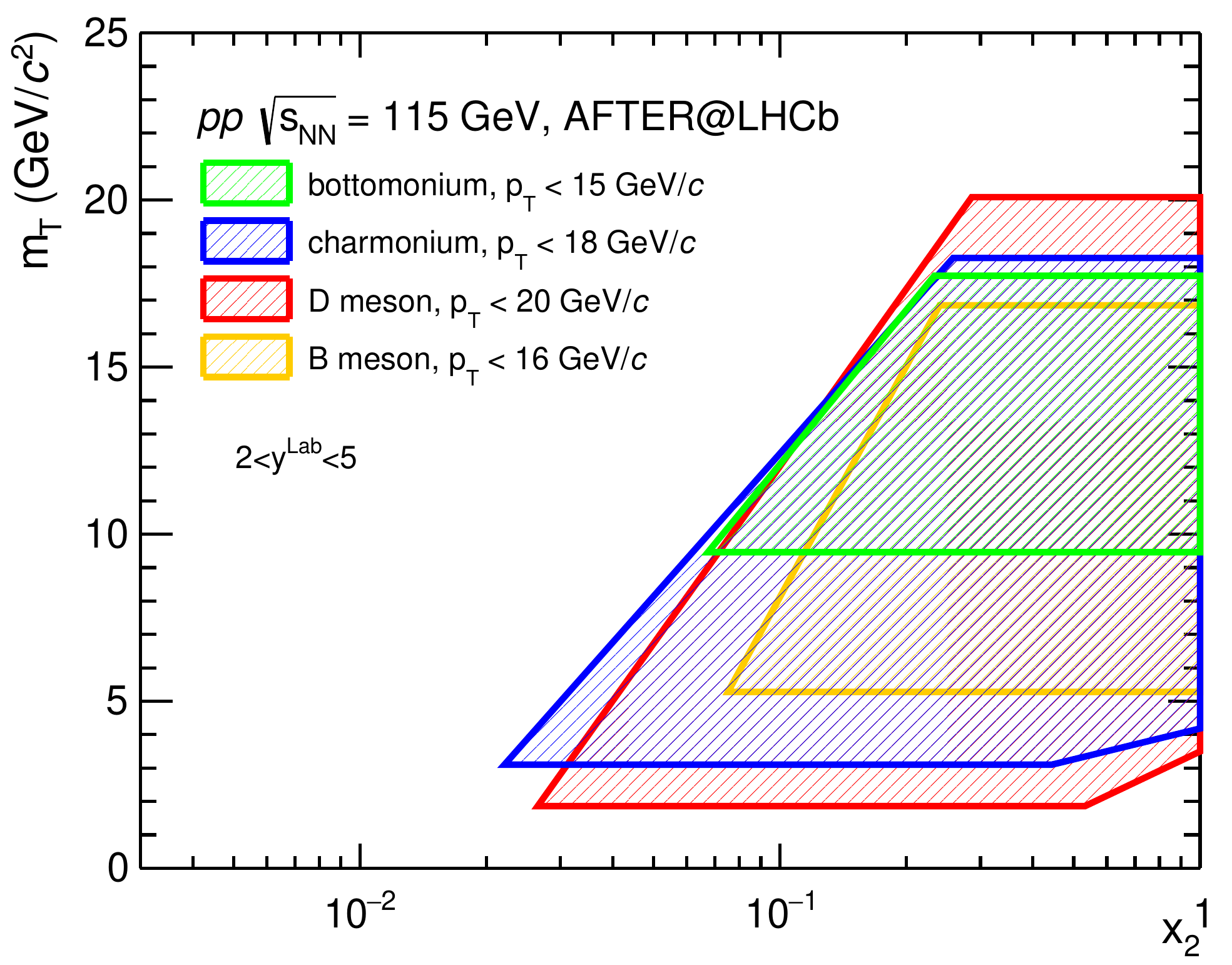}
\caption[Schematic comparison of kinematical reach for heavy-quark
production in $pp$ collision at $\sqrt{s}=115$ GeV for ALICE-FT and
LHCb-FT.]{Schematic comparison of kinematical reach for heavy-quark
  production in $pp$ collision at $\sqrt{s}=115$ GeV for ALICE-FT
  (left) and LHCb-FT (right).  (Figures from
  Ref.~\cite{Hadjidakis:2018ifr}.)}
\label{fig:xQ2-LHCb-HQ}
\end{figure}

Data on the production of mesons containing charm or bottom quarks in
LHC-FT kinematics would be valuable both to constrain the gluon
distribution at high $x$ and to investigate the presence of intrinsic
charm in the nucleon.  Figure~\ref{fig:xQ2-LHCb-HQ} shows the broad
kinematic range for heavy flavor accessible to \revised{ALICE and
  LHCb} in fixed-target mode.

It has been known for a long time that parton distributions are
modified by the nuclear environment when probed in nuclei.  The
features observed in the different \(x\) regions are shadowing,
anti-shadowing, and the "EMC effect".  A bulk of the data presently
available comes from lepton scattering by a variety of nuclei, which
restricts the flavor separation of nuclear PDFs (nPDFs).  The \DY
process can provide complementary flavor information, undisturbed by
the nuclear environment as the final-state leptons to not interact
(strongly) with the nuclear medium, in contrast to semi-inclusive DIS
in which the propagation of the struck parton and final-state hadrons
might be affected beyond the modification of the parton distribution.
\revised{Just as proton PDFs,} nPDFs are hardly constrained by data at
large \(x\).  Nuclear PDFs are intriguing by themselves but are also
important as input for the description of the initial state of
heavy-ion collisions. Furthermore, nuclear targets are often employed,
e.g., for the production of neutrinos.  A precise description of the
neutrino flux depends on the level of understanding of nuclear
modifications, especially of the gluon distribution.

Switching to polarised distribution, the three-dimensional mapping of
the nucleon structure has become an intense field of
research. Spin-dependent transverse-momentum distributions probe in a
unique way the gauge structure of the strong interaction. The
observation of the Sivers effect~\cite{Sivers:1989cc} in
semi-inclusive DIS~\cite{Airapetian:2004tw} called immediately for the
verification of the fundamental QCD prediction of a sign change when
probing the Sivers effect in the Drell--Yan
process~\cite{Collins:2002kn}. By now an extensive data set on the
Sivers and other transverse-momentum distributions has been collected
at various facilities (see, e.g., Ref.~\cite{Avakian:2016rst} for a
review), but a similar data set on polarised Drell--Yan (or the
related $W$ and $Z$ boson production) lags behind. Sparse data exist
from COMPASS on Drell--Yan~\cite{Aghasyan:2017jop} and STAR on $W$ and
$Z$ boson production~\cite{Adamczyk:2015gyk} are not yet conclusive,
and while additional data has been taken, a similar precision as in
lepton-nucleon scattering is not to be reached in the near future.

\begin{figure}
\centering
\includegraphics[width=0.65\textwidth,angle=0]{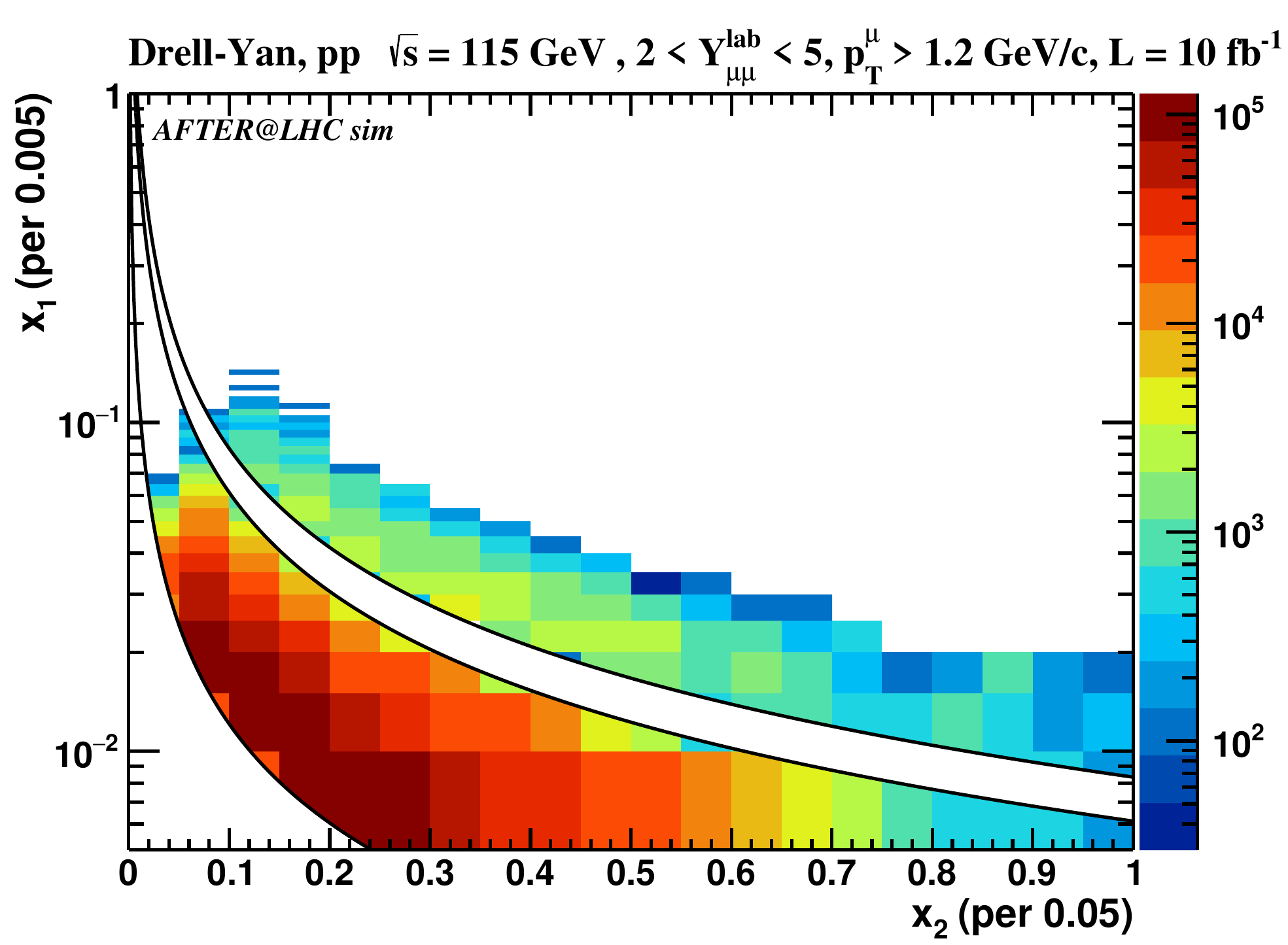}
\caption[Kinematical reach in beam and target momentum fractions for
Drell--Yan lepton-pair production in $pp$ collisions at $\sqrt{s}=115$
GeV with an acceptance of $2<\eta^{\rm lab}_\mu<5$ and $p_{T,\mu}>1.2$
GeV.]  {Kinematical reach in beam (\(x_{1}\)) and target (\(x_{2}\))
  momentum fractions for Drell--Yan lepton-pair production in $pp$
  collisions at $\sqrt{s}=115$ GeV with an acceptance of
  $2<\eta^{\rm lab}_\mu<5$ and $p_{T,\mu}>1.2$ GeV.  Each coloured
  cell contains at least 30 events. (Figures taken
  from~\cite{Hadjidakis:2018ifr}.)}
\label{fig:DYacceptance-LHCb}
\end{figure}

Drell--Yan and heavy-quark production in a fixed-target setup using
the LHC beam will greatly improve the knowledge of large-\(x\) parton
distributions.  The combination of a forward acceptance and large
centre-of-mass energy, \({\sqrt{s}}\), samples a phase space of
large-\(x\) in the target and small enough \(x\) in the beam hadron to
have copious gluons and sea quarks
(cf.~figure~\ref{fig:DYacceptance-LHCb}). This allows to study
large-\(x\) valence-quark (via Drell--Yan) and gluon (via heavy-quark
production) distributions.

\begin{figure}
	\centering
		\includegraphics[width=0.55 \textwidth]{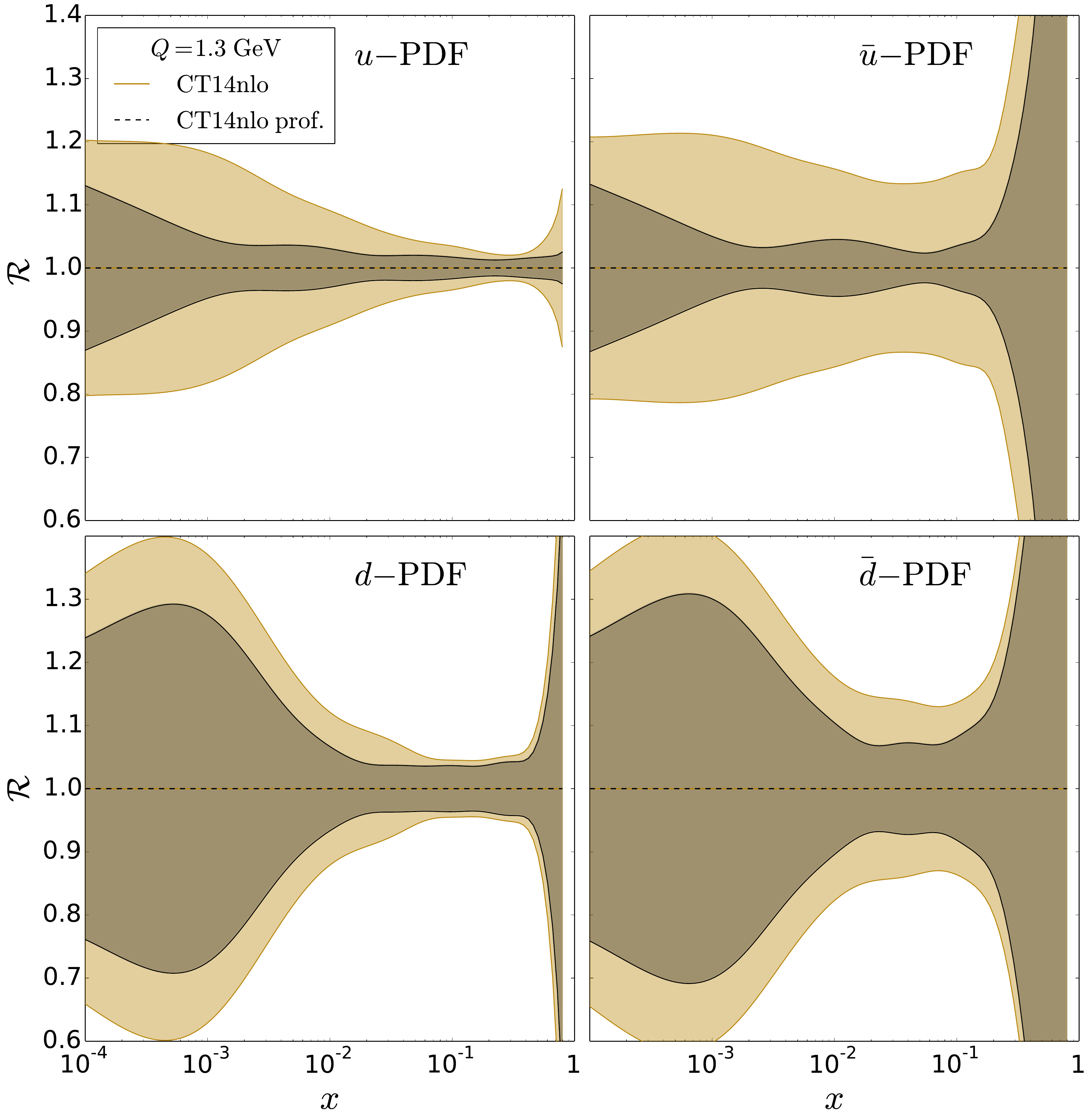}
		\includegraphics[width=0.4 \textwidth]{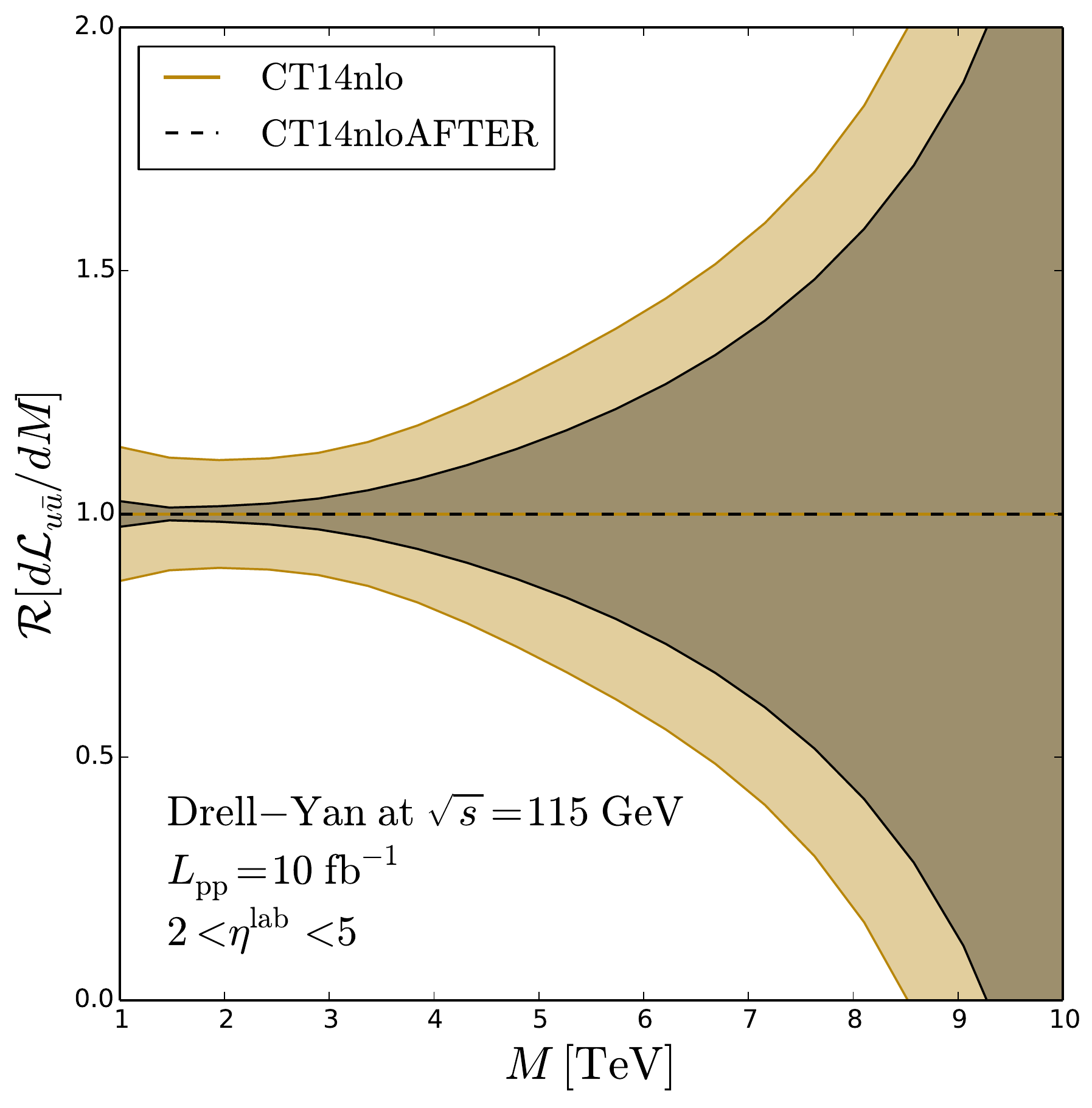}
                \caption[Proton PDF and \(u\bar{u}\) luminosity
                profiling using 10 fb$^{-1}$ of \DY data in \(pp\)
                collisions for the LHCb fixed-target conditions.]{ (Left panels)
                  Proton PDF profiling using 10 fb$^{-1}$ of \DY data
                  in \(pp\) collisions for the LHCb fixed-target conditions. The
                  smaller darker band indicates the reduction in the
                  CT14~\cite{Dulat:2015mca} NLO PDF uncertainty when
                  the pseudo-data are accounted for.  (Right)
                  Corresponding improvement of \(u\bar{u}\)
                  luminosities.  (Left figures taken
                  from~\cite{Hadjidakis:2018ifr}, right figure
                  from~\cite{Kusina:20181128}.)  }
	\label{fig:quarkPDFforLHCb}
\end{figure}

\begin{figure}
  \centering \includegraphics[width=1.0\textwidth]{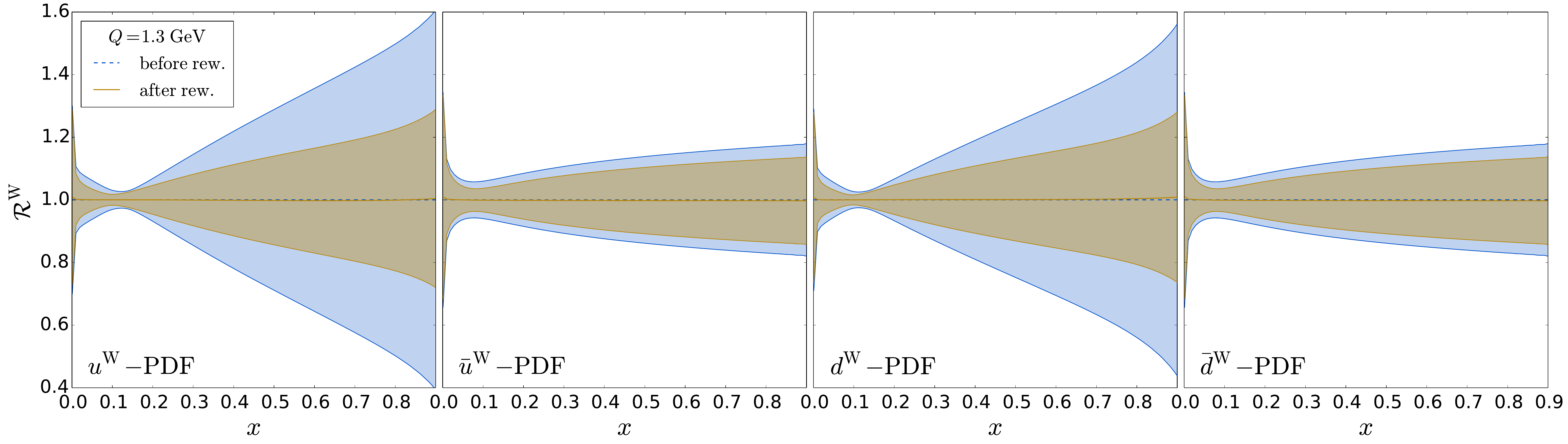}
	\caption[Impact of 10 fb$^{-1}$ of $pp$ and 100 pb$^{-1}$ of \(p\)Xe collisions on nuclear-modification factors 
	\(R_{q}^{\text{W}}\) (\(q=u,~\bar{u},~d,~\text{and}~\bar{d}\)) from Drell--Yan data for LHCb fixed-target conditions.]{
	Impact of 10 fb$^{-1}$ of $pp$ and 100 pb$^{-1}$ of  \(p\)Xe collisions on nuclear-modification factors 
	\(R_{q}^{\text{W}}\) (\(q=u,~\bar{u},~d,~\text{and}~\bar{d}\)) from Drell--Yan data for LHCb fixed-target conditions.
	(Figures taken from~\cite{Hadjidakis:2018ifr}.)}
	\label{fig:W-quarkPDFforLHCb}
\end{figure}

\begin{figure}
	\centering
	\includegraphics[width=0.75 \textwidth]{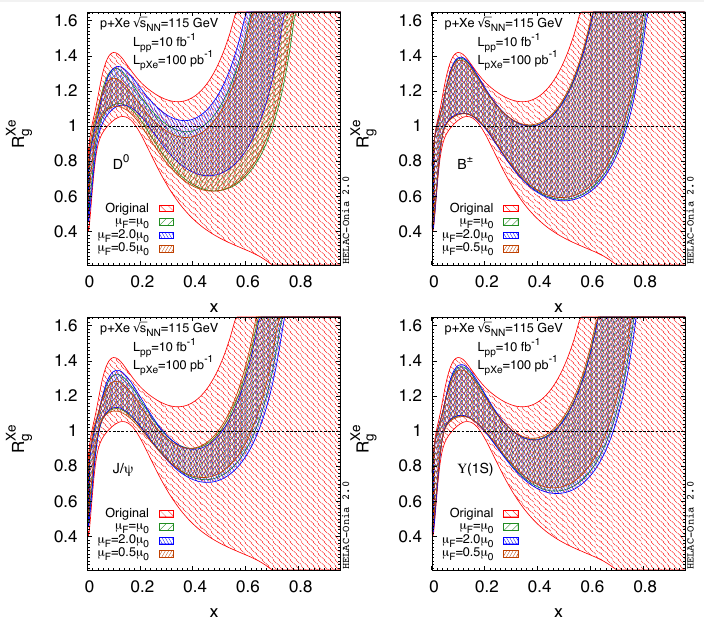}
	\caption[Impact of 10 fb$^{-1}$ of $pp$ and 100 pb$^{-1}$ of  \(p\)Xe collisions on nuclear-modification factors 
	\(R_{g}^{\text{Xe}}\) through various gluon-sensitive probes for LHCb fixed-target conditions.]
	{Impact of 10 fb$^{-1}$ of $pp$ and 100 pb$^{-1}$ of  \(p\)Xe collisions on nuclear-modification factors 
	\(R_{g}^{\text{Xe}}\) through gluon-sensitive probes (\(D^{0}\), \(B^{\pm}\), \(J/\psi\), and \(\Upsilon (1S)\) production)
	for LHCb fixed-target conditions. (Figures taken from~\cite{Hadjidakis:2018ifr}.)}
	\label{fig:Xe-gluonPDFforLHCb}
\end{figure}

As an example, figure~\ref{fig:quarkPDFforLHCb} demonstrates the
impact on the proton PDFs using Drell--Yan data in \(pp\) collision.
Likewise, Figs.~\ref{fig:W-quarkPDFforLHCb} and
\ref{fig:Xe-gluonPDFforLHCb} illustrate the improvement in precision
on the nuclear modification of quark and gluon distributions,
expressed as the nuclear-modification factor $R_{q/g}^{\text{A}}$ of
nuclei (A) to proton PDFs, using Drell--Yan and heavy-quark (here
\(D^{0}\), \(B^{\pm}\), \(J/\psi\), and \(\Upsilon (1S)\)) production
in $pA$ collision, respectively~\cite{Hadjidakis:2018ifr}.  Note that
for the quark distributions, tungsten was used in above simulation for
convenience.  Prompt-photon production and $W$ production may provide
further constraints on the gluon and anti-quark distributions,
respectively, both in protons and nuclei.  In above examples, a
realisation of a fixed-target experiment using the LHCb
detector~\cite{Alves:2008zz} was assumed (see also
Section~\ref{sec:LHCb-FT}).  Such a program could be pursued with
moderate modifications already during LHC run 3.

\begin{figure}
\centering
\includegraphics[width=0.59\textwidth,angle=0]{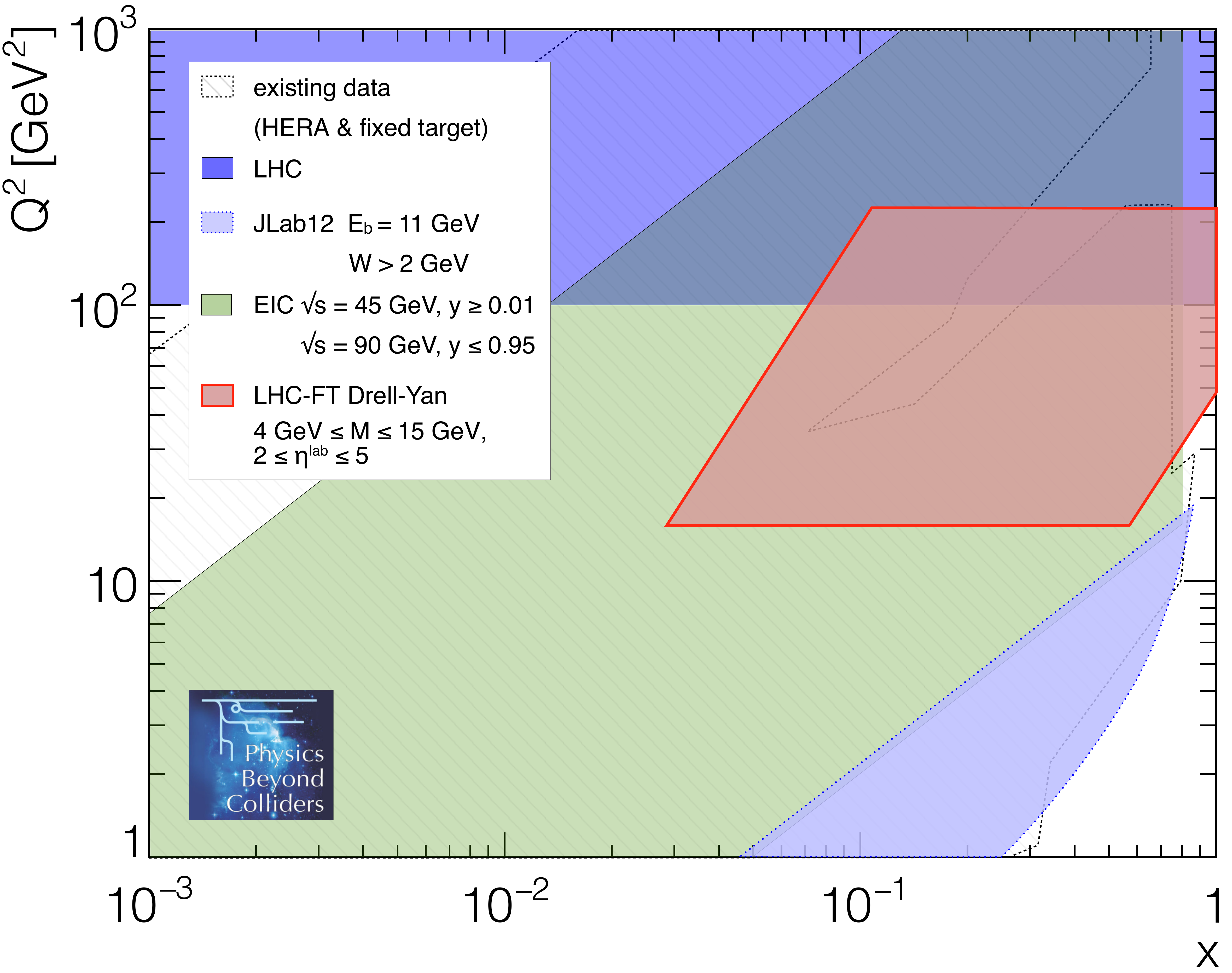}
\includegraphics[width=0.59\textwidth,angle=0]{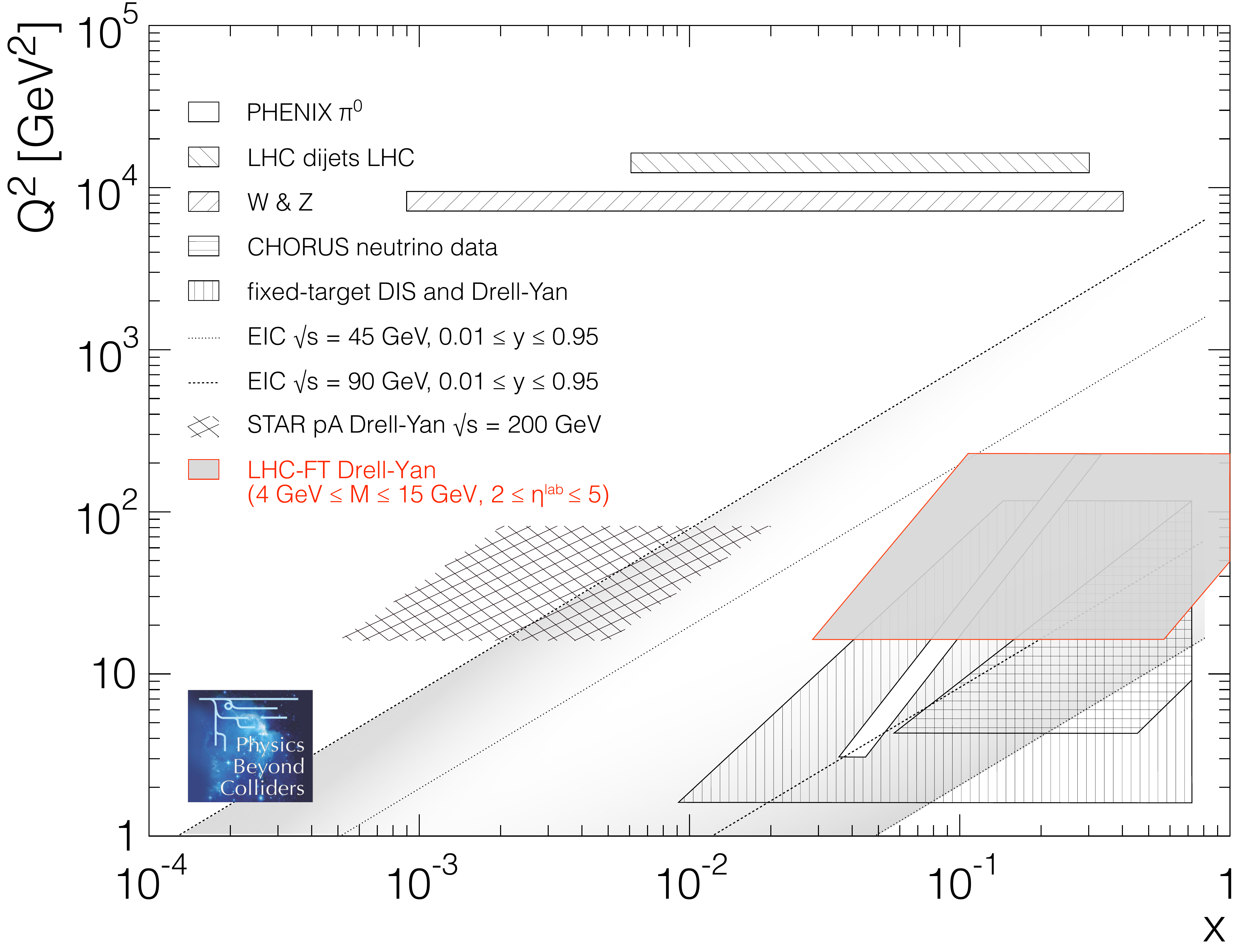}
\caption[Schematic comparison of kinematical reach in \(x\) and
\(Q^{2}\) for Drell--Yan lepton-pair production in $pp$ and \(p\)Xe
collisions at $\sqrt{s}=115$ GeV with typical data sets used in proton
and nuclear PDF fits.]{Schematic comparison of kinematical reach in
  \(x\) and \(Q^{2}\) for Drell--Yan lepton-pair production
  ($4~\text{GeV}< M<15~\text{GeV}$) in $pp$ and \(p\)Xe collisions at
  $\sqrt{s}=115$ GeV with an acceptance of $2<\eta^{\rm lab}<5$ with
  typical data sets used in proton (top) and nuclear (bottom) PDF
  fits. (bottom figure inspired by~\cite{Eskola:2016oht}.) }
\label{fig:xQ2-LHCb}
\end{figure}

Figure~\ref{fig:xQ2-LHCb} compares schematically the kinematic reach
(here \DY) for a fixed-target setup using the LHCb detector for both
proton (left) and nuclear (right) PDFs with the kinematic coverage of
relevant current and future data sets. \revised{As highlighted
  already, LHC-FT can access the large $x$ region with high
  statistics.  Measurements at JLab 12 are also targeting large $x$,
  but at smaller $Q^2$, so that the information from the respective
  data will be complementary in PDF fits.  Vigorous PDF studies are
  foreseen for the HL-LHC \cite{Khalek:2018mdn} and for future
  lepton-hadron colliders like EIC and LHeC.\footnote{See e.g.\ the
    presentations by E.~C.~Aschenauer and P.~Newman at
    \url{http://www.int.washington.edu/talks/WorkShops/int_18_3}.}
  Given the respective timescales of these programs, the projected
  performance of LHC-FT measurements is highly competitive.}

\begin{figure}
	\centering
	\includegraphics[width=0.95 \textwidth]{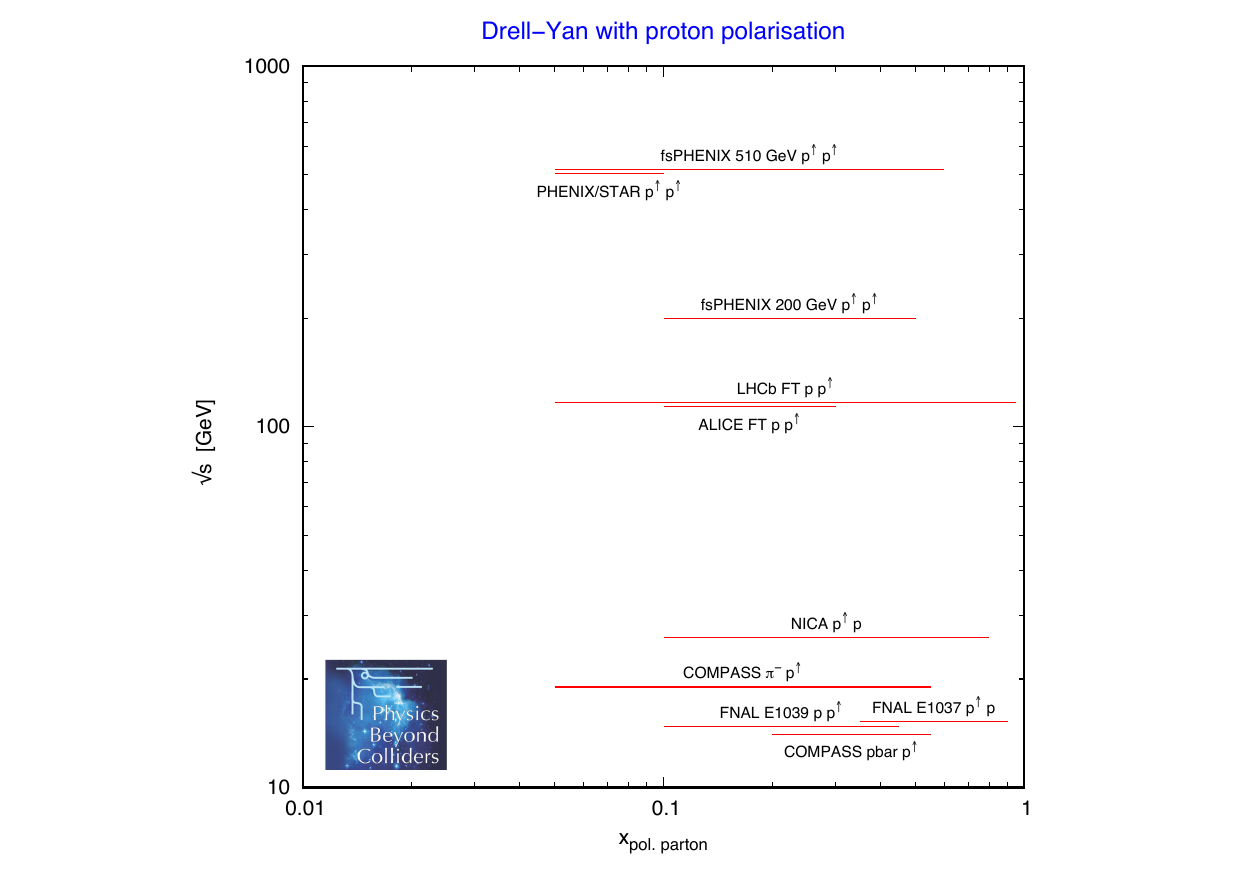}
	\caption[Kinematic reach of future or planned polarised \DY
        experiments.]{Kinematic reach of future or planned polarised
          \DY experiments (from Table~16 in
          Ref.~\cite{Hadjidakis:2018ifr}).}
	\label{fig:DY-SSA-projects}
\end{figure}

\begin{figure}
	\centering
	\includegraphics[width=0.52 \textwidth]{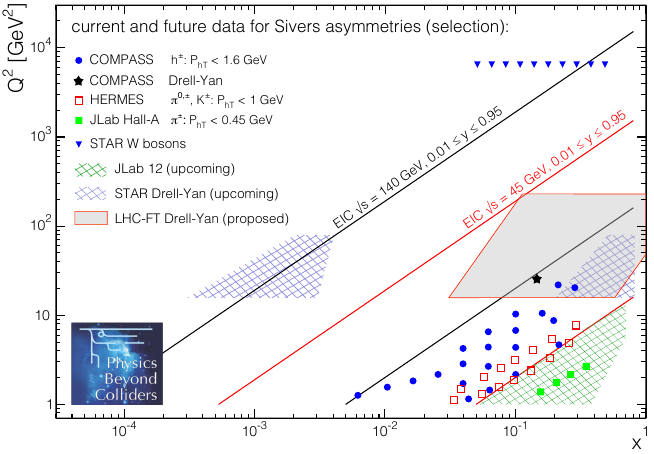}
	\includegraphics[width=0.47 \textwidth]{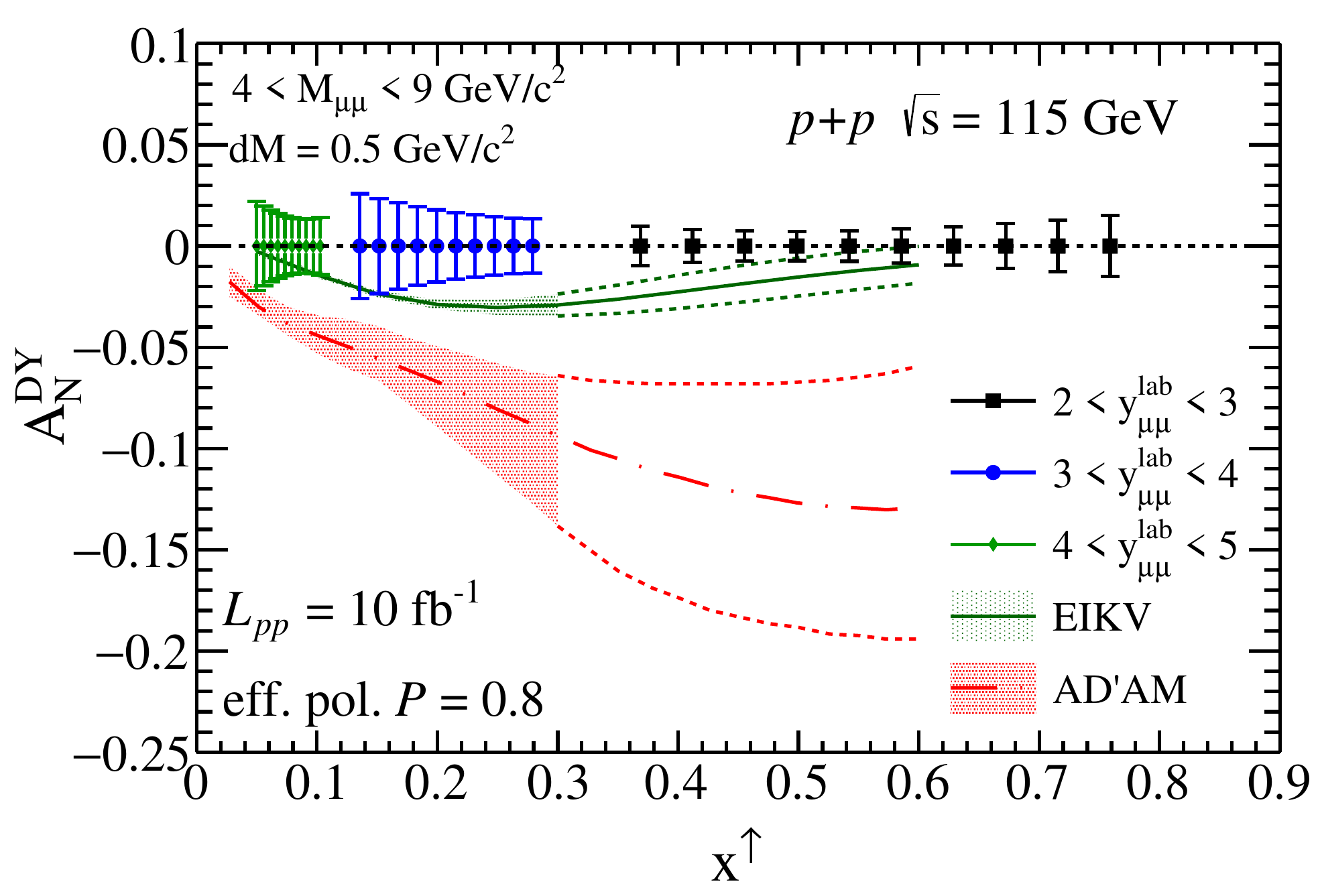}
	\caption[Schematic comparison of kinematical reach in \(x\)
        and \(Q^{2}\) for Drell--Yan lepton-pair production in
        single-polarised $pp$ collisions at $\sqrt{s}=115$ GeV with
        selected current and upcoming data sets on Sivers asymmetries
        and expected precision on Sivers Drell--Yan asymmetries for
        LHCb fixed-target conditions.]{(left) Schematic comparison of kinematical reach in
          \(x\) and \(Q^{2}\) for Drell--Yan lepton-pair production
          ($4~\text{GeV}< M<15~\text{GeV}$) in single-polarised $pp$
          collisions at $\sqrt{s}=115$ GeV and an acceptance of
          $2<\eta^{\rm lab}<5$ with selected current and upcoming data
          sets on Sivers asymmetries.  (right) Expected precision on
          Sivers Drell--Yan asymmetries for LHCb-FT (\rev{LHCSpin}).
          Data points are for different ranges in laboratory rapidity
          (different symbols, as labelled) and increasing with \(x\)
          for increasing invariant mass of the lepton pairs
          ($4~\text{GeV}< M<9~\text{GeV}$).  Phenomenological curves
          are based on
          Refs.~\cite{Echevarria:2014xaa,Anselmino:2015eoa}.
          (\rev{Left} figure adapted
          from~\cite{Accardi:2012qut,Aschenauer-INT18-3}, \rev{right}
          one taken from~\cite{Hadjidakis:2018ifr}.)}
	\label{fig:Sivers-DY-A_N-forLHCb}
\end{figure}

Measurements with a polarised target will require more substantial
modifications and are thus not expected to commence before LHC run 4.
Even then, the physics impact is still expected to be large in view of
the world-wide landscape (see, e.g.,
figure~\ref{fig:DY-SSA-projects}).  As an example, the expected
precision for the Drell--Yan transverse single-spin asymmetry is
compared in figure~\ref{fig:Sivers-DY-A_N-forLHCb} to predictions
based on two different phenomenological fits to lepton-nucleon
scattering data. Note that for a significant part of the \(x\) region
probed in such an experiment no data are yet available. This is
expected to improve with data to come from the upgraded Jefferson Lab,
albeit at relatively low \(Q^{2}\). Furthermore, Drell--Yan with
polarised beams at RHIC and at an even longer time-scale the
Electron-Ion Collider will also contribute.

Beyond Drell-Yan measurements, heavy flavour production can be used to
explore the Sivers asymmetry in the gluon sector.  Last but not least,
a wide range of azimuthal correlations and hadronic final states allow
studies of a variety of transverse-momentum distributions and the
transversity distribution at large \(x\)~\cite{Hadjidakis:2018ifr}.

\subsubsection{LHCb-FT}\label{sec:LHCb-FT}

LHCb is planning to pursue its fixed-target program during Run3 after
improving the gas target setup (known currently as SMOG).  Its
upgrade, SMOG2 \revised{\cite{CERN-PBC-Notes-2018-007}}, aims at
containing the injected gas inside a storage cell.  Because of that
the gas density can be increased by up to two orders of magnitude
without changing the impact on the LHC operations.  \revised{This has
  recently been approved within the Collaboration to be installed
  during the LHC Long Shutdown II.}  The goal is also to extend the
choice of gas species that can be injected, including notably hydrogen
and deuterium.  The $pp$ collisions in fixed-target mode would be a
reference for all $pA$ collision samples. Moreover, H and D targets
would provide additional inputs to the study of cosmic ray propagation
in the interstellar medium (cf.~section~\ref{sec:cosmics}) and allow
to extend the physics case to the study of the three-dimensional
structure of the nucleon through spin-independent observables.
Another \revised{important} advantage of the new target setup is the
accurate control of the injected gas density, so that the acquired
luminosity can be determined to a better precision with respect to the
present setup.

\revised{It is clear that the physics reach strongly depends on the
  achievable integrated luminosity. In an ambitious program of
  collecting several fb$^{-1}$ (as assumed for many of the performance
  studies shown in this section), a breakthrough advance in the
  understanding of parton distribution functions for gluons,
  antiquarks and heavy quarks at $x>0.5$, where they are now almost
  unconstrained, can be expected.
  Such integrated luminosities could be collected in principle by
  running continuously with the maximum instantaneous luminosities of a
  few $10^{32}$ cm$^{-2}$\,s$^{-1}$ allowed by the SMOG2 target.  By
  contrast, the running scenario currently envisaged by the LHCb
  Collaboration is aiming at fixed-target data of the order of
  0.1~fb$^{-1}$ per year.  This can be realised with minimal
  interference with the core LHCb program, given the common trigger
  and DAQ bandwidth and, particularly for heavy targets, the
  contamination of the LHC vacuum.
  Samples of this size would already allow copious production of
  Drell-Yan and heavy-flavor states, including $b{\overline b}$ mesons
  (see, for instance, Table~\ref{tab:SMOG2charm} or
  Ref.~\cite{Bursche:2649878}). LHCb-FT with nuclei could in principle
  be used for heavy-ion physics (cf.~section~\ref{sec:HIC-Summary})
  and corresponding studies are to be performed.}

\paragraph{Implementation:}
The main challenges are related to the optimisation of the target
design and to the integration of the data acquisition and offline
computing strategies with the core LHCb program.  The latter requires
a substantial effort that will be implemented in progressive stages.

\paragraph{Estimated cost and time scale:}
The cost of the proposed new target setup is a minor fraction of the
approved LHCb upgrade project.  The collaboration is also evaluating
more ambitious proposals that could be realised at LHCb on the time
scale of LHC Run 4, including the polarised target
(cf.~section~\ref{sec:LHC-SPIN}) and the crystal target
(cf.~section~\ref{sec:Crystal}) proposals.

\subsubsection{\rev{LHCSpin}}\label{sec:LHC-SPIN}

The project has several ambitious goals rewarding new-era quantitative
searches in QCD, through the study of nucleons internal dynamics.  By
using a transversely polarised gaseous target, one of the main
measurements will allow to study quark transverse-momentum
distributions (TMDs) in proton-proton collisions at the unique
kinematic conditions described above through 
single transverse spin asymmetries.

The LHCb detector has been designed and optimised for heavy-flavor
physics, so final states with $c$ or $b$ quarks, carrying information
on the gluon dynamics inside nucleons, will be efficiently
reconstructed and will allow to extend the measurements of spin
asymmetries to the hardly known gluon sector.

Due to the possibility of parallel data taking with the LHCb main
stream physics, it is possible to project statistical uncertainties
for different data taking periods.  As an example, for Drell-Yan
\revised{transverse single-spin asymmetries} in 4, 12, and 24 months
of data taking, considering 8h/day and 20 days/month data of duty
time, absolute statistical uncertainties of 0.015, 0.01, and 0.008 can
be reached, respectively.

\paragraph{Implementation:}
The proposed target is based on the technology that has been applied
successfully at the 27.6~GeV HERA electron ring in 1995 to 2005 for
the HERMES experiment~\cite{Airapetian:2004yf}.  The target consists
of a chamber with an open T-shaped tubular structure, the Storage Cell
(SC), into which a thermal beam of H atoms from an Atomic Beam Source
(ABS) is injected and radially trapped.  The LHC beam passes the cell
through the straight beam tube.  The SC can be opened during injection
and kept at different temperatures in order to optimise density and
polarisation.  A transverse magnetic field ($\sim$300 mT) defines the
direction of polarisation.  A sample of gas is continuously extracted
from the SC and analysed on-line by a polarimeter with respect to
nuclear polarisation and atomic fraction, a key parameter related to
the degree of polarisation. This allows to determine the polarisation
of target atoms as seen by the beam along the SC to a precision of a
few \%.  The SCÕs straight beam tube is assumed to be 10 mm in
diameter and 300 mm in length, coated with \revised{Amorphous Carbon}.
By running the SC at about 100 K, an ice layer is formed providing an
optimum surface which, according to the experience at HERMES,
suppresses atomic recombination quantitatively, a major cause of
depolarisation.  At 100 K and with an injected flow rate of
6.5$\times$10$^{16}$ H/s, an areal density of more than 10$^{14}$
H/cm$^{2}$ can be obtained.  Assuming 1A for the HL-LHC proton-beam
current an instantaneous luminosity of close to 10$^{33}$
cm$^{-2}$\,s$^{-1}$ could be achieved~\cite{Steffens:2015kvp}, which
corresponds to about 10 \revised{fb$^{-1}$} per year.  Whether LHCb
will be able to cope with the corresponding event rate in addition to
the collider data is an issue to be investigated.  An advantage of a
gaseous target is, however, the absence of diluting material. This is
important for event-rate limited situations, as the event rate from
the polarised target does not explode artificially due to presence of
unpolarized scattering centres.

A different limitation might be the presently foreseen location
upstream of the nominal LHCb IP.  It will---in the present detector
configuration---reduce substantially the coverage in rapidity and thus
at high-$x$ of the partons in the polarised (target) nucleon and needs
to be investigated further.

Various effects are already being studied that may disturb a proper
functioning of the polarised gas target, including wake fields and
wall impedance, instabilities due to electron clouds, and
depolarisation of the target gas by the bunch fields. No showstoppers
have been found to date. Work on a conceptual design is in progress
and discussions with machine experts are ongoing. The aim of these
studies is to have a prototype target chamber ready for installation
during LS3, enabling in-situ tests with gas injection. This set-up
could be extended in steps by adding the ABS and the polarimeter
during shorter stops of the machine.

A detailed R\&D program, \revised{which benefits from the experience
  achieved with the installation and running of the unpolarised fixed
  target SMOG2,} is going on and will take at least three years for
addressing all the important issues connected to the implementation of
the target system into the LHC. After this process, and if no
showstopper has been identified, LHC Spin will be proposed for
installation during the LHC LS3.

\paragraph{Collaboration strength:}
The main proponents form a group of experts with long experience on
the polarised physics, HEP detectors and data analysis. A growing list
of institutes includes, at the moment: INFN and University of Cagliari
(Italy), INFN and University of Ferrara (Italy), INFN-Frascati
(Italy), J{\"u}lich Laboratory (Germany), High Energy Physics
Laboratory Gatchina (Russia), University of Erlangen (Germany),
University of Virginia (US). Other groups from INFN Italy and in EU
expressed their interest. For this purpose, a parallel discussion
session \revised{has been} organised during the SPIN2018 international
conference organised by the group in Ferrari. \revised{LHCSpin had}
also been invited, in October 2018, to present the project at the
Institute of Nuclear Theory, Seattle for involving the relevant
American community, already present with few groups, and a dedicated
meeting will be organised in Russia due to the expressed significant
interest of the Russian colleagues.

\paragraph{Estimated cost and time scale:}
A rough estimation of the cost for the hardware R\&D and the final
construction is of the order of \revised{2 M\euro.}

\subsubsection{ALICE-FT}\label{sec:ALICE-FT}

One of the main strengths of ALICE in fixed-target mode would be its
large rapidity coverage. Assuming a target location at $Z=0$, the
ALICE muon spectrometer (and future Muon Forward Tracker) would access
the mid- to backward rapidity in the centre-of-mass frame
($-2.3 < y_{\text{cms}} < -0.8$).\footnote{This is considering an
  incident proton beam on the target. For an incident lead beam, the
  muon spectrometer rapidity coverage is
  $-1.8 < y_{\text{cms}} < -0.3$ and central barrel coverage
  $-5.2 < y_{\text{cms}} < -3.4$.  \rev{Here, rapidities are computed
    for massless particles.}}
In addition, the long absorber in front of the muon tracking station
is an asset for background rejection and Drell--Yan studies at low
energy. The ALICE central barrel offers a complementary coverage to
the muon arm by accessing the very backward rapidity region
($-5.7 < y_{\text{cms}} < -3.9$), reaching the end of phase space for
several \revised{processes}.  Thanks to its excellent particle
identification capabilities, particle detection and identification
down to low $p_T$, unique measurement of soft probes and open heavy
flavors can be pursued.  Another asset of the ALICE apparatus is the
capability to operate with good performance in a high particle density
environment.  Access to most central AA collisions at
$\sqrt{s_{NN}} = 72$ GeV should be possible if the interaction rate
remains low.  In addition, the ALICE Collaboration could potentially
devote a significant data taking time to a fixed-target program
(especially with the proton beam), allowing the collection of large
integrated luminosities and the investigation of several target
species.  Two main solutions are being investigated to deliver
fixed-target collisions to ALICE: an internal gaseous target or an
internal solid target (coupled to a bent crystal to deflect the beam
halo).  On the one hand, a gas-jet or a storage cell (with levelled
gas pressure) would allow to deliver about 45 pb$^{-1}$ of
proton-polarised hydrogen collisions to ALICE per year (\revised{260
  pb$^{-1}$} in case of proton-unpolarised H$_{2}$ collisions), and 8
nb$^{-1}$ of Pb-Xe collisions.  With such a system the target can be
polarised, but requires large space to be installed (most likely
outside the ALICE barrel magnet), requiring thus the need of
additional detectors for vertexing and additional studies of the
tracking performance of the TPC in such conditions.  A simple
unpolarized storage cell might potentially be used \revised{closer to
  the ALICE IP.}  On the other hand, the usage of an internal solid
target coupled to a bent crystal has the advantage of more
portability, allowing one to install the target closer to the nominal
ALICE IP and thus benefiting of the optimal performance of the current
ALICE apparatus. With such a device 37 pb$^{-1}$ (6 pb$^{-1}$) of
$p$-C ($p$-W) collisions, and 5 nb$^{-1}$ (3 nb$^{-1}$) of Pb-C (Pb-W)
collisions could be registered in ALICE per year.  \revised{Studies
  are ongoing on the integration and compatibility of a fixed target
  with the ALICE apparatus and its operation in collider mode.}


\subsection{LHC-FT: crystals}
\label{sec:Crystal}

The LHC-FT crystal proposal discussed in the QCD working group aims at
measuring the magnetic dipole moment (MDM) of short-lived baryons that
contain a charm or a bottom quark.  The key experimental element is to
use a setup with bent crystals inside the LHC.  Such a setup might
also be used for measuring the electric dipole moments of the same
hadrons, which would be a probe of $CP$ violation.  Both possibilities
are described in \cite{Bagli:2017foe}.  Recently, the possibility to
use the same technology for measuring the MDM of the $\tau$ lepton has
been discussed in \cite{Fomin:2018ybj}.

\begin{figure}
\begin{center}
\includegraphics[height=0.43\textwidth,trim=65 0 65 0,clip]{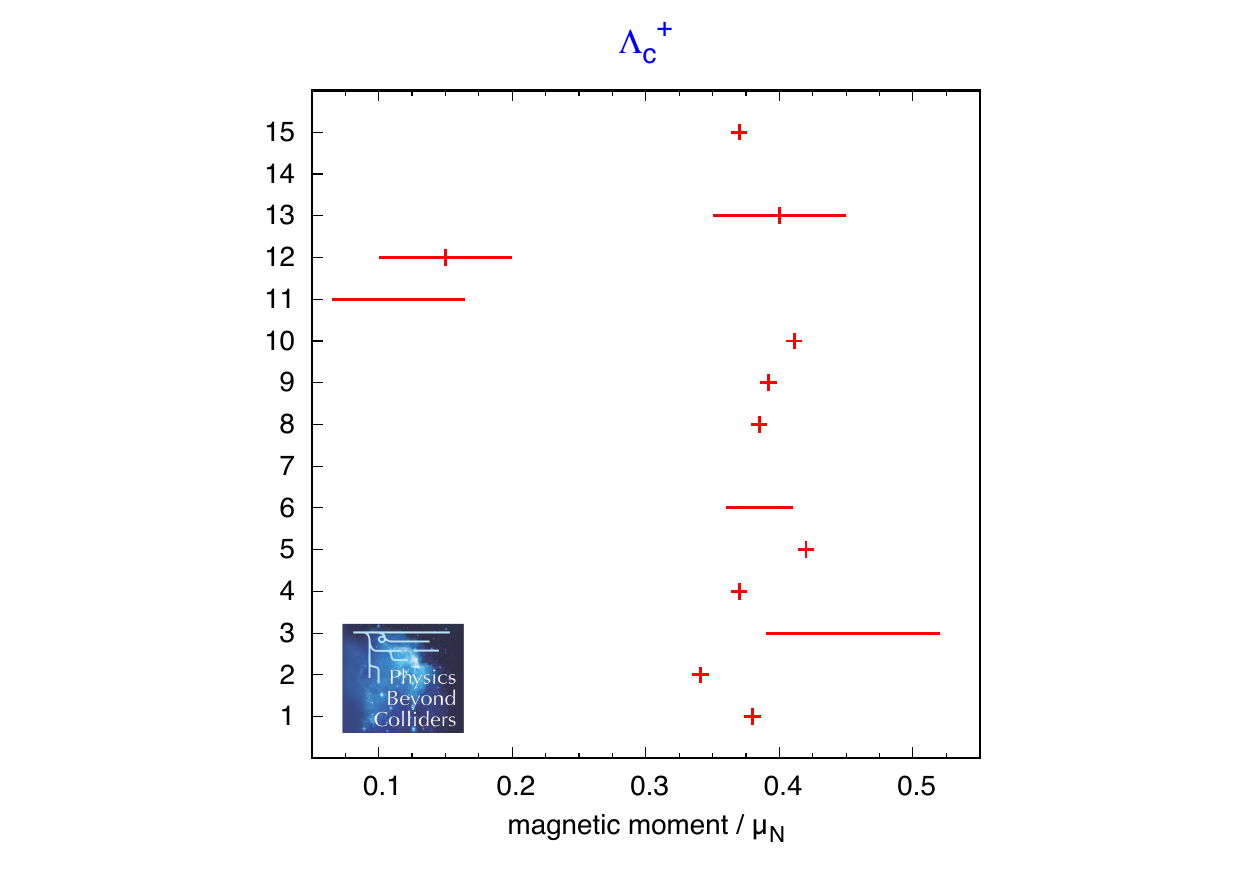}
\includegraphics[height=0.43\textwidth,trim=0 0 30 0,clip]{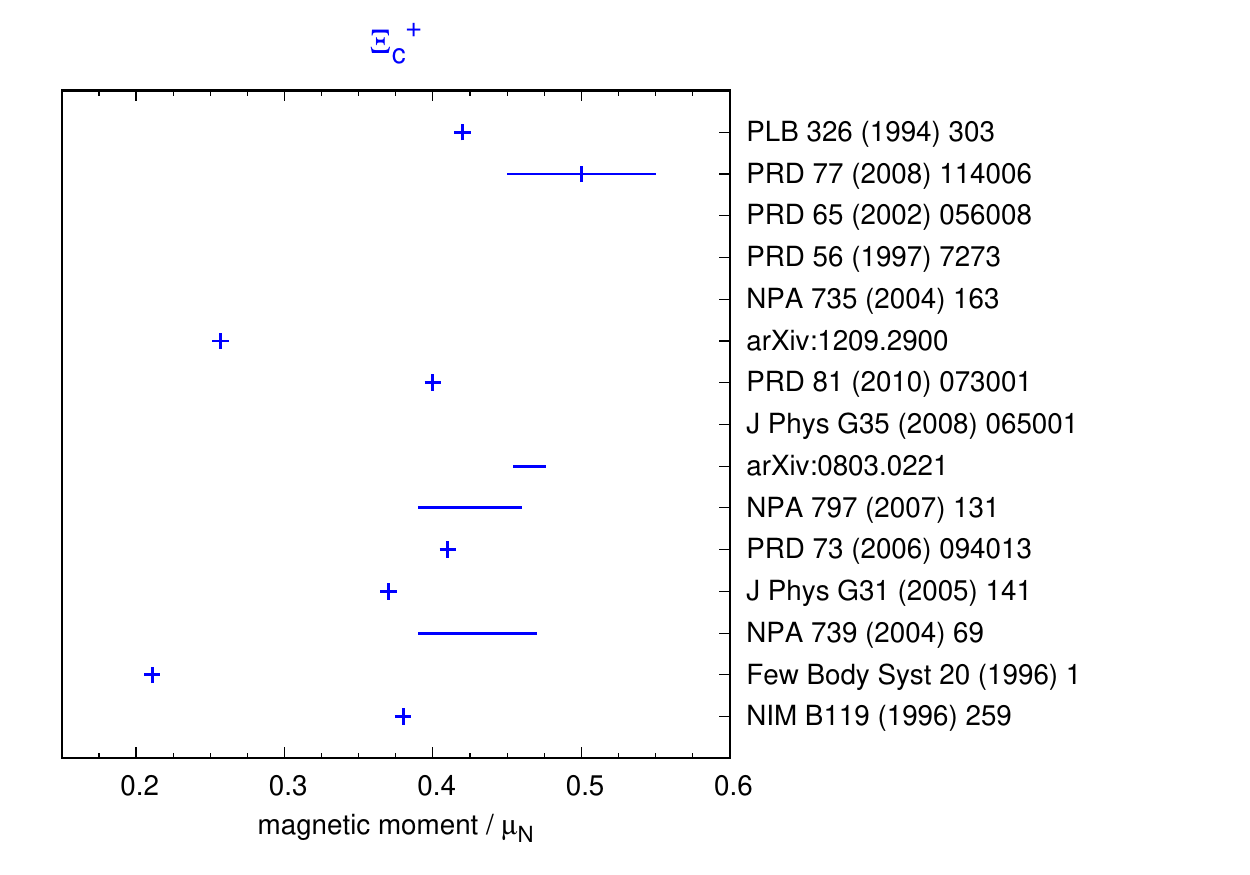}
\includegraphics[height=0.43\textwidth,trim=65 0 65 0,clip]{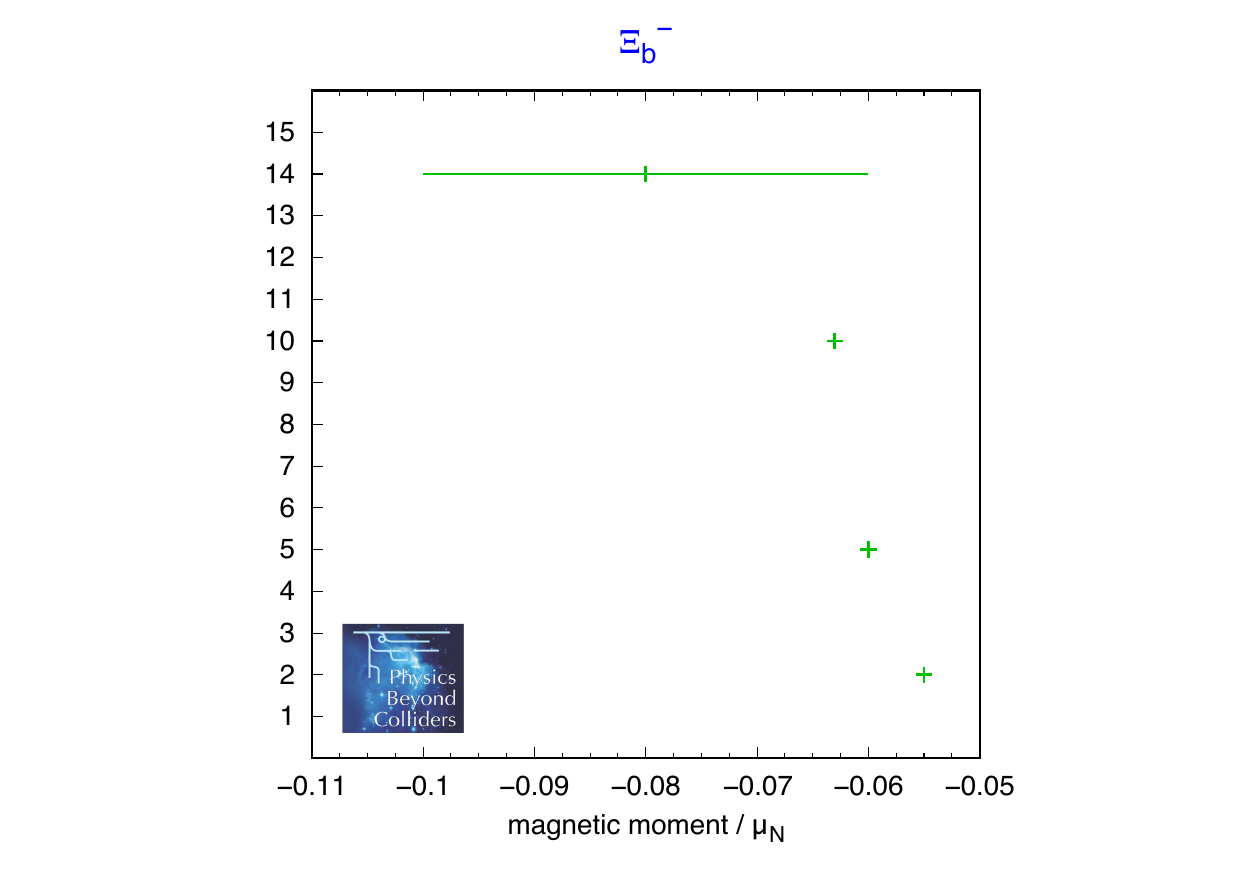}
\includegraphics[height=0.43\textwidth,trim=0 0 30 0,clip]{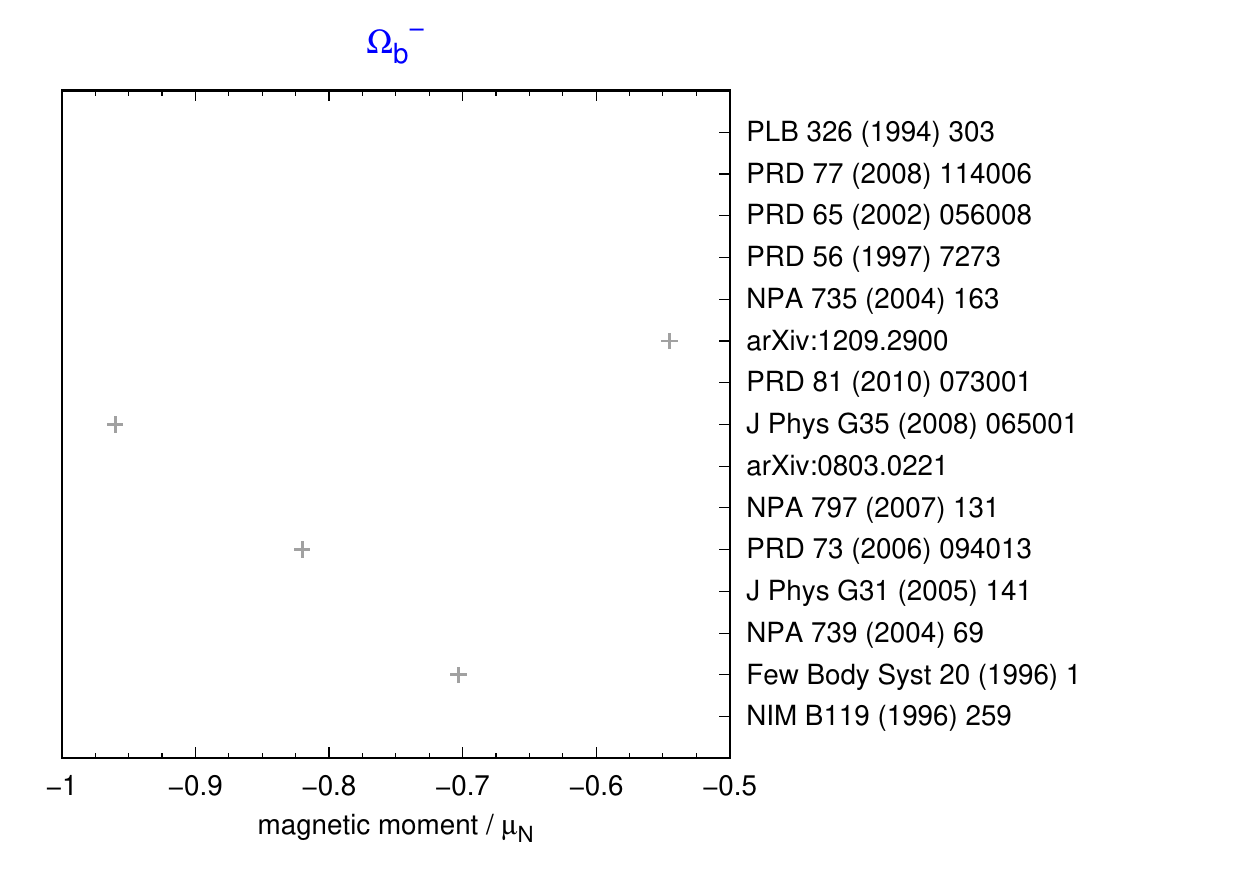}
\caption[Spread of theoretical predictions for the magnetic moments of
heavy baryons.]{\label{fig:magmom-thy} Spread of theoretical
  predictions for the magnetic moments of heavy baryons.  Values in
  rows 1 to 10 are from different quark models, 11 from a soliton
  approach, 12 to 14 from sum rules, and 15 from chiral perturbation
  theory.}
\end{center}
\end{figure}

\paragraph{Physics motivation.}
Hadrons consisting of both light and heavy quarks play a special role
in QCD: their binding is dominated by large-distance, non-perturbative
dynamics, whilst the presence of a heavy quark mass brings in a
simplifications due to heavy-quark symmetry.  Next to their masses,
the magnetic moments of baryons are among the most prominent
quantities characterising their static properties.  The significant
range of model predictions for the magnetic moments of selected heavy
baryons is shown in figure~\ref{fig:magmom-thy}.  Further model
comparison and discussion can for instance be found in
\cite{Faessler:2006ft,Sharma:2010vv,Bernotas:2012nz}.  \revised{Of
  particular interest} are modern effective theory approaches based on
the heavy quark limit, e.g., the prediction in row 15 of the figure,
which was obtained in heavy-baryon chiral perturbation theory.  The
prospect of having experimental data for magnetic moments might
trigger further theory activity in this direction.

\paragraph{Experimental setup.}

The possibility to measure the MDM of heavy and strange baryons at the
LHC has been explored in recent
years~\cite{Baryshevsky:2016cul,Burmistrov:2194564,Bezshyyko:2017var,Bagli:2017foe}
and relies on the precession of their spin in external electromagnetic
fields. The experimental setup is based on three main elements: i) a
source of polarised particles with known direction and polarisation,
ii) an intense electromagnetic field able to induce a sizeable spin
precession during the lifetime of the particle, iii) a detector to
measure the final polarisation vector by analysing the angular
distribution of the decay particles.  For short-lived,
positively-charged charm and beauty baryons, e.g., $\Lambda^+_c$ ,
$\Xi^+_c$, $\Xi^+_b$ and $\Omega^+_b$, an intense electromagnetic
field, such as the one between the atomic planes of bent crystals, is
necessary to induce spin precession. For this purpose, a fixed-target
experiment is proposed to extract protons from the LHC beam halo with
a bent crystal. These protons will then hit a dense target and produce
charged heavy baryons that will then be channelled in bent crystals
positioned in front of the detector.

The LHC interaction point IP8, where the LHCb detector sits, has been
identified as a most suitable location of the experiment. This is
motivated by the fixed-target like geometry of the apparatus, the only
one operating at LHC that is fully instrumented in the forward region
($2 < \eta < 5$).  A main challenge of the experiment is represented
by the limited coverage of the LHCb detector acceptance in the very
forward region, approximately greater than 15 mrad, and the limited
particle identification information for high-momentum tracks above 100
GeV. Detector simulations have been conducted to demonstrate the
feasibility of the experiment under realistic operational conditions
and to estimate
sensitivities~\cite{Bagli:2017foe,Bezshyyko:2017var}. The clear signal
signature combined with precise kinematical information compensates
the lower vertex reconstruction performance from the upstream
configuration.

\paragraph{Implementation:}
The location of the target and bent crystals is upstream of the vertex
locator detector (VELO) vacuum tank and of a new sector valve that
will be installed during LS2 to isolate from the LHC vacuum the region
upstream of the VELO. This means about 1.16 m upstream of the position
of the nominal proton-proton collision point at the LHCb interaction
point.  The proposal is to run the fixed-target configuration in
parallel with standard proton-proton collisions in order to maximise
the time of running, collecting both types of collisions at the same
time. It has been shown that with the performance of the Phase-I
upgrade LHCb detector, a W target of 2 cm thickness hit by a proton
flux of $10^7$ protons/s is the upper limit for a parallel parasitic
operation not affecting LHCb data taking of proton-proton collisions
(i.e., an increase of less than 10\% in occupancies due to the
fixed-target collisions).

The proposed fixed-target setup can also be used for other
flavor-physics studies complementary to those of the LHCb core physics or
as a standard fixed target experiment, reaching luminosities similar
to those discussed by the AFTER@LHC
collaboration~\cite{Brodsky:2012vg,Hadjidakis:2018ifr}.

\begin{figure}
\begin{center}
\includegraphics[width=0.76\textwidth]{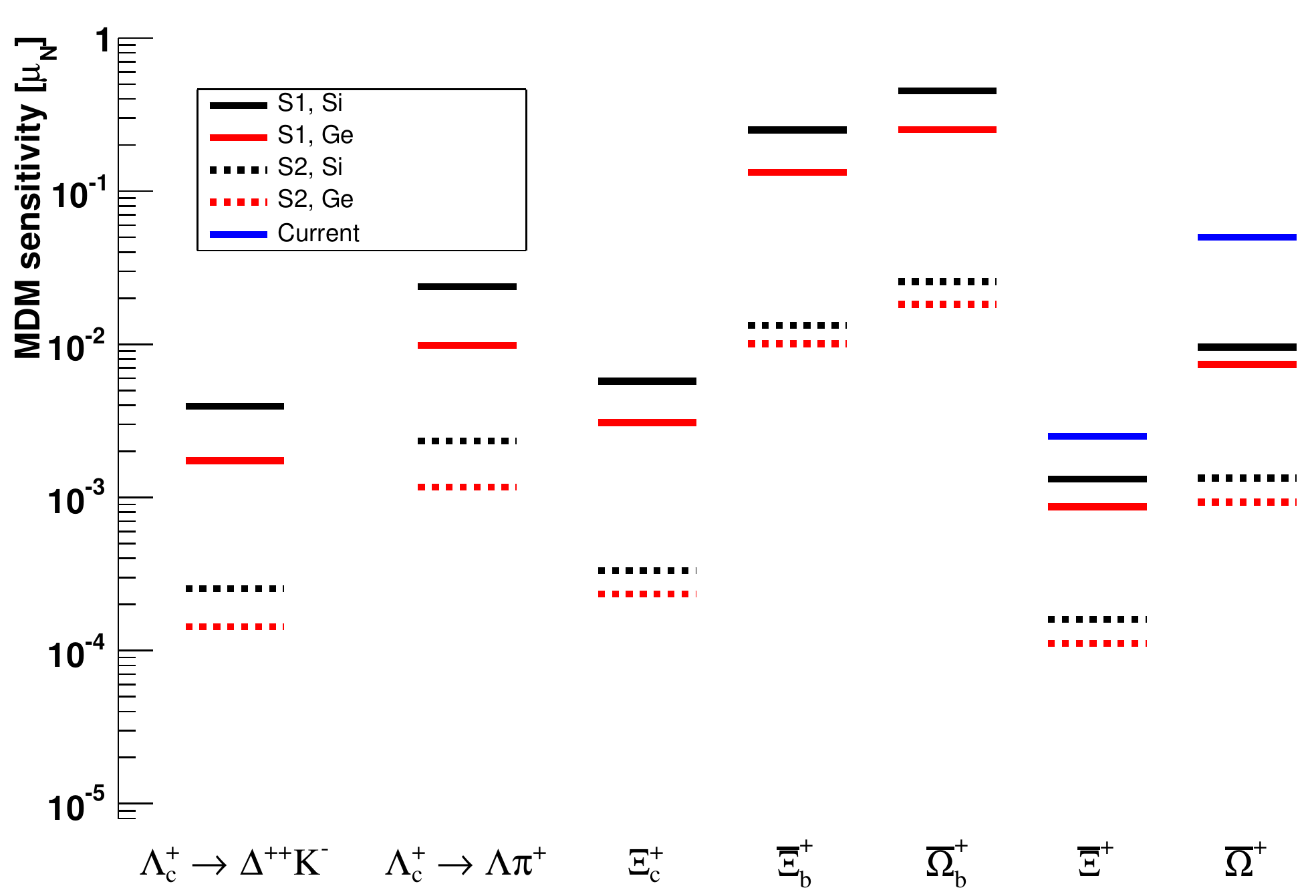}
\caption[Estimated magnetic dipole moment sensitivities for heavy
baryons using bent crystals inside the LHC at
LHCb.]{\label{fig:magmom-projection} Estimated MDM sensitivities, for
  Si and Ge with crystal parameters optimised for S1 and S2 scenarios
  (see text). A total of 10$^{15}$ and 10$^{17}$ protons on a 0.5 cm
  thick target have been considered for S1 and S2, respectively. For
  comparison purposes, the \rev{$\Lambda_{c}^{+}$} case has been
  studied in the $\Delta^{++} K^{-}$ and $\Lambda \pi^{+}$ final
  states. Blue lines show the sensitivity of the current \(\Xi^{-}\)
  and \(\Omega^{-}\) MDM measurements. (Figure from
  Ref~\cite{Bagli:2017foe}.)}
\end{center}
\end{figure}

\paragraph{Estimated performance and time scale:}
About $2.4 \times 10^{14}$ protons on target (PoT) with a target
thickness of 2 cm could be reached with three years of data taking,
either with two weeks per year of dedicated detector running at
$10^8 \, p/$s or with parallel detector operation at $10^7 \,
p/$s. This would lead to MDM sensitivities of about
$10^{-3} \, \mu_N$, $10^{-1} \, \mu_N$, and $10^{-3} \, \mu_N$ level
\cite{Bagli:2017foe} for charm, beauty and strange baryons,
respectively (where $\mu_N$ is the nuclear magneton).
\revised{Figure~\ref{fig:magmom-projection} shows the sensitivities
  for a scenario (labelled S1) with $10^{15}$ PoT and a target
  thickness of 0.5 cm (which is equivalent to $2.4 \times 10^{14}$ PoT
  and a 2 cm target).}  Extending the detector coverage down to 10
mrad would significantly reduce the crystal bending angle
required. The enhanced channelling efficiency and particle
identification for tracks of 300 GeV would improve the reconstruction
of signal events. Combined with longer running time and an increase of
the proton flux in Run 5, this would improve sensitivities by about
one order of magnitude.  \revised{This is shown as scenario S2 in
  figure~\ref{fig:magmom-projection} for an integrated 10$^{17}$ PoT
  and a 0.5 cm target (equivalent to $2.4 \times 10^{16}$ PoT and a 2
  cm target).}

\paragraph{Challenges:}
Studies and R\&D are ongoing to assess other challenges of the
proposal. These include, on one hand, the compatibility with the
machine, its operation mode, maximum reachable proton flux and the
design of the collimation system (absorber) downstream the detector,
along with the operation mode of the LHCb detector (parasitic as
described above or dedicated short runs with only fixed-target
collisions). On the other hand, the feasibility of the $\approx $15
mrad bent crystal, mandatory to optimise the acceptance coverage of
the LHCb detector, will be a key milestone for the realisation of the
experiment. In 2017, the double crystal scheme has been demonstrated
at SPS energies and with a second crystal of lower bending without
target.\footnote{S. Montesano, Testing the double-crystal setup for
  physics beyond colliders experiments in the UA9-SPS experiment,
  \url{http://ipac2018.vrws.de/papers/tupaf043.pdf, 2018}.}  In 2016,
the channelling of 6.5 TeV protons extracted from the LHC halo was
demonstrated~\cite{Scandale:2016krl}. The viability of this technique,
although in another energy and lifetime regime, was proved by the E761
Collaboration~\cite{Chen:1992wx} through the measurement of the MDM of
$\Sigma^+$ baryons produced using a 800 GeV proton beam impinging on a
Cu target.

The sensitivity reach for measuring MDMs strongly relies on the
self-polarisation of baryons in high-energy hadron
collisions. \revised{It is clear that only in case of large
  polarisation a meaningful measurement (as depicted in
  figure~\ref{fig:magmom-projection}) of the MDMs can be performed.
  While it has been known for a long time that $\Lambda$ and other
  hyperons are produced transversely polarised, the magnitude and
  kinematic dependence of polarisation is poorly known for many of the
  baryons considered here. Corresponding systematic uncertainties have
  not been worked out.  Valuable input to that can come from the SMOG
  data already recorded during LHC Run 2 and more so from future SMOG2
  data at the same centre-of-mass energy of
  $\approx 115 \gev$.\footnote{\rev{See for instance
      Ref.~\cite{LHCb-CONF-2017-001} for a first sample of
      reconstructed $\Lambda_{c}^{+}$.}}  Production cross sections,
  estimated initially from $c\bar{c}$ production at PHENIX at 200 GeV
  and scaled to 115 GeV, are consistent with the first measurement of
  $\jpsi$ and $D^0$ production in fixed-target configuration at the
  LHC \cite{Aaij:2018ogq}.}


\subsection{COMPASS++}
\label{sec:compass}

For more than four decades the M2 beam line of the SPS has provided secondary beams of muons and hadrons to the EHN2 experimental area.\footnote{Low-intensity electron beams have been used as well, mainly for calibration purposes.}
Several experiments dedicated to QCD utilised those beams, starting with EMC~\cite{Aubert:1980rm} in the 1970s.
Currently, COMPASS~\cite{Baum:1996yv} is in the process of completing its COMPASS-II program~\cite{Gautheron:2010wva} with a remaining one-year muon run on a transversely polarised $^{6}$LiD target in 2021.
The versatility of the beam line with respect to particle species and energies has allowed a broad physics program on hadron structure spectroscopy, including polarisation dependence and modifications in the nuclear environment.

The COMPASS++ project of a ``New QCD facility at the M2 beam line of the CERN SPS''~\cite{Denisov:2018unj} continues in this tradition.
Among others, it addresses such open questions as
\begin{itemize}
\item the proton charge radius,
\item quark structure of light mesons, in particular of the pion,
\item kaon polarisabilities from the Primakov reaction,
\item strange-meson spectroscopy,
\end{itemize}
and further complementary measurements, e.g., as input to the interpretation of cosmic-ray measurements (cf.~Section~\ref{sec:cosmics}).
A summary of topics (including details on running time and on beam, target, as well as detector requirements) is presented in Table~\ref{tab:COMPASS++programs} and will be discussed in parts below. It should be noted that some of the programs (e.g., the last two in above list) require extensive modifications of the M2 beam line in order to provide RF-separated hadron beams. The feasibility is discussed in the parallel PBC  ``Conventional Beams'' activity~\cite{PBC-convbeam}.

\begin{table}[h]
  \begin{center}
    \resizebox{\textwidth}{!}{\setlength{\tabcolsep}{5pt}%
      \begin{tabular}{c|c|c|c|c|c|c|c|c}
        & Physics & Beam & Beam & Trigger & Beam & & Earliest & Hardware \\
        Program & Goals& Energy&Intensity&Rate&Type&Target&start time,&additions\\
        &  & [GeV]&[s$^{-1}$]&[kHz]&&&duration\\\hline \hline

        muon-proton & Precision&& & & &  high-&& active TPC,\\
        elastic&proton-radius&100&$4\cdot 10^6$&100&\color{blue}{$\mu^\pm$}&pressure&2022&SciFi  trigger, \\
        scattering&measurement&&&&&H2&1 year&silicon veto,\\\hline

        Hard&&&&&&&&recoil silicon,\\
        exclusive &GPD $E$ & 160 & $2\cdot 10^7$ & 10 & \color{blue}{$\mu^\pm$} & NH$_3^\uparrow$ & 2022& modified polarised  \\
        reactions&&&&&&&2 years&target magnet \\\hline

        Input for Dark&$\overline{p}$ production&20-280&$5\cdot 10^5$&25& \color{ForestGreen}{$p$}&LH2,&2022&liquid helium\\
        Matter Search & cross section&  & & & & LHe & 1 month & target \\\hline

        &&&&&&&&target spectrometer:\\
        $\overline{p}$-induced & Heavy quark& 12, 20 & $5\cdot 10^7$ &25& \color{ForestGreen}{$\overline{p}$} & LH2 & 2022& tracking, \\
        spectroscopy&exotics&&&&&&2 years& calorimetry\\\hline

        Drell-Yan & Pion PDFs& 190 & $7\cdot 10^7$ & 25 & \color{ForestGreen}{$\pi^{\pm}$}  &  C/W & 2022 &\\
        &&&&&&&1-2 years&\\\hline\hline

        &&&&&&&&\\
        Drell-Yan & Kaon PDFs \& & $\sim$100 & $10^8$& 25-50 & \color{red}{$K^{\pm}$, $\overline{p}$} & NH$_3^\uparrow$, &2026 &''active absorber'', \\
        (RF)&Nucleon TMDs& &&&&C/W&2-3 years&vertex detector \\ \hline

        &Kaon polarisa-&&&&&&\footnotesize{non-exclusive}& \\
        Primakov & bility \& pion& $\sim$100 & $5\cdot 10^6$ & $>10$ & \color{red}{$K^-$} & Ni &2026&  \\
        (RF)&life time&&&&&&1 year&\\\hline

        Prompt&&&&&&&\footnotesize{non-exclusive}&\\
        Photons & Meson gluon& $\geq 100$ & $5\cdot 10^6$ & 10-100 & \color{red}{$K^{\pm}$} & LH2, & 2026 &  hodoscope\\
        (RF)&PDFs &&&&\color{ForestGreen}{$\pi^{\pm}$}&Ni&1-2 years&\\\hline

        $K$-induced&High-precision&&&&&&&\\
        Spectroscopy &strange-meson& 50-100 & $5\cdot 10^6$ & 25 & \color{red}{$K^-$} &LH2 & 2026 & recoil TOF, \\
        (RF)&spectrum&&&&&&1 year&forward PID\\\hline

        &Spin Density&&&&&&&\\
        Vector mesons &Matrix& 50-100 & $5\cdot 10^6$ & 10-100 & \color{red}{$K^{\pm},\pi^{\pm}$} & from H  & 2026 &  \\
        (RF)&Elements&&&&&to Pb&1 year&\\

      \end{tabular}%
    }
  \end{center}
  \caption[Summary of physics programs within the COMPASS++ project, including the physics goals, beam and target requirements, running time, and detector upgrades (where applicable).]{Summary of physics programs within the COMPASS++ project (from~\cite{Denisov:2018unj}, including the physics goals, beam and target requirements, running time, and detector upgrades (where applicable). The table is segmented into projects requiring {\color{blue}muon beams} (in blue), {\color{ForestGreen} conventional hadron beams} (in green), and {\color{red}RF-separated hadron beams} (in red), the latter also separated by double horizontal lines from the presently available beams. }
  \label{tab:COMPASS++programs}
\end{table}


\subsubsection{$\mu p$ elastic scattering and the proton charge radius}

The hydrogen atom is the most simple and abundant atom around.
As such it has received enormous attention and served as testing ground for various benchmark calculations in QED.
Nevertheless, it still confronts the field with open questions, in particular about its nucleus, the proton.
As discussed before (cf.~Sec.~\ref{sec:LHC-FT}), the investigation of the proton structure, e.g., its description in terms of quark and gluon degrees of freedom in the framework of QCD, is a wide field of intense experimental and theoretical research.
But even such a seemingly simple property as the charge radius has still not been measured precisely despite decades of activities.
Two complementary approaches have traditionally been used to provide insights into the size of the proton, employing on one side hydrogen spectroscopy by measuring Lamb shifts (or more generally atomic energy-level splittings) and on the particle physics side using elastic lepton-proton scattering.
The results of these approaches had seemed to agree and converged to a value of around 0.88~fm~\cite{Mohr:2012tt},
until---close to ten years ago---new spectroscopy results became available using muonic hydrogen~\cite{Pohl:2010zza}.
They disagreed by more than 5$\sigma$ with the earlier measurements of the proton radius, both using elastic scattering but also ordinary-hydrogen spectroscopy, and suggested a smaller radius of about 0.84~fm.
Further excitement was spurred by the results using muonic deuterium that also suggested a smaller charge radius.
This triggered an ongoing debate about possible signs of lepton universality.

On the other hand side, electron-scattering results do not provide a coherent picture either.
In particular, different fitting strategies to available scattering data (even using the same data) result in different proton radii. One should remember that the radius is not measured directly in elastic scattering, but rather deduced from extrapolating the slope of the electromagnetic form factors to zero four-momentum transfer squared, i.e.,
\begin{equation}
  \frac{1}{6} r_p^2  = - \left.\frac{d }{d Q^2}\right|_{Q^2=  0} G_E(Q^2)
\label{eq:Rp}
\end{equation}
for which the electric form factor \(G_E(Q^2)\) can be obtained from the elastic-scattering cross-section
\begin{equation}
  \frac{d\sigma^{\mu p \to \mu p}}{d Q^2} =  \frac{\pi\alpha^2}{Q^4\,m_p^2\, \vec{p}_\mu^{\,2}}
\left[\left(G_E^2+\tau G_M^2\right)
  \frac{4E_\mu^2 m_p^2-Q^2(s-m_\mu^2)}{1+\tau}
-G_M^2\frac{2m_\mu^2Q^2-Q^4}{2}\right], \label{eq::elasticmup}
\end{equation}
where $Q^2=-t=-(p_\mu-p_{\mu'})^2$, $\tau=Q^2/(4m_p^2)$ and $s=(p_\mu+p_p)^2$.
The squared centre-of-momentum energy $s$ is given in the laboratory system
by $s=2E_\mu m_p + m_p^2+m_\mu^2$ with $E_\mu$ being the energy and $\vec{p}_\mu$
the three-momentum of the incoming muon when colliding with a proton at rest.

\begin{figure}
\begin{center}
\includegraphics[width=0.99\textwidth,trim=45 0 90 0,clip]{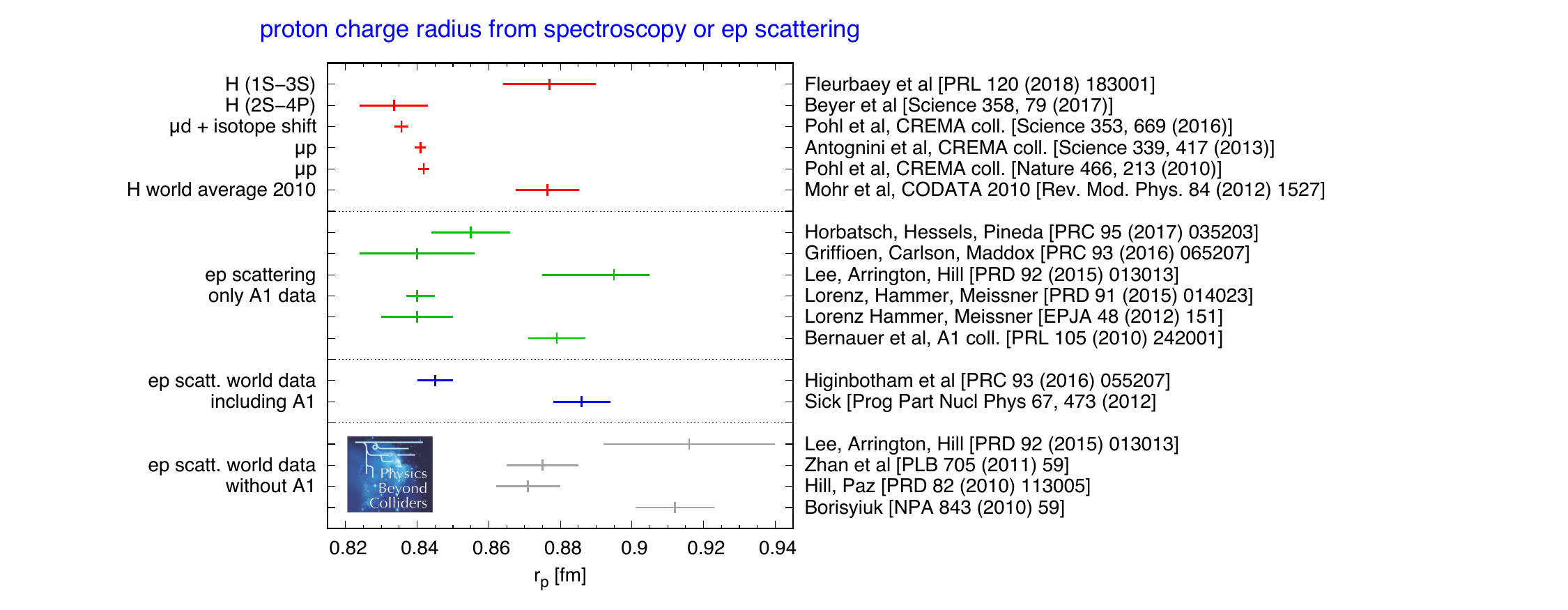}
\caption[Overview of recent determinations of the proton charge radius $r_p$ from spectroscopy of
	ordinary or muonic hydrogen and from analyses of $ep$ scattering data.]{\label{fig:proton-rad}
	Overview of recent determinations of the proton charge radius $r_p$ from spectroscopy of
	ordinary or muonic hydrogen (upper part) and from analyses of $ep$ scattering data (lower three parts).
	The latter are grouped in analysis using only the MAMI A1 data (green), using all the world data (blue),
	and using the world data excluding the A1 data (grey).
	Some analyses use dispersion relations and thus additional data from the time-like region
	(Lorentz, Hammer, Meissner and Hill, Paz).
	The analysis of Horbatsch, Hessels, Pineda uses additional input from $\chi$PT.
              }
\end{center}
\end{figure}

The situation has not improved over the past years. In contrary, new results feed even more the confusion with hydrogen spectroscopy now also suggesting a lower radius~\cite{Beyer79},\footnote{In that work a strong correlation between the values of Rydberg constant and the proton radius in spectroscopy measurements was pointed out.}
though also the larger value for the radius has been reproduced in spectroscopy~\cite{Fleurbaey:2018fih}.
The situation is illustrated in figure~\ref{fig:proton-rad}. Recently, preliminary results from hydrogen spectroscopy and from the Jefferson Lab PRad Collaboration support again a small radius (not included in the figure).\footnote{A radius of  \(0.830 ʱ\pm 0.008_{\text{stat}} \pm 0.018_{\text{syst}} \) fm was reported by PRAD as their {\em Preliminary Result} at the recent 5$^{\text{th}}$ Joint Meeting of the APS Division of Nuclear Physics and the Physical Society of Japan.}

One thus faces a complicated experimental situation both within spectroscopy and from elastic $ep$ scattering experiments. In that respect, $\mu p$ scattering at high energy at COMPASS++  is placed in a unique position and could be of particular interest. Using a muon beam it intrinsically has smaller QED corrections compared with $ep$ and the contribution from the magnetic form factor $G_M(Q^2)$ to the cross section is suppressed due to the muon mass. As such extracting  $G_E(Q^2)$ from the cross section is facilitated, and COMPASS++ has the potential to clarify theoretical and systematic uncertainties of the $G_E$ extraction at low $Q^2$.
Still, the ultimate precision on $r_p$ extracted from the slope of $G_E$ at $Q^2=0$ remains to be quantified. The range in $Q^2$, \(0.001~\text{GeV}^{2}  < \text{Q}^{2} < 0.02~\text{GeV}^{2}\),  is indeed not larger than those for existing $ep$ measurements, which actually yield very different radii depending on the choice of the fitting range and fitting functions, as was shown in figure~\ref{fig:proton-rad}. Technically, the low \(Q^{2}\) to be accessed requires novel measurement methods. The plan is to use an active target by reconstructing the momentum transfer from the recoiling proton in a time-projection chamber (TPC) with pressurised pure hydrogen that also acts as the target.
Such a target has been developed by PNPI~\cite{Vorobyov:1974ds,Ilieva:2012vhs},
and is in the testing phase for a similar experiment using electron scattering at Mainz.

\begin{figure}
\begin{center}
\includegraphics[width=0.97\textwidth]{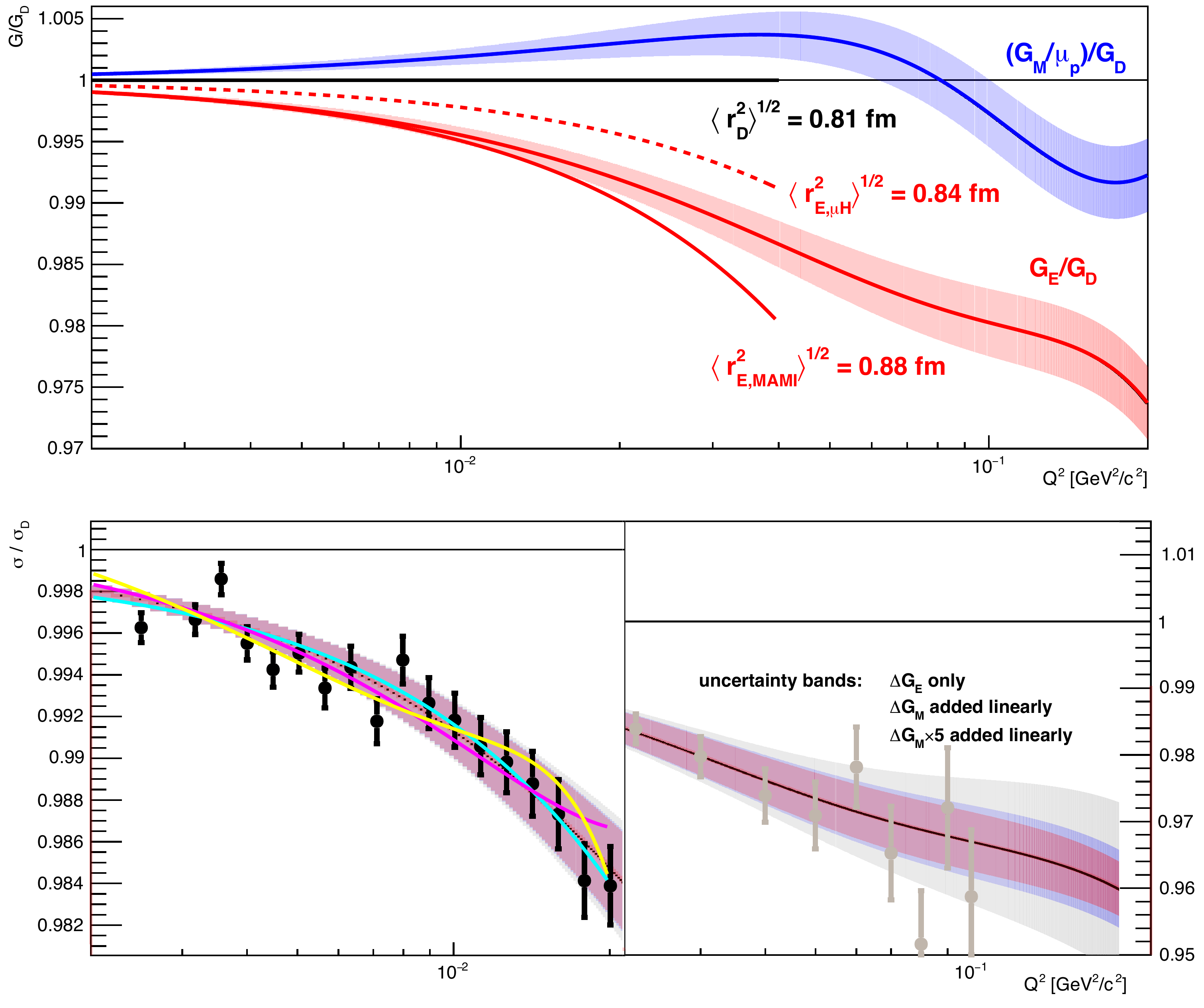}
\caption[Proton form factors $G_E$ and $G_M$ from fits to MAMI data~\cite{Bernauer:2010wm},
compared to the curves for proton charge radii of 0.81, 0.84, and 0.88 fm, and ratio of the cross section 
(using the MAMI form-factor parameterisations) over the one using the dipole form factor,
overlaid with pseudo-data that reflect the envisaged statistical precision of the COMPASS++ measurement and various polynomial fits to the pseudo-data.]{\label{fig:rp_projections}
(Top) Proton form factors $G_E$ and $G_M$ (as labelled) from fits to MAMI data~\cite{Bernauer:2010wm},
shown as ratios to the dipole form factor $G_D$. Indicated as well are the curves for proton charge radii of 0.81, 0.84, and 0.88 fm.
(Bottom) Ratio of the cross section (using the MAMI form-factor parameterisations) over the one using the dipole form factor. The innermost (red) uncertainty band corresponds to the effect of the uncertainty of $G_E$ only, while for the (blue) middle band the uncertainty from $G_M$ was added linearly, and for the outer (grey) band the contribution from $\Delta G_M$ was increased by a factor of five.
Pseudo-data (points) were sampled according to the form factors from Ref.~\cite{Bernauer:2010wm} and reflect the envisaged statistical precision of the COMPASS++ measurement. Only the low-$Q^{2}$ points in black were used in the various fits (polynomial in \(Q^{2}\)) to the pseudo-data shown as magenta (linear), purple (quadratic) and yellow (3$^{\text{rd}}$ order)  curves.
Pseudo-data points in grey require a different detector setup and are shown here for completeness.
Only statistical uncertainties are shown as expected to dominate the systematic point-to-point uncertainty.
}
\end{center}
\end{figure}

\paragraph{Performance expectations:}
The design goal is to measure the electric form factor to such precision as to be sensitive to the proton charge radius at a level of 0.01~fm. Figure~\ref{fig:rp_projections} (top) shows the electric and magnetic form factors from Ref.~\cite{Bernauer:2010wm} scaled to the specific dipole form factor \(G_{D} = [1+r_{D}^{2}Q^{2}/12]^{-2}\), one frequently used parameterisation of the form factor corresponding to an exponential charge distribution of the proton. The figure also shows in the bottom the elastic cross section (again normalised to the one for the dipole form), including the present uncertainty band resulting from the precision of the MAMI data, in comparison to the expected statistical precision of the COMPASS++ data (here from a Monte Carlo simulation that used the same form factor parameterisation as used for the curves).

The linear, purple, and yellow curves in figure~\ref{fig:rp_projections} show polynomial fits to the pseudo-data, in particular fits that are linear, quadratic, and cubic in $Q^{2}$, respectively. \revised{From these fits, the proton radius could indeed be extracted with high statistical precision: the quadratic fit in $Q^2$ is preferred by the data and gives $r_p = 0.877 (13) \fm$, consistent with the value $r_p = 0.879 \fm$ used to generate the pseudo-data.}
However, it is clear from the spread of results in figure~\ref{fig:proton-rad} from the same data set but using different fitting Ans\"atze that the ultimate goal for the precision on the radius extraction requires careful systematic studies using a similar selection on fitting approaches. Furthermore, the insensitivity of the proton radius fits to experimental systematics, e.g., from the luminosity normalisation (including all efficiency corrections) needs to be demonstrated.

\paragraph{Implementation considerations:}
As low values of momentum transfer have to be reconstructed, the proposal foresees usage of a variable high-pressure TPC with pure hydrogen that acts as a target and at the same allows reconstruction of $Q^{2}$ from the recoiling proton. Variation of gas pressure allows coverage of various regions in $Q^{2}$. Such a setup is currently developed for a future measurement at MAMI. Usage of a TPC, however, puts stringent requirements on the maximum beam intensity. The currently tested upper limit of 2$\times$10$^{6}$ $\text{s}^{-1}$ is an order of magnitude below the full available muon-beam intensity. Further studies are needed on whether the wider beam profile of the M2 beam compared to the MAMI beam can be exploited in favour of larger beam intensities.

A further challenge concerns triggering, which currently also limits the manageable beam intensity. Triggering on the recoiling proton is not sufficient, but the present muon trigger is not suitable for muon trajectories with very small scattering angles imposed by the \(Q^{2}\) range probed. The substantial background rate from very-small-angle muon scattering poses a challenge that requires additional tracking instrumentation (e.g., additional SciFi or silicon pixel detectors) and/or a trigger-less readout, the timeline of which is, however, not yet well defined.

Last but not least, the efficient use of the M2 muon beam might require coexistence of several PBC projects, e.g., MUonE (cf.~Sect.~\ref{sec:muE}) and NA64$\mu$ (cf.~Ref.~\cite{PBC-BSM}).
Parallel installation of the proton-radius project and MUonE is discussed in more detail in Sect.~\ref{sec:compass-muone}.

\paragraph{Timelines:} The earliest possible scenario for test runs is in 2021, with physics data taking in 2022, feasible if decision is taken
	sufficiently early.

\paragraph{Worldwide landscape:}
The discrepancies in the proton radius measurements spurred a whole range of activities, both on the spectroscopy as well as on the elastic-scattering side. Here, only the latter is considered as relevant to the COMPASS++ proposal.
\begin{itemize}
	\item  The MUSE collaboration at PSI~\cite{Gilman:2017hdr} will use low-energy \(\mu^{\pm}\) and \(e^{\pm}\) scattering to reduce systematics but
	      also to compare them directly for hints of lepton-flavor violation.
	      Full commissioning is foreseen for this Fall (2018) and first data taking of 20 weeks
	      between May and December 2019. Further data taking planned in 2020, and analysis as well as publication foreseen for 2021/22.
	      The goal is to achieve sub-\% relative precision over a \(Q^{2}\) range of 0.002---0.07 GeV$^{2}$ to extract
	      the proton radius to a precision of 0.007 fm.
	\item  The PRad collaboration at Jefferson Lab~\cite{Gasparian:2011} already took the presently lowest $Q^{2}$
	      data in \(ep\) elastic scattering and presented preliminary results this year. The analysis of the slope
	      favours a generally lower (preliminary) value of the proton radius of
	      \(0.830 ʱ\pm 0.008_{\text{stat}} \pm 0.018_{\text{syst}} \)~fm.
	      The experiment exploits a simultaneous measurement of the well-known
	      M{\o}ller scattering as reference to reduce systematics. Final results are expected to come out in 2019.
	\item  At MAMI a similar approach as proposed here using a TPC as an active target will be employed in a future
	      measurement,\footnote{\url{https://www.blogs.uni-mainz.de/fb08-mami-experiments/files/2017/11/ep_-proposal.pdf}}
	      likely in 2020. Ongoing are runs with low beam-energy and via initial-state
	      radiation~\cite{Mihovilovic:2016rkr}. Furthermore, MAGIX/MESA\footnote{\url{http://magix.kph.uni-mainz.de/physics.html}}
	      plans to run from 2021/22 on.
	\item  Data at very-low $Q^{2}$ are expected to come from the ULQ$^{2}$ (``Ultra-Low Q$^{2}$'')
	      and ProRad (``Proton Radius'') experiments at Tohoku Low-Energy Electron Linac (Japan)
	      and the PRAE facility in Orsay (France), respectively.
	      ULQ$^{2}$, to start in 2019, aims at an absolute cross section measurement with 10$^{-3}$ precision to obtain
	      Rosenbluth separated $G_{E}(Q^{2})$ and $G_{M}(Q^{2})$ within $0.0003 \le Q^{2} [\text{GeV}^{2}] \le  0.008$,
	      using a 10--60 MeV $e^{-}$ beam.\footnote{\url{https://www.jlab.org/indico/event/280/session/0/contribution/31/material/slides/0.pdf}}
	      ProRad~\cite{Hoballah:2018szw} will utilise a 30--70 MeV $e^{-}$ beam and aims at a 0.1\% precision on the elastic cross section.
\end{itemize}


\subsubsection{Pion PDFs from Drell-Yan production}\label{sec:pionPDF}

The pion is the lightest hadron and plays a fundamental role in QCD. Despite its importance
little is known about its internal structure and the reason for its small mass.
The simplest possible process for measuring pion PDFs, deep-inelastic lepton-pion scattering, is experimentally not feasible.
As such other direct or indirect means are needed to constrain the pion PDFs, also to match the
ongoing improvements in lattice calculations.
Two approaches are highlighted here. In the so-called Sullivan process~\cite{Sullivan:1971kd},
one uses deep-inelastic lepton scattering off the pions radiated from a proton by tagging the recoil neutron.
Its interpretation in terms of pion PDFs is however complicated by the fact that the pion is off shell and that one needs to know the pion flux in the proton.
A direct approach avoiding these problems is
pion-induced \DY, where typically a secondary pion beam is steered
onto a solid fixed target to compensate for the limited beam intensity.

\begin{figure}
\begin{center}
\includegraphics[width=0.49\textwidth,trim=45 0 55 0, clip]{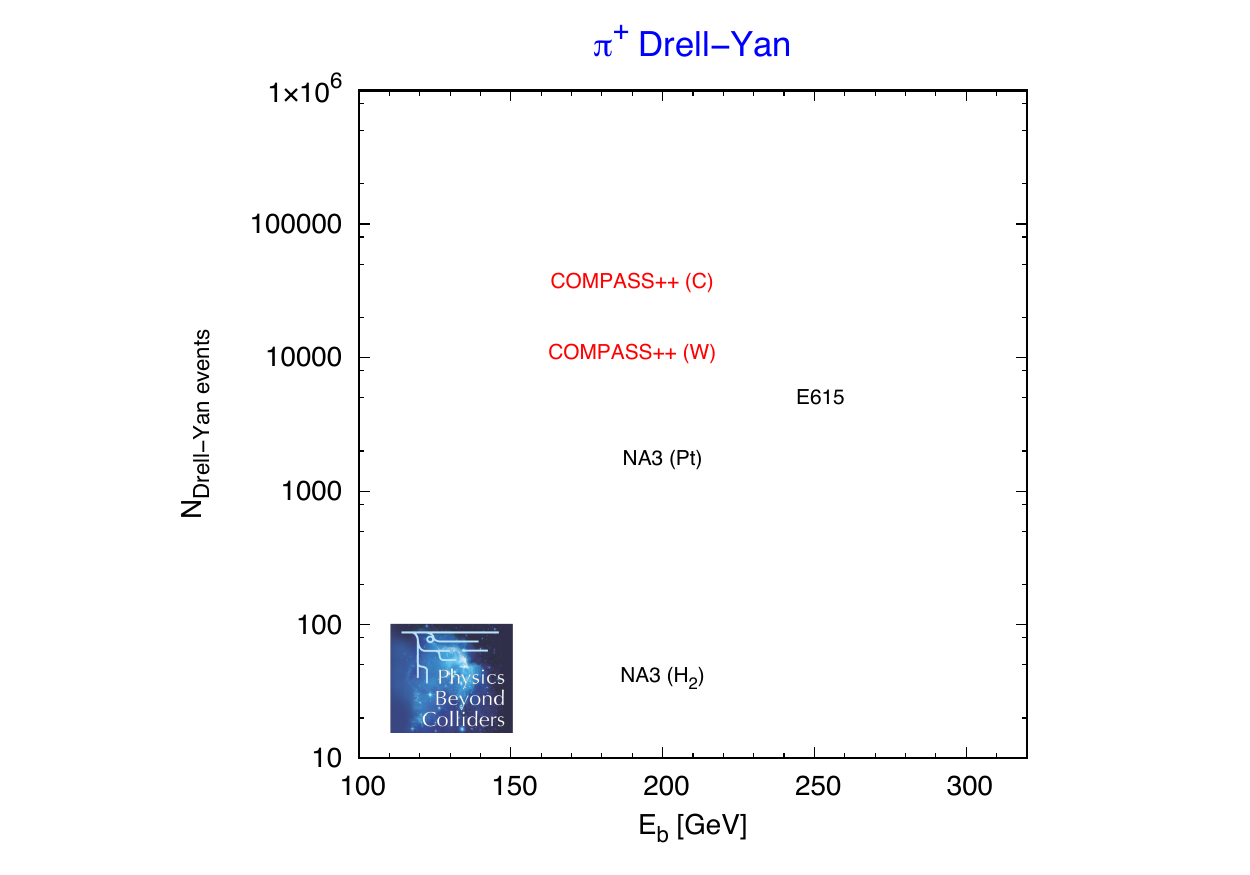}
\includegraphics[width=0.49\textwidth,trim=45 0 55 0, clip]{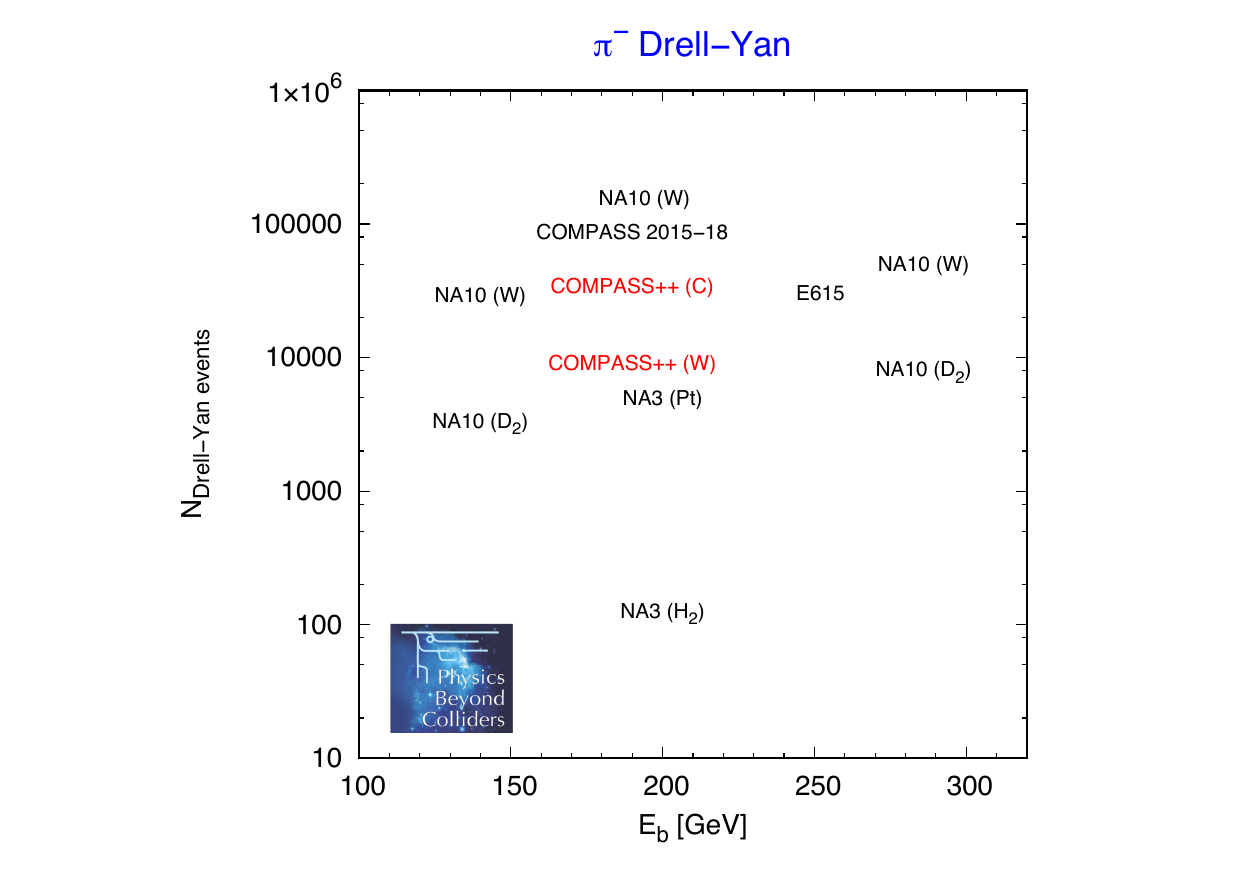}
\caption[Overview of pion-induced \DY event yields for \(\pi^{+}\) and \(\pi^{-}\) from
		past experiments and the proposed COMPASS++ running.]{\label{fig:pion-dy}
		Overview of pion-induced \DY event yields for \(\pi^{+}\) (left) and \(\pi^{-}\) (right) from
		past experiments and the proposed COMPASS++ running of 280 days at a beam energy of 190 GeV.
		In addition to the pion beam energies, the various target nuclei at the different experiments are indicated.
		(Numbers compiled from Table~3 in the LoI \protect\cite{Denisov:2018unj})
		}
\end{center}
\end{figure}

The theoretical framework of using \DY for PDF determinations is well established and
used extensively for nucleon and nuclear PDFs.
Nevertheless, data on pion-induced \DY are sparse, especially for \(\pi^{+}\) beams,
as can be seen in figure~\ref{fig:pion-dy}. That in turn affects the flavor separation of pion PDFs.
Accordingly, phenomenological work is rather limited with important work dating back to the 1990s:
SMRS~\cite{Sutton:1991ay}, GRV~\cite{Gluck:1991ey}, and GRS~\cite{Gluck:1999xe}.
However, only the valence and gluon distributions (the latter mainly through prompt-photon production)
were constrained in those works.
A recent analysis~\cite{Barry:2018ort}, including also data from
leading-neutron production in deep-inelastic scattering at HERA,
results in much improved sea-quark and gluon distributions, the latter contributing about 1/3 of the
pions momentum. Although claimed to be small, the model dependence inherent to employing the
Sullivan process will make it desirable to have complementary data that can directly constrain the
sea-quark contribution.

The differences in cross sections for \(\pi^{+}\) and \(\pi^{-}\) induced \DY are highly sensitive to the pion's sea.
More specifically, for an iso-scalar target, assuming charge conjugation,
SU(2)$_{f}$ symmetry for valence quarks and SU(3)$_{f}$ symmetry for sea quarks, the leading-order
sea-quark distribution can be uniquely constrained through the two linear combinations~\cite{Londergan:1995wp}
\begin{align}
\label{pion-dy-combinations}
\Sigma_{\text{val}}^{\pi D} &= - \sigma^{\pi^+ D} + \sigma^{\pi^- D} \,,
&
\Sigma_{\text{sea}}^{\pi D} &= 4 \sigma^{\pi^+ D} - \sigma^{\pi^- D} \,.
\end{align}

\(\Sigma_{\text{sea}}^{\pi D}\) contains only contributions of sea-valence and sea-sea terms, i.e.,
no valence-valence contribution, of pion and proton quark distributions.
In contrast, \(\Sigma_{\text{val}}^{\pi D} \) receives only contributions from valence-valence combinations.
The proton PDFs and the pion valence-quark PDFs are
known well enough in order to extract the pion sea-quark PDF.

\begin{figure}
\begin{center}
\includegraphics[width=0.8\textwidth]{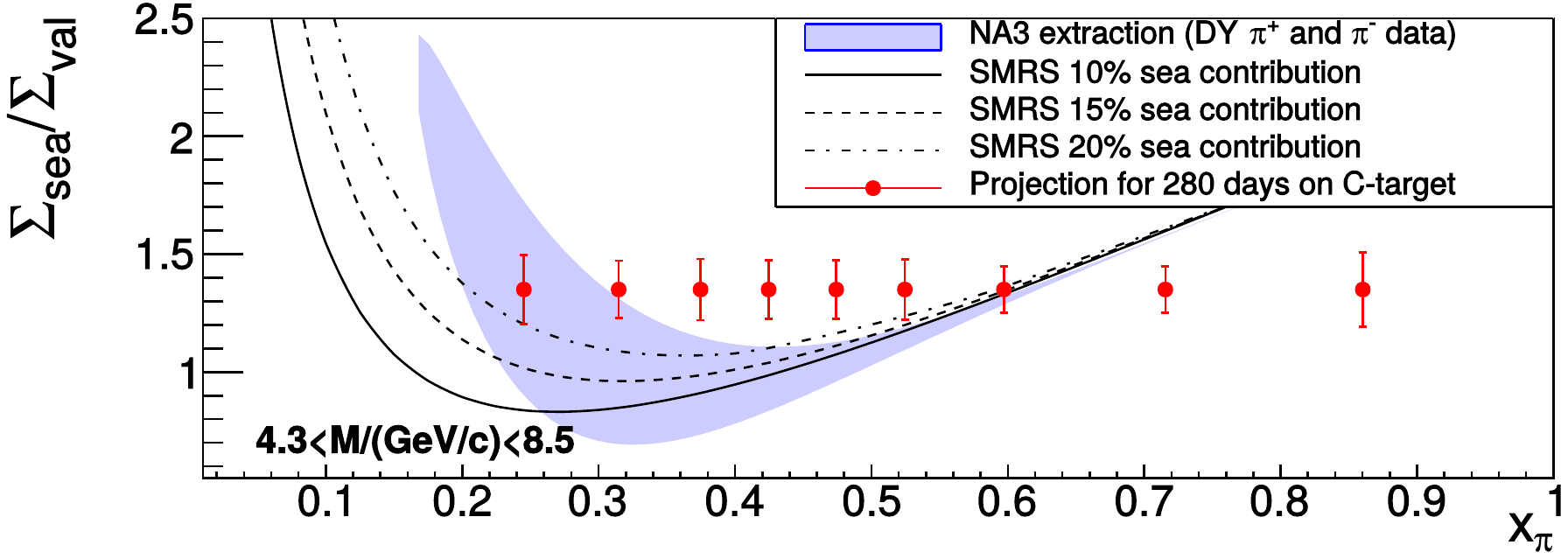}
\caption[COMPASS++ projections for ratio of certain cross-section combinations
 	compared to three different scenarios for the sea-quark distributions and to
	a prediction based on the extraction of the pion valence and sea distributions
	by NA3.]{\label{fig:pi-dy-projection}
	COMPASS++ projections for ratio of the cross-section combinations in \protect\eqref{pion-dy-combinations},
	for 280 days of running on a carbon target at a beam energy of 190 GeV.
	The SMRS curves correspondent to three different sea-quark distributions from Ref.~\cite{Sutton:1991ay},
	while the shaded area represents a prediction based on the extraction of the pion valence and sea distributions
	by NA3~\cite{Badier:1983mj}.
	(Figure from \protect\cite{Denisov:2018unj})
	}
\end{center}
\end{figure}

\paragraph{Performance expectations and worldwide landscape:}
COMPASS++ proposes to use the M2 pion beams at 190 GeV on an iso-scalar carbon target.
Two years (2$\times$140 days) of data taking at high
beam intensity would allow an improved extraction of the pion PDFs for \(x_{\pi}>0.1\).
Figure~\ref{fig:pi-dy-projection} illustrates the impact of the COMPASS++ data
on the \(\Sigma_{\text{sea}}/\Sigma_{\text{val}}\) ratio.\footnote{The SMRS curves are ad hoc assumptions for the sea distributions as data could not constrain them in that analysis.}
Such measurement would be a unique possibility for separating valence and sea quarks in the pion for
$x_{\pi} \gsim 0.1$, with strong complementarity to the pion-structure studies in $ep$ scattering
(off virtual pions via the Sullivan process) foreseen at JLab12 and at a future EIC.\footnote{See, e.g., the presentation by T.~Horn at \url{https://www.jlab.org/conferences/ugm/program.html}.}
There is presently no competition in this sector as no other facility provides high-intensity
high-energy pion beams.
It should be noted that parallel to the data on the \DY process a rather large data set
on \(J/\psi\) production will be collected, which would allow detailed studies of charmonium production.

\subsubsection{Kaon polarisability from the Primakov reaction}\label{sec:KPrimakoff}

The electric and magnetic polarisabilities of a meson, $\alpha$ and
$\beta$, characterise its response to a quasistatic electromagnetic
field.  For pions and kaons, they can be computed in chiral
perturbation theory ($\chi$PT), which makes them prominent observables
for the quantitative investigation of QCD in the low-energy limit.
The currently most precise measurement of the electric pion
polarisability is
$\alpha_\pi =
(2.0\pm0.6_{stat}\pm0.7_{syst})\times10^{-4}\,\text{fm}^3$, which is
in good agreement with the predictions of $\chi$PT and dispersion
relations. This result was obtained by COMPASS \cite{Adolph:2014kgj}
in the so-called Primakov reaction
$\pi^- Z \rightarrow \pi^- \gamma Z$ using a $190\,\text{GeV}$
negative pion beam and a Ni target, making the assumption
$\alpha_{\pi} + \beta_{\pi} = 0$.  The latter is motivated by $\chi$PT
\cite{Gasser:2006qa}, where $\alpha_{\pi} + \beta_{\pi} = 0$ is
nonzero only at two-loop level.

The prediction of one-loop $\chi$PT for the charged kaon
polarisability is
$\alpha_K = - \beta_K =
(0.64\pm0.10)$$\times$10$^{-4}\,\text{fm}^3$ \cite{Guerrero:1997rd}.
Experimentally, only an upper limit $\alpha_{K} <
200$$\times$10$^{-4}\,\text{fm}^3$ (CL=90\%) was established from the
analysis of $X$-ray spectra of kaonic atoms \cite{Backenstoss:1973jx}.

A measurement of the kaon polarisability via the reaction
$K^- Z \rightarrow K^- \gamma Z$ along the lines of the measurement of
the pion polarisability performed by COMPASS, is challenging. The kaon
component in a conventionally produced hadron beam is too small at
high beam energies to collect the required amount of data on a
reasonable timescale. Also, it is difficult to identify beam particles
with high enough purity. To this end, a RF-separated hadron beam, in
which kaons are enriched, would provide a unique opportunity to
perform the first measurement of the kaon polarisability. Additional
difficulties for the kaon polarisability measurement are the small
kinematic gap between the threshold in the invariant mass
$M_{K^- \gamma}$ and the first resonance $\text{K}^*(892)$, when
compared to the pion case with the $\rho(770)$ resonance, and the one
order of magnitude smaller Primakov cross section compared to the case
of the pion.

\begin{figure}
\begin{center}
  \includegraphics[width=0.49\textwidth,trim=10 0 40
  0,clip]{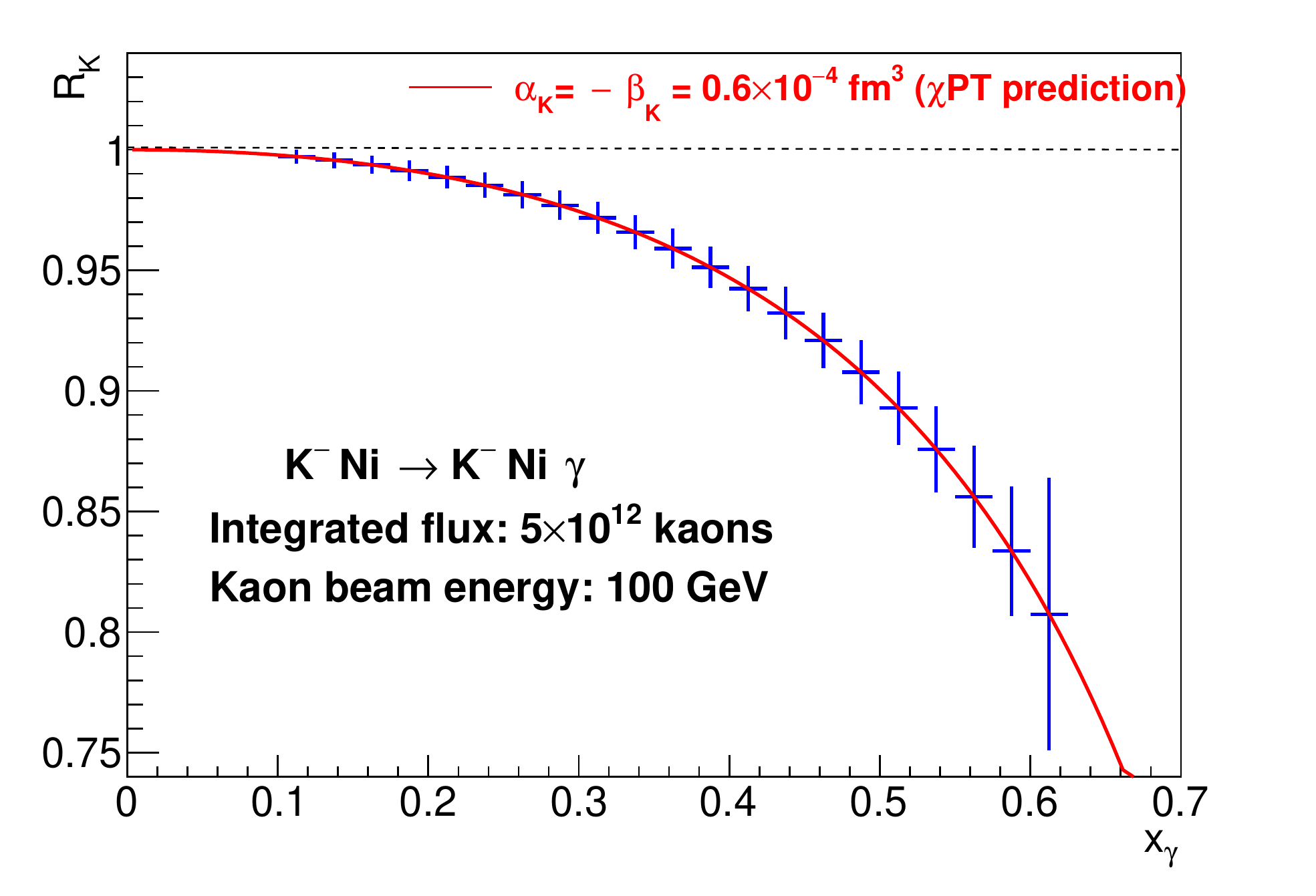}
  \includegraphics[width=0.49\textwidth,trim=10 0 10
  0,clip]{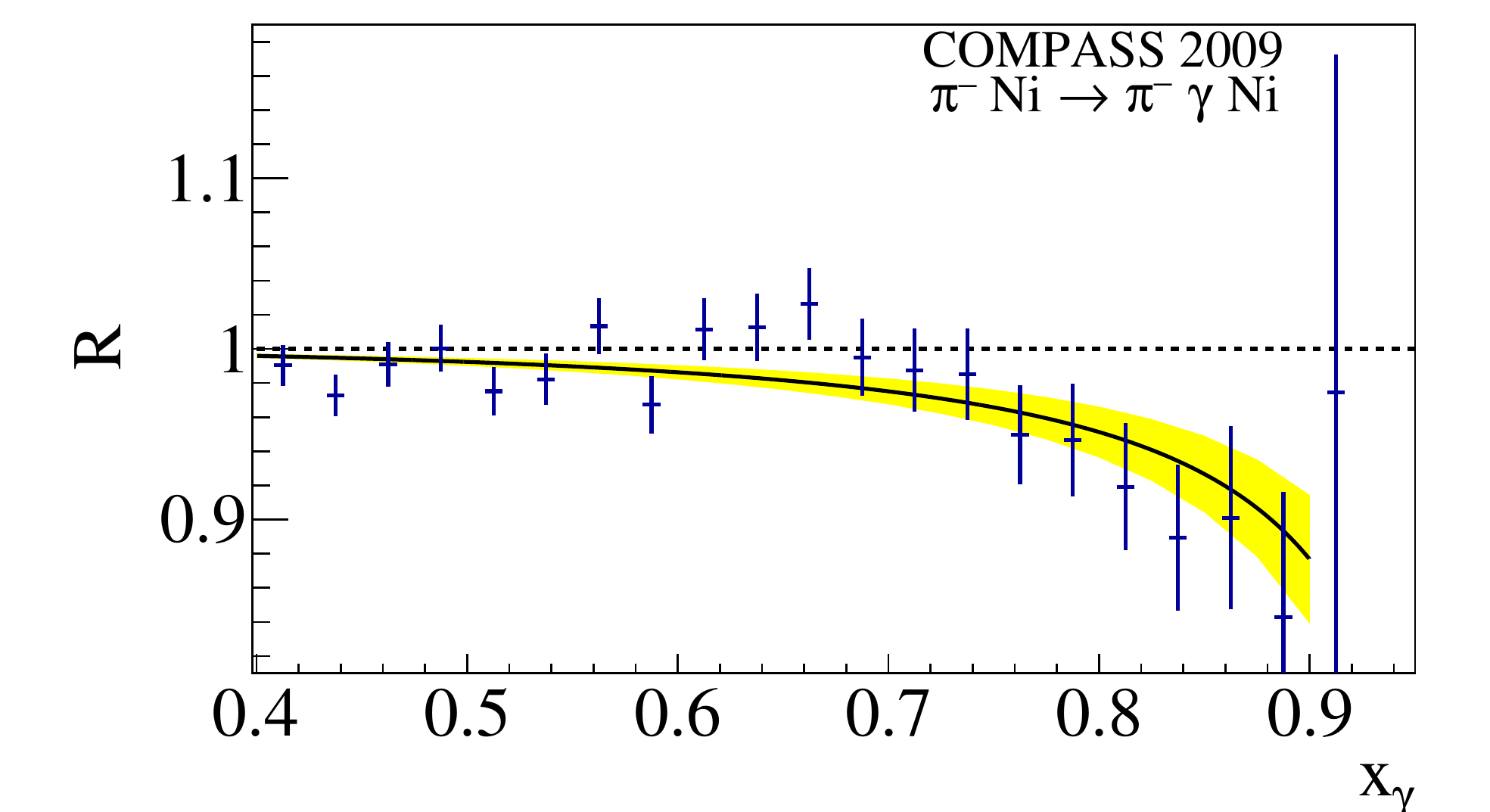}
\caption[Projected COMPASS++ data for the scaled Primakov cross
section $R_K$ for a kaon beam, compared to the prediction from
$\chi$PT, and the corresponding (existing) measurement of $R_\pi$ for
the pion.]{\label{fig:primakov} Left: projected data for the scaled
  Primakov cross section $R_K$ with a 100 GeV kaon beam (figure 37 of
  \protect\cite{Denisov:2018unj}).  $x_{\gamma}$ is the energy of the
  produced photon normalised to the beam energy.  The red curve shows
  the prediction from $\chi$PT.  Right: the corresponding measurement
  of $R_\pi$ for the pion, published in \protect\cite{Adolph:2014kgj}.
}
\end{center}
\end{figure}

\paragraph{Performance expectations:}
For a kaon polarisability measurement with a $100\,\text{GeV}$
RF-separated kaon beam with intensity 5$\times$10$^{6}$ s$^{-1}$, an
estimate of the achievable precision has been given in the COMPASS++
LoI \cite{Denisov:2018unj}.  Projected data for the ratio $R_{K}$ of
the differential cross section for the physical kaon over the expected
cross section for a hypothetical point-like kaon are shown in
figure~\ref{fig:primakov}, along with the published results for a pion
beam \cite{Adolph:2014kgj}.  Assuming an integrated flux of
5$\times$10$^{12}$ kaons after one year of data taking, this estimate
finds the achievable statistics to be about 6$\times$10$^5$
$K^- \gamma$ events in the kinematic range $0.1<x_{\gamma}<0.6$ and
$M_{K^- \gamma}<0.8\,\text{GeV}$, where, $x_{\gamma}$ is the energy of
the produced photon normalised to the beam energy.  The relation
between $R_K$ and $\alpha_{K}$ under the hypothesis
$\alpha_K + \beta_K = 0$ approximately reads
\begin{equation}
\label{R-K-alpha}
R_{K} - 1 = - \frac{3 m_K^3}{2 \alpha_{\text{em}}} \,
  \frac{x_{\gamma}^2}{1-x_{\gamma}} \, \alpha_{K}^{3} \,,
\end{equation}
where $\alpha_{\text{em}}$ is the fine structure constant.  Note that
polarisation effects in case of the kaon are amplified by the factor
$(m_{K}/m_{\pi})^3 \approx 44$ when compared to the pion.  The
statistical accuracy of the $\alpha_K$ extraction under the assumption
$\alpha_{K}+\beta_{K}=0$ is estimated to be
0.03$\times$10$^{-4}\,\text{fm}^{3}$ if the relation \eqref{R-K-alpha}
is used.  This corresponds to a 5\% error w.r.t.\ the predicted value
from one-loop $\chi$PT.  The experimental systematic uncertainty is
expected to be smaller than the statistical
one~\cite{Denisov:2018unj}.  An additional theory uncertainty will
come from corrections to \eqref{R-K-alpha} from higher-order
contributions in the chiral expansion, which remain to be calculated.

\paragraph{Challenges and timelines:}
No specific additions to the experiments are foreseen. Potential
difficulties might be the above-mentioned small kinematic gap between
the threshold in the invariant mass $M_{K^- \gamma}$ and the first
resonance $\text{K}^*(892)$ and the smaller Primakov cross section
compared to the case of the pion.  The main experimental challenge for
this measurement lies in the availability of RF-separated beams, which
puts the time window for the required 1-year data-taking period at
earliest after LS3.

\paragraph{Worldwide landscape:}
This would be a unique benchmark measurement for low-energy QCD in the
three-flavour sector, and we are not aware of any other existing or
planned facility that would be able to perform it.


\subsubsection{Strange meson spectroscopy with kaon beams}

\begin{figure}
\begin{center}
\includegraphics[width=0.7\textwidth]{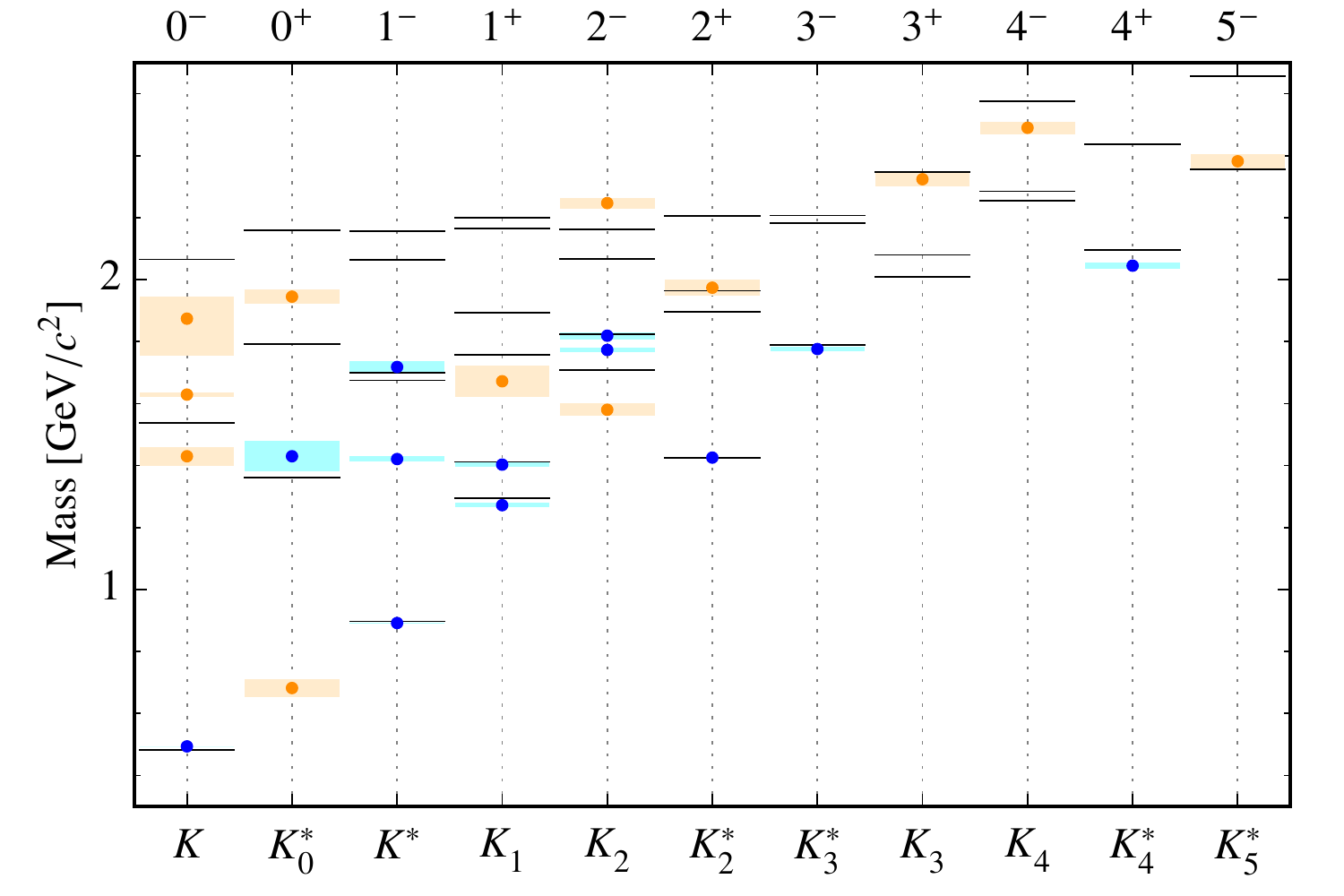}
\caption[Excitation spectrum of strange mesons grouped by their $J^P$
quantum numbers.]{\label{fig:kaon-spectrum} Excitation spectrum of
  strange mesons grouped by their $J^P$ quantum numbers.  Known states
  from the PDG (points and shaded boxes representing the central value
  and uncertainty of the measurements, respectively) are compared to
  the relativistic quark-model prediction of Ref.~\cite{Ebert:2009ub}
  (black lines).  States included also in the PDG summary table are
  shown in blue.  (Figure from \protect\cite{Denisov:2018unj}) }
\end{center}
\end{figure}

Baryon and meson spectroscopy has been a vital tool for the
understanding of the QCD bound states.  However, while the spectrum of
non-strange mesons has seen tremendous progress, also thanks to the
activities of the COMPASS collaboration, the spectrum of mesons
carrying strangeness with masses beyond 1~GeV is known with much less
precision, not to say {\em terra incognita} when going beyond the
ground states.  This is illustrated in figure~\ref{fig:kaon-spectrum},
which shows the known strange mesons in comparison to the prediction
of states from a relativistic quark-model~\cite{Ebert:2009ub}.  A more
precise knowledge of the kaon spectrum would have a broad impact, as
excited kaons appear in many processes in modern hadron and particle
physics, e.g., in the study of $CP$ violation in heavy-meson decays,
which are studied at LHCb and Belle II.

COMPASS has already developed a powerful partial-wave analysis
framework during its previous spectroscopy program with non-strange
mesons~\cite{Adolph:2015tqa}.  The M2 hadron beam, in fact, has a kaon
component. However, it is too weak to efficiently exploit it for a
detailed spectroscopy program for strange mesons.  An alternative
could be the use of high-intensity RF-separated hadron beams of at
least 50~GeV beam energy, whose feasibility is studied in a parallel
PBC activity~\cite{PBC-convbeam}.

\paragraph{Performance expectations:}
The goal of a kaon spectroscopy program with RF-separated beams would
be to map out the complete spectrum of excited kaons with
unprecedented precision~\cite{Denisov:2018unj}.

\paragraph{Challenges and timelines:}
The main experimental challenges---beyond the already mentioned
ambitious requirement of RF-separated beams---are high-precision
vertex reconstruction, photon detection with electromagnetic
calorimeters to reconstruct neutral hadrons, and final-state particle
identification (for instance, kaons have to be distinguished from
pions with high efficiency in the kinematic range of 1 GeV up to the
beam momentum).  As for the Primakov measurement
(section~\ref{sec:KPrimakoff}), these requirements situate this 1-year
measurement earliest after LS3.

\paragraph{Worldwide landscape:}
There are proposals and plans for future measurements of strange mesons at other facilities.
\begin{itemize}
\item The investigation of \(\tau\) decays, in which strange mesons
  can appear in subsystems, will be pursued at Belle2, BES III and
  LHCb to study strange mesons.  The largest possible mass of the
  strange subsystem is limited by the rather low \(\tau\) mass of
  1.8~GeV, so that many of the observed or predicted kaon states are
  out of reach.
\item GlueX will study strange mesons in photo-production at JLab12.
  Achievable photon energies are limited by the maximum energy of 12
  GeV of the electron beam delivered to Hall D.  Measurements using a
  secondary \(K_{L}\) beam are considered as
  well~\cite{Amaryan:2017ldw}.
      \item At J-PARC, a new beam line with a separated kaon beam
        is under discussion, which aims for a $K^{-}$
        intensity of 10$^{7}$ per spill at much lower beam momentum of
        2--10 GeV.  At such low momenta, the separation between beam
        and target excitations would become difficult and
        might lead to larger systematic uncertainties.
\end{itemize}

In comparison, the COMPASS++ measurement would profit from a
high-intensity high-energy kaon beam that allows for excellent
precision and clear separation of the strange-meson from the recoiling
system. It is as such unique and complementary to other world-wide
activities.


\subsubsection{Selected other COMPASS++ measurements and summary}

The realisation of RF-separated beams opens up the possibility of a
wide program of \DY measurements. Similarly to the pion PDFs
(sect.~\ref{sec:pionPDF}), kaon PDFs can be studied in unpolarized
\DY. Prompt-photon production will further help constraining the gluon
distribution of the kaon. As discussed already for LHC-FT
(sect.~\ref{sec:LHC-FT}) a transversely polarised target permits a
series of measurements related to transverse-momentum distributions
like the Sivers function. Especially attractive in that respect would
be a high-intensity \(\bar{p}\) beam. The kinematics (see, e.g.,
figure 30 in LoI~\protect\cite{Denisov:2018unj}) would be such that
the region $x \gsim 0.1$ in both beam and target would be probed,
e.g., valence quarks in the polarised proton would annihilate with
(valence) anti-quarks in the anti-proton.

Nevertheless, the overall rate is rather low. Table 5 in
LoI~\protect\cite{Denisov:2018unj} quotes about 5 to 8 $\times 10^{4}$
DY events with $M > 4 \gev$ for 140 days of beam time, thus remaining
unclear whether it would be competitive with $pp$ proposals at larger
$\sqrt{s}$, which profit from the abundance of anti-quarks at low
\(x\) in the unpolarized proton.
In the area of low-energy observables, a parasitic running of one year
along the kaon Primakov measurement would allow a direct measurement
of the neutral pion lifetime.  COMPASS++ measurements of $\bar{p}$
cross sections for air showers, possible with the existing M2 proton
beam, are discussed in Sec.~\ref{sec:cosmics}.

Altogether, the COMPASS++ proposal comprises a diverse physics program
centred around its main pillar of hadron structure. The skeleton of
the {\em Future QCD Facility} is an upgraded COMPASS spectrometer with
a number of new elements for different physics programs (see
Table~\ref{tab:COMPASS++programs}).  The whole program would require
up to 11 or 12 years of running time, but in view of competing
experiments the overall beam time request might be reduced after
further prioritisation of measurements.  The earliest starting year of
the program is 2022 as a one year long approved extension of the
COMPASS-II program is scheduled to run in 2021.

For the moment it is quite difficult to give a solid cost estimates
for new detectors and for upgrade and renewal of the COMPASS
spectrometer, \revised{a total budget in the range of 10 to 20 M CHF
  seems to be reasonable.}
According to the previous experience of the COMPASS Collaboration, in
order to carry out the COMPASS++ program, a new collaboration of
approximately 250 physicists has to be established.

The main challenge for the long-term program is a newly designed and
constructed RF-separated hadron beam, \revised{which would become a
  very versatile and world-wide unique beam line for the foreseeable
  future.  For that, studies of feasibility and estimated beam
  parameters have started} in the PBC Conventional Beam
WG~\cite{PBC-convbeam}.

\subsection{MUonE}
\label{sec:muE}


\paragraph{Physics motivation.}

\begin{figure}
\begin{center}
\includegraphics[width=0.9\textwidth,trim=80 0 145 0,clip]{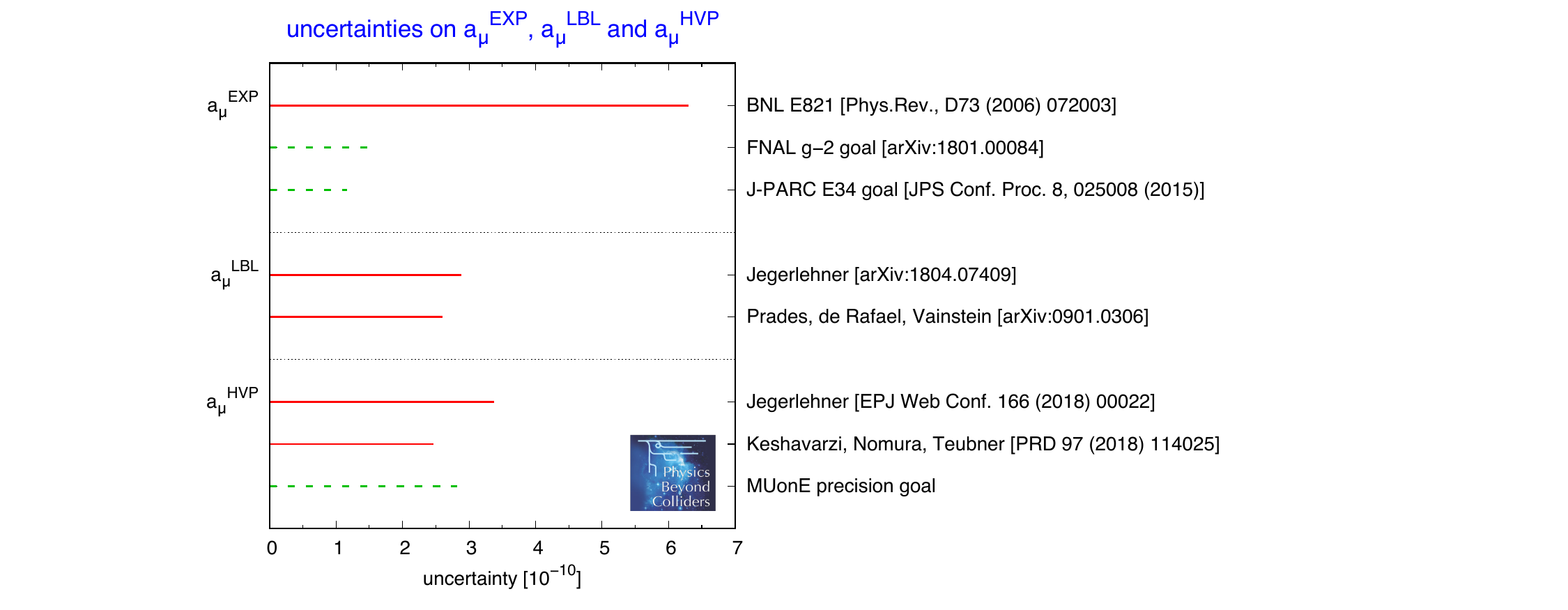}
\caption[Current experimental precision on the anomalous magnetic
moment $a_\mu$ of the muon (and the accuracy goals of the planned
experiments at FNAL and J-PARC), of two current determinations of the
contribution to $a_\mu$ from light-by-light scattering, of two
current determinations of the contribution of hadronic vacuum
polarisation to $a_\mu$, and the ultimate precision aimed at by
MUonE on the latter contribution.]{\label{fig:a-mu-acc} Top part:
  current experimental precision on $a_\mu$ and the accuracy goals of
  the planned experiments at FNAL and J-PARC.  Central part: accuracy
  of two current determinations of the contribution to $a_\mu$ from
  light-by-light scattering (LBL).  Bottom part: accuracy of two
  current determinations of the contribution of hadronic vacuum
  polarisation (HVP) to $a_\mu$, as well as the ultimate precision
  aimed at by MUonE.  The latter is given as a 0.3\% statistical error
  and a systematic error of the same size, added in quadrature.
  Numbers have in part been taken from the
  review~\protect\cite{Jegerlehner:2018zrj}.}
\end{center}
\end{figure}

The long-standing 3 to 4$\sigma$ discrepancy between the experimental
value and theoretical prediction for $(g-2)_\mu$ is among the most
prominent problems of the Standard Model.  Future measurements at FNAL
and J-PARC aim at improving the accuracy of the E821 experiment at BNL
by a factor of 4.  At the same time, it is essential to consolidate
the theoretical prediction and, if possible, further decrease its
uncertainty.  Reviews of the current state of these efforts can for
instance be found in \cite{Jegerlehner:2018zrj,Meyer:2018til}.  By far
the most important uncertainties on $a_{\mu} = (g-2)_\mu / 2$ are due
to the leading order hadronic vacuum polarisation (termed HVP in the
sequel) and to the hadronic contribution to light-by-light scattering.
Figure~\ref{fig:a-mu-acc} gives an overview of the situation.

\begin{figure}
\begin{center}
\includegraphics[width=0.9\textwidth,trim=70 0 110 0,clip]{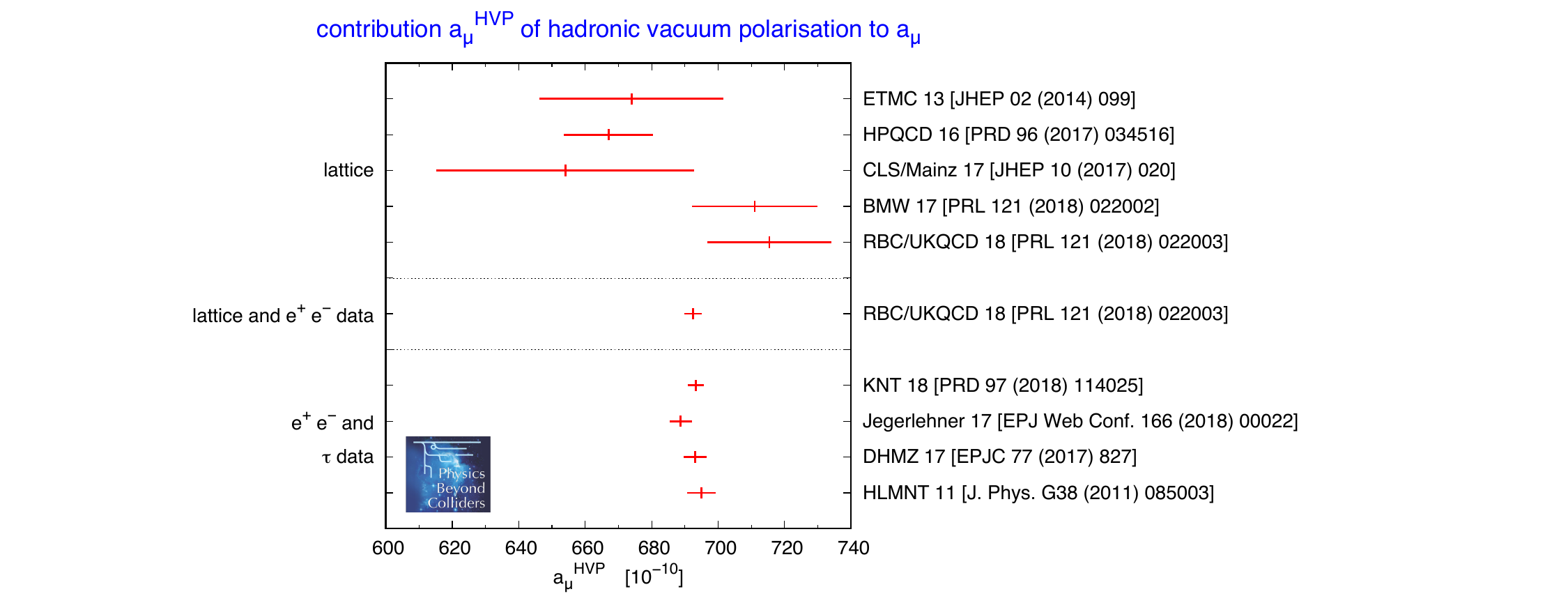}
\caption[Selected current determinations of the contribution of the
hadronic vacuum polarisation to the anomalous magnetic moment of the
muon.]{\label{fig:a-mu-hvp} Selected current determinations of the
  contribution $a_\mu^{\text{HVP}}$ to the anomalous magnetic moment
  of the muon.  Detailed information can e.g.\ be found in the
  reviews~\protect\cite{Jegerlehner:2018zrj,Meyer:2018til}.}
\end{center}
\end{figure}

The currently most precise way of calculating the leading hadronic
contribution is to use a time-like dispersion relation, which relies
on the knowledge of the $e^+ e^-$ annihilation cross section into
hadrons from the $\pi^0$ mass threshold upwards. The low energy range
is affected by large non perturbative effects, and experimental data
from $e^+e^-$ machines are needed. At present, large amounts of data
have been collected at different machines for many exclusive and
inclusive channels, through direct energy scan or radiative return
methods. The overall data set allows a determination of leading
hadronic contribution with an uncertainty of the order of
0.4--0.5\%. In the near future new precise data will be available,
\revised{notably from BES III and Belle 2, as well as CMD2 and SND at
  Novosibirsk.} However, the estimate of the systematic uncertainties,
of both experimental and theoretical origin, related to each channel
and to the global combination (e.g. the correlation among different
channels) is very delicate. In this situation, corroboration from
independent methods, like the space-like
approach~\cite{Calame:2015fva} with muon-electron scattering data of
the MUonE project~\cite{Abbiendi:2016xup} would be extremely
valuable. Concerning the evaluation of $a_\mu^{\text{HVP}}$ in lattice
QCD, in the last few years there has been a lot of progress on the
leading hadronic contribution to $(g-2)_\mu$ (as well as the hadronic
light-by-light one). In February 2018, a dedicated workshop was held
at KEK, Japan, where the recent developments of these determinations
were presented.\footnote{All presentations can be found at
  \url{https://kds.kek.jp/indico/event/26780}.}
Although their progress is very interesting and promising, the lattice
calculations of the leading hadronic contribution to $a_\mu$ are not
yet competitive with the dispersive ones.  This is clearly visible in
figure~\ref{fig:a-mu-hvp}, where the only lattice determination with
competitive errors [RBC/UKQCD18] is using a hybrid method with
$e^+ e^-$ data as input.

The review in Ref.~\cite{Jegerlehner:2018zrj} concluded that
``Therefore, the very different Euclidean approaches, lattice QCD and
the proposed alternative direct measurements of the hadronic shift
$\Delta\alpha(-Q^2)$~\cite{Abbiendi:2016xup}, in the long term will be
indispensable as complementary cross-checks.''


\paragraph{Hadronic vacuum polarisation from elastic $\mu e$ scattering.}

The MUonE experiment aims to measure the leading order hadronic
contribution to $(g-2)_\mu$ with a statistical precision of 0.3\%. The
goal in accuracy and the novel technique will make this measurement an
important ingredient to reinforce the SM prediction on $(g-2)_\mu$ and
eventually to clarify any possible hint of new physics when compared
with the expected measurements at FNAL and J-PARC.

\begin{figure}
\begin{center}
\includegraphics[width=0.84\textwidth]{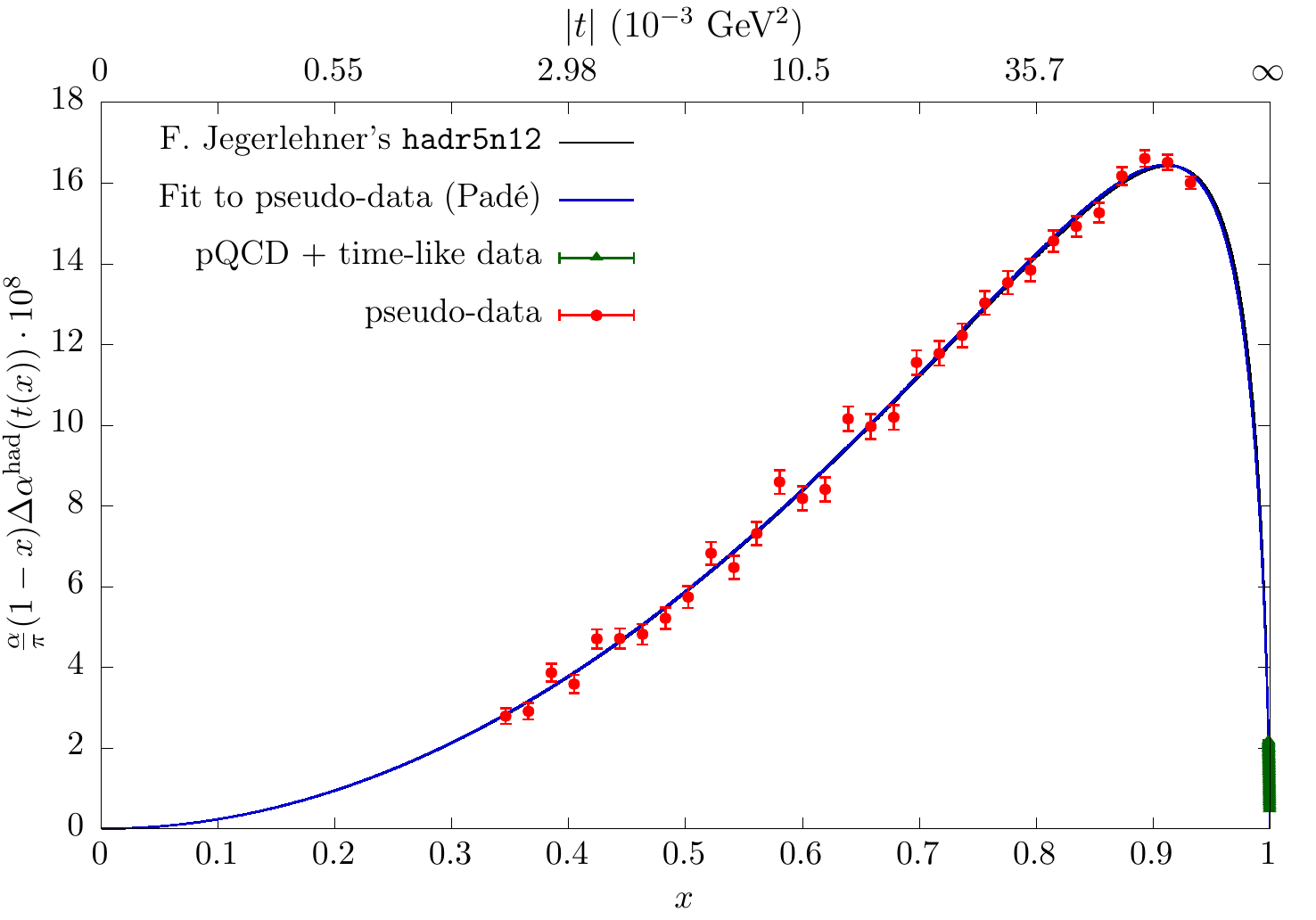}
\caption[Integrand of the sum rule that relates the leading-order
hadronic contribution to $a_\mu$ to the running fine structure
constant, together with MUonE pseudodata.]{\label{fig:muone-sumrule}
  Integrand of the sum rule \protect\eqref{muone-sum}, together with
  pseudodata and a fit described in the text below.}
\end{center}
\end{figure}

The method is based on a sum rule, which relates the leading order
hadronic contribution to $a_\mu$ to the running fine structure
constant $\alpha(t)$ as \cite{Abbiendi:2016xup}
\begin{align}
\label{muone-sum}
a_\mu^{\text{HVP}} &= \frac{\alpha(0)}{\pi} \int_0^1 dx\, (1-x)\, \Delta\alpha_{\text{had}}\bigl( t(x) \bigr) \,,
\end{align}
where $\Delta\alpha(t) = 1 - \alpha(0) / \alpha(t)$ and
$\Delta\alpha_{\text{had}}(t)$ is its hadronic component, which is
obtained by subtracting the purely leptonic part.  The scaling
variable $x$ is related to the timelike momentum transfer by
\begin{align}
t(x) &= - \frac{x^2\, m_\mu^2}{1-x} \,.
\end{align}
The integrand of the sum rule \eqref{muone-sum} is shown in
figure~\ref{fig:muone-sumrule}.  The peak of the spectrum is at
$x \approx 0.914$, which corresponds to $t \approx -0.108 \gev^2$.

In order to have a systematic error compatible with the statistical
one, a systematic uncertainty at 10$^{-5}$ level is required on the
knowledge of the $\mu e$ cross section. Although this is not a request
on the knowledge of the absolute cross section (i.e. the running of
$\alpha$ will be obtained by the ratio between a signal and a
normalisation region both of them obtained by $\mu e$ data itself) it
poses severe requests on the knowledge of the following quantities:
\begin{description}
\item[Multiple scattering:] Multiple scattering (MS) effects in the
  target affect the reconstructed angle and if not properly taken into
  account result in a systematic error. Preliminary studies indicate
  that an accuracy of the order of 1\% is required on the knowledge of
  the MS effects. Such accuracy will require dedicated test beams
  where the results of data will be compared with the prediction from
  (GEANT) simulation. Preliminary results from a test beam in 2017 are
  encouraging.~\footnote{See the presentation
    \url{https://indico.cern.ch/event/686555/contributions/2971020}.}
  A corresponding publication is being prepared.  It is important to
  stress that the MS effects (especially the tails) will be monitored
  during the run by looking at specific observables like acoplanarity.
\item[Tracking uniformity, alignment and reconstruction of angles:] It
  is important to keep the systematic error arising from
  non-uniformity of the tracking efficiency and angle reconstruction
  at the 10$^{-5}$ level.  The use of state-of-the-art silicon
  detectors is sufficient to ensure the required uniformity. The
  relative alignment of the silicon detectors will be monitored with
  the high statistics provided by the muon beam; absolute calibration
  on the transverse plane can be achieved taking advantage of the
  constrained kinematic of two-body elastic scattering. The
  longitudinal position of the silicon detector must be monitored in
  real time at the 10 microns level; this precision can be achieved
  with commercial laser systems based on interferometry.
\item[PID:] A $\mu$-e separation at 10$^{-5}$ is required in the 2-3
  mrad region where there is an ambiguity between the outgoing
  electron and muon. Given the high momentum of the emitted particle
  in this range (70 GeV) a high granularity electromagnetic
  calorimeter followed by a muon detector should be sufficient to
  reach this accuracy.
\item[Knowledge of the Beam:] A high intensity muon beam of the
  energy of 150 GeV as used by COMPASS at M2 with a few percent
  momentum spread would be suited for the measurement.  There are
  discussions with the beams experts about the possibility to have the shape
  of the M2 beam adapted to the MUonE apparatus, i.e.\ a width of several
  centimetres and parallel.  A 0.8\% percent accuracy of the beam
  momentum, as obtained by the spectrometer used by COMPASS, would be
  useful to over-constrain the $\mu-e$ kinematics and control the
  systematic effects arising from beam spread, reconstruction and MS.
\item[Background:] Background events will be rejected by cutting on
  the numbers of the hit recorded in the tracker, and on the
  elasticity band. A rejection of more than 10$^{5}$ has been obtained
  by Monte Carlo simulation.
\item[Theory:] A complete fixed order calculation of QED NNLO
  radiative corrections, consistently matched with resummation of
  higher-orders and implemented into a Monte Carlo event generator,
  will be required to control the theoretical systematics at the
  10$^{-5}$ level and, in turn, to extract the leading order hadronic
  contribution to $(g-2)_\mu$ with the aimed accuracy. A core of
  theoretical Italian groups in Padova, Parma and Pavia, together with
  international collaborators, have started to work on the MUonE
  project.\footnote{See the workshops
    \url{https://agenda.infn.it/internalPage.py?pageId=0&confId=13774}
    at Padova and \url{https://indico.mitp.uni-mainz.de/event/128} at
    Mainz.}
  \revised{Results of this activity are presented in
    \cite{Mastrolia:2017pfy,DiVita:2018nnh,Fael:2018dmz,Alacevich:2018vez}.}
\end{description}

The systematic error on the normalisation as well as uncertainty on
the model dependence of the multiple scattering, on the beam spectrum
and on the alignment of the detector will be correlated in $x$. Other
systematic errors like PID and background subtraction are expected to
affect only some specific $x$ values.

The integral in the region $0.93 < x < 1$, accounting for 13\% of the
total $a_{mu}^\text{HVP}$ integral, cannot be reached by the proposed
MUonE experiment. However, it can be determined using timelike data
and perturbative QCD, and/or lattice QCD results, with a tiny
estimated uncertainty of about one per mil.

In an exercise to validate the method at the level of statistical
errors and fitting, pseudodata was generated and subsequently fitted,
as shown in figure~\ref{fig:muone-sumrule}.  The fit result is
$a_{\mu}^{\text{HVP}} = (686.9 \pm 2.3) \times 10^{-10}$, compared
with an input value $a_{\mu}^{\text{HVP}} = 688.54 \times 10^{-10}$
used to generate the pseudo-data.  The error reflects only the
statistical uncertainty on the pseudodata and corresponds to
$0.33 \%$, consistent with the ultimate accuracy goal of the
experiment.  The additionally used information from timelike data and
perturbative QCD is represented by the points close to $x=1$ in the
figure.


\paragraph{Collaboration, estimated timeline and cost}

Several groups from Italy (Bologna, Ferrara, Milano Bicocca, Padova,
Pavia, Pisa, Trieste), Poland (Kracow), Russia (Novosibirsk), UK
(Liverpool and London) and USA (Virginia University) have started to
work on MUonE. Many of them are experts in the field of precision
physics. Years 2018-2019 will be devoted to detector optimisation
studies, simulation, Test Run and theory improvement with a final goal
to present a Letter of Intent to the SPSC. The detector construction
is expected during LS2 and the plan is to install the detector (or a
staged version of it) and start data taking in the period 2021-2024. A
very preliminary estimate of the cost gives less than \revised{10 M
  CHF.}

According to current estimates (which will be refined in the course of
further studies), a running time of around 3 years will be required to
achieve the 0.3\% statistical accuracy goal quoted above.  \revised{A
  staged approach is being envisaged.}

\subsection{NA60++}
\label{sec:NA60+}

\paragraph{Physics motivation.}
Experiments at high-energy heavy-ion colliders (RHIC, LHC) investigate
the properties of a Quark-Gluon Plasma state at high energy density
and initial temperature, but at low baryochemical potential
$\mu_{\rm B}$ (typically a few MeV). In this region of the QCD phase
diagram the transition between quark-gluon plasma (QGP) and hadronic
matter is a rapid cross-over with no latent heat. It occurs at a
pseudo-critical temperature $T_{\rm c}\sim 155$ MeV. High-energy
collider experiments cannot access the large-$\mu_{\rm B}$ region of
the phase diagram, whose study requires lower collision energies. In
this area, QCD-inspired model computations suggest that a first-order
phase transition may occur. This implies a second order critical end
point (CP) of the cross-over line. Experimentally, information on this
region is relatively poor, and its study is of chief interest for our
understanding of ultra-relativistic heavy-ion physics. The global
landscape of heavy-ion experiments is mapped out in
section~\ref{sec:HIC-Summary}, see i.e.\ the
figures~\ref{fig:PhaseStructure}, \ref{fig:InteractionRates} there. In
section~\ref{sec:HIC-Summary} the uniqueness of NA60++ as compared to
the other experiments is also discussed.

The NA60++ project aims at measuring the properties of the QGP and the
nature of the phase transition by performing an energy scan in the
range accessible to the CERN SPS (beam energies in the range 20-160
GeV/nucleon, corresponding to $\sqrt{s_{\rm NN}}\sim 6-17$ GeV). Such
an experiment could address several fundamental and still open physics
questions, and in particular 
\begin{itemize} 
\item[(i)] can a signal of the first-order
  phase transition to QGP be detected?
  \item[(ii)] can we observe a signal
    corresponding to the restoration of chiral symmetry?
    \item[(iii)] can we
study the transport properties of the high-$\mu_{\rm B}$ QGP?
\end{itemize}

An experiment based on a muon tracking system and a sophisticated
vertex spectrometer gives access to these questions. With such a
set-up, the dimuon invariant mass spectrum can be studied from
threshold up to the J/$\psi$ mass region and beyond, discriminating
between prompt and non-prompt sources. The vertex spectrometer
can also be used to detect hadronic decays of particles containing
strange and heavy quarks. Its layout is conceptually similar to that
of the past NA60 experiment, which took data in 2003-2004 with indium
and proton beams on various nuclear targets at top SPS energy and
performed accurate measurements of muon pair production.

\begin{figure}[t]
\begin{center}
\includegraphics[width=1\textwidth]{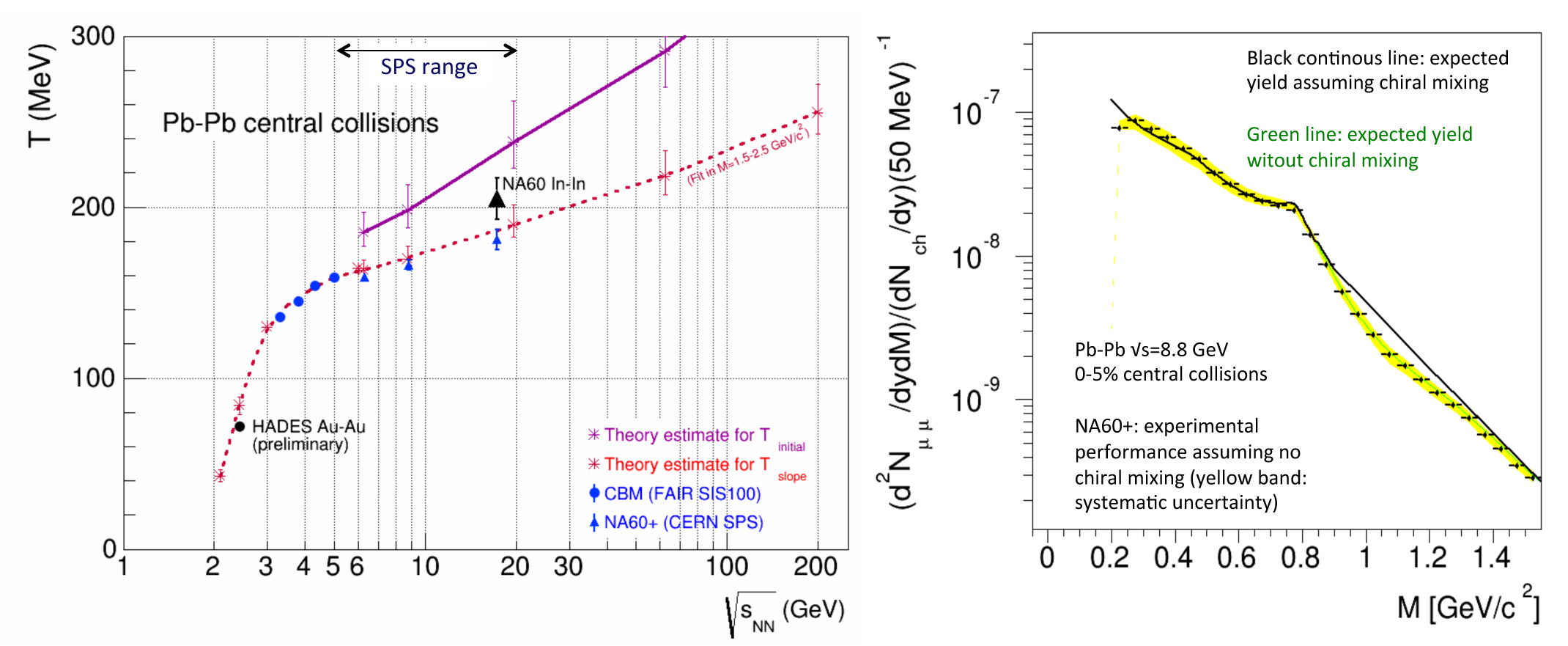}
\caption[Medium temperature evolution vs.~$\sqrt{s_{NN}}$ in central
Pb-Pb collisions with expected performance from NA60++ and CBM and the
only existing measurements (NA60 and HADES), as well as NA60++
projections for the acceptance-corrected thermal dimuon mass spectrum
at $\sqrt{s_{NN}}=8.8$ GeV in case of no chiral mixing compared to the
theoretical expectation.]{Left: medium temperature evolution
  vs.~$\sqrt{s_{NN}}$ in central Pb-Pb collisions.
  $T_\textrm{initial}$ (magenta points) and $T_\textrm{slope}$ (red
  points) are theoretical estimates for the initial medium temperature
  and the temperature from dilepton spectra
  respectively~\cite{RAPP2016586}, and a coarse graining approach in
  URQMD~\cite{PhysRevC.92.014911,GALATYUKQM2018}. Blue triangles and
  squares are the expected performance from NA60++ and CBM. The only
  existing measurements at present are from NA60 In-In
  data~\cite{Arnaldi:2008er,Specht:2010xu} and from HADES preliminary
  Au-Au data~\cite{HADESQM2018}. Right: NA60++ projection for the
  acceptance corrected thermal dimuon mass spectrum at
  $\sqrt{s_{NN}}=8.8$ GeV in case of no chiral mixing (the yellow band
  is the systematic uncertainty from combinatorial background
  subtraction) compared to the theoretical expectation (green dashed
  line). The black line above 1 GeV is the expectation from full
  chiral mixing~\cite{RAPP2016586}. The experimental precision makes
  the experiment fully sensitive to the yield increase of $\sim30\%$
  expected in 1 GeV $<M<$ 1.4 GeV in case of chiral mixing. }
\label{fig:T-Flattening+ChiralMixing}
\end{center}
\end{figure}

This brings us to the observables whose measurement can answer the
physics questions posed above. The nature of the phase transition can
be investigated by studying the evolution of the temperature of the
system, for various collision energies, as a function of its energy
density. The temperature can be precisely extracted by a measurement
of the mass spectrum of the thermal muon pairs in the mass region
1.5-2.5 GeV, while the energy density is estimated via charged
multiplicity measurements in the vertex spectrometer. In this way a
``caloric curve'' for the phase transition can be determined and the
flattening related to a first order phase transition experimentally
detected, see left plot in
figure~\ref{fig:T-Flattening+ChiralMixing}. The feasibility of the
temperature measurement via thermal dimuons was clearly demonstrated
by NA60, which obtained at top SPS energy average temperatures of the
order of 200 MeV for the created strongly interacting system.

%

The approach to the restoration of chiral symmetry, expected in the
proximity of the phase transition to the QGP, should modify in a
detectable way the spectral function of the vector/isovector mesons
($\rho$-$a_1$). A strong broadening of the $\rho$ was already observed
by NA60, and it is considered a manifestation of chiral
restoration. Additionally, while the $a_1$ cannot be directly observed
in the dilepton spectrum, chiral mixing of the vector and axial vector
spectral functions is expected to increase the dilepton yield by
$\sim30\%$ in the mass region 1-1.5 GeV, see right plot in
figure~\ref{fig:T-Flattening+ChiralMixing}. This is a direct
manifestation of chiral restoration and NA60++ aims at measuring the
related yield excess for the first time. Such a study is particularly
intriguing at low beam energies where the contribution of the QGP
phase, which would constitute a background to this measurement,
becomes less important.

\begin{figure}[t]
\begin{center}
\includegraphics[width=0.43\linewidth]{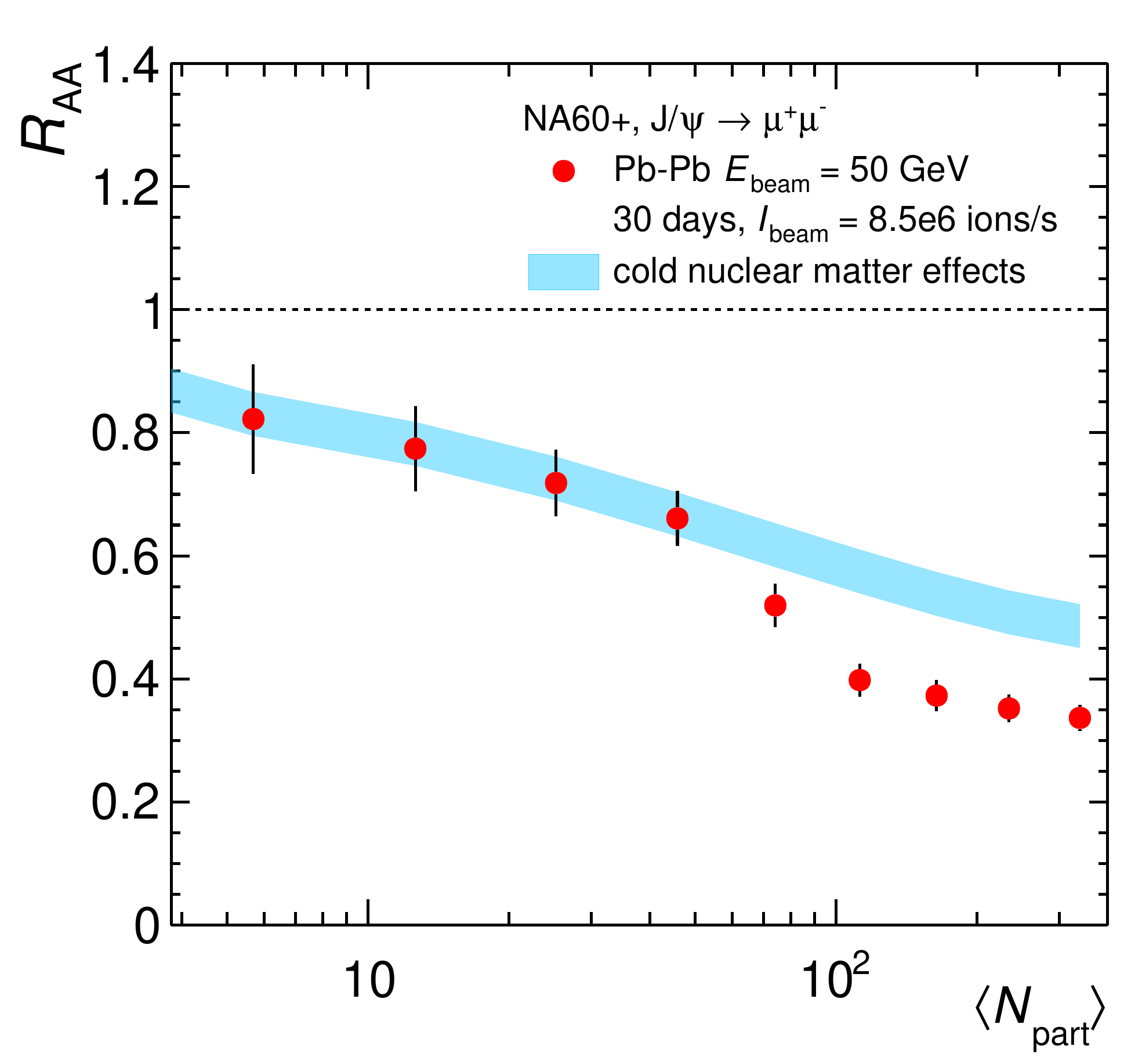}
\caption[NA60++ projected performance for the nuclear modification
  factor of J/$\psi$ production in Pb--Pb collisions at
  $E_{\rm lab}=50$ GeV as a function of the number of participants
  $N_{\rm part}$, compared with expectations from an estimate for cold
  nuclear matter effects similar as at full SPS energy.]{NA60++ 
  projected performance for the nuclear modification
  factor of J/$\psi$ production in Pb--Pb collisions at
  $E_{\rm lab}=50$ GeV as a function of the number of participants
  $N_{\rm part}$, compared with expectations from an estimate for cold
  nuclear matter effects similar as at full SPS energy, shown as a
  blue band. An extra anomalous suppression of $J/\psi$ in the quark
  gluon plasma on the level of $20\%$ is assumed to settle in for
  $N_{part}>50$.}
\label{fig:Jpsiresult}
\end{center}
\end{figure}

When decreasing the incident beam energy, the initial energy density
of the created strongly interacting system also decreases, while the
net baryon content increases, leading to a high-$\mu_{\rm B}$ QGP if
the critical temperature is attained. At high energy, it has been
shown that such a system can selectively melt the quarkonium states
according to their binding energy, via a colour screening mechanism in
the QGP. At low energies, one should therefore observe the suppression
effect to become weaker and then disappear when the initial
temperature of the system does not exceed the critical temperature. By
measuring the J/$\psi$ or $\psi(2S)$ decay to $\mu^+\mu^-$, and
possibly the $\chi_{\rm c}\rightarrow {\rm J}/\psi\,\gamma$ process,
one could therefore track the onset of the deconfinement transition
See figure~\ref{fig:Jpsiresult} for the evaluation of the J/$\psi$
nuclear modification factor.

\begin{figure}[t]
\begin{center}
\includegraphics[width=0.45\linewidth]{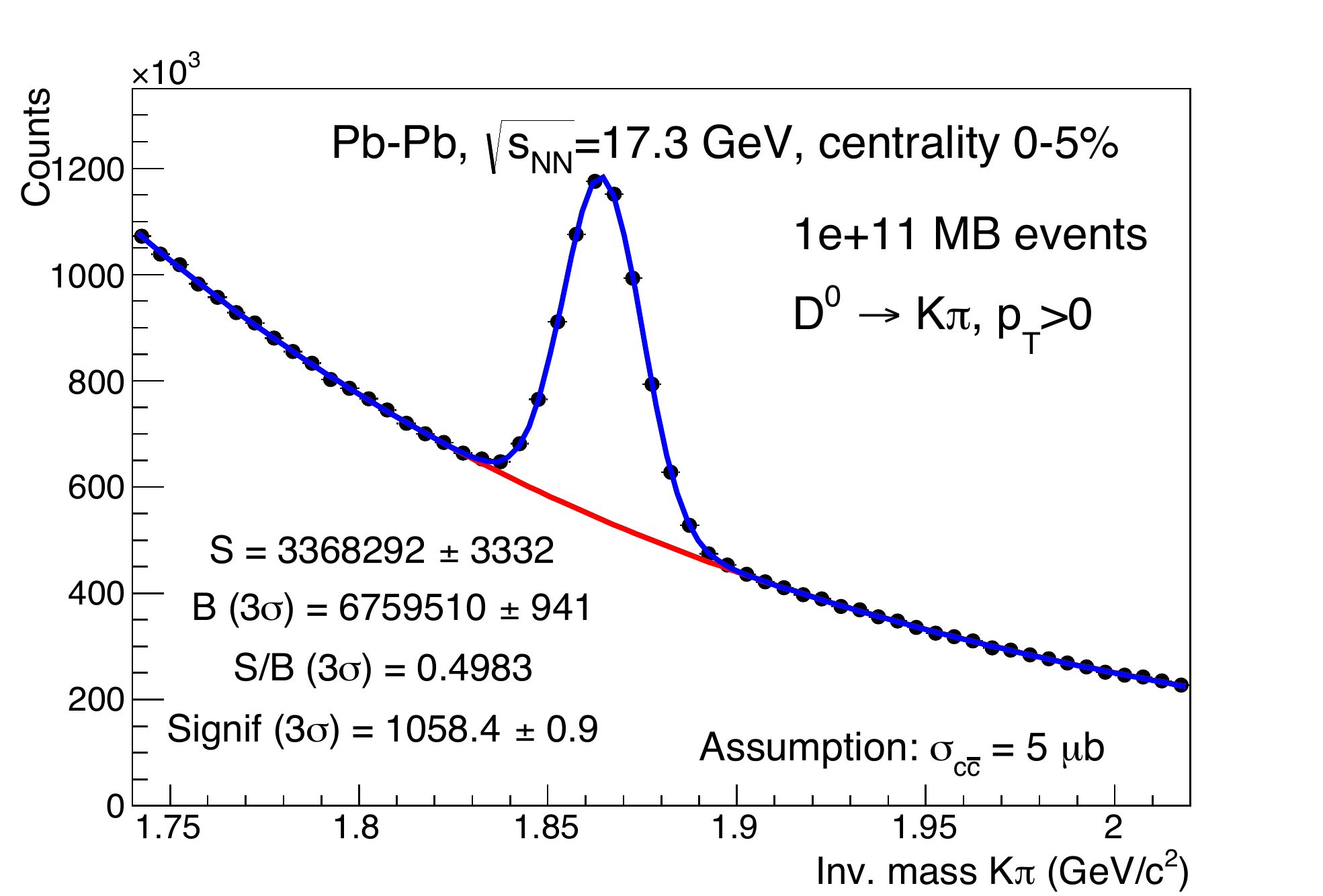}
\includegraphics[width=0.45\linewidth]{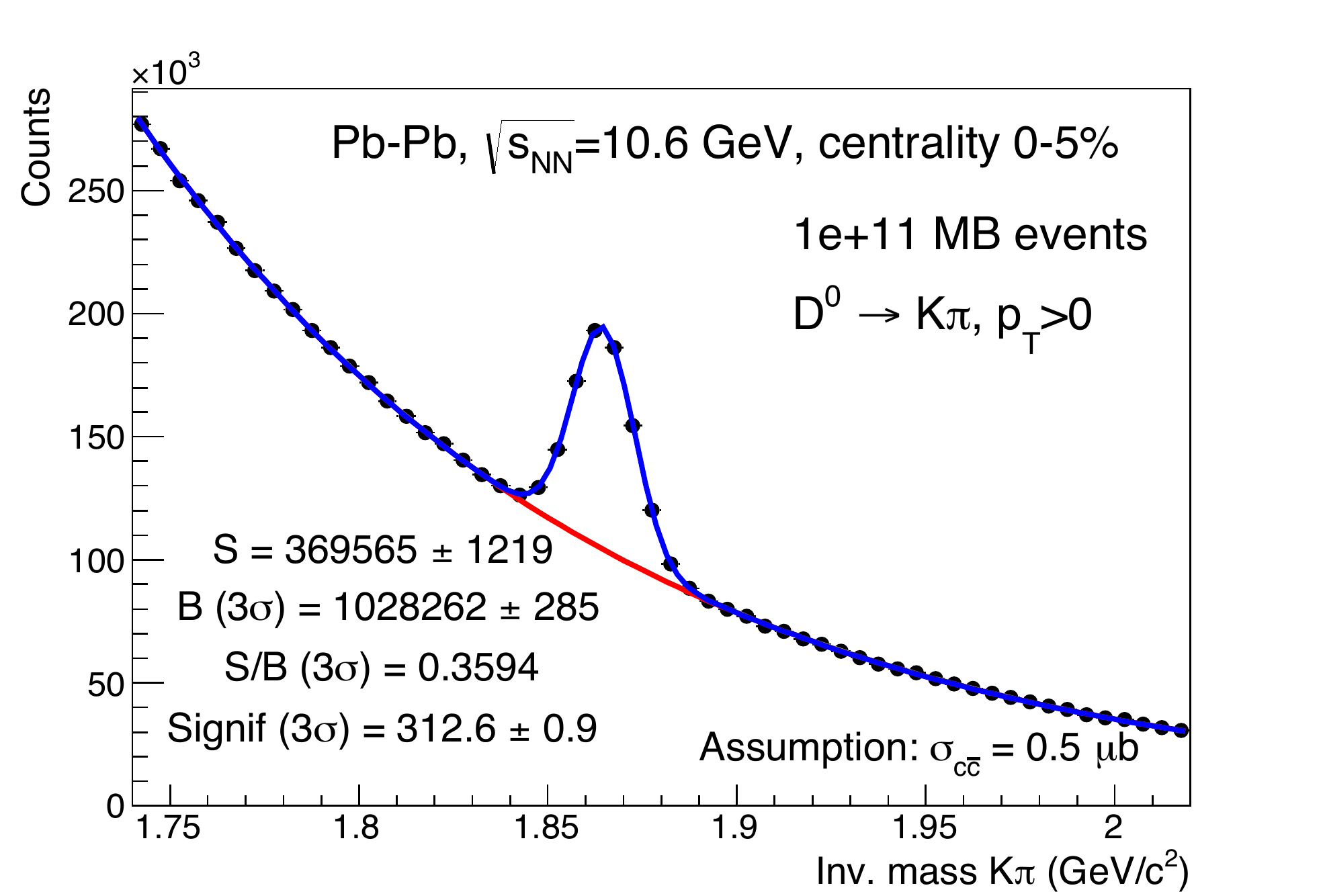}
\caption[NA60++ projection for the ${\rm D}^0$ invariant-mass
  distribution from $5\cdot 10^{9}$ central Pb--Pb events at beam
  energies of $\sqrt{s_{\rm NN}} = 17.3$~GeV and 10.6~GeV
  with a vertex detector based on MAPS.]
  {NA60++ projection for the ${\rm D}^0$ invariant-mass
  distribution from $5\cdot 10^{9}$ central Pb--Pb events at beam
  energies of $\sqrt{s_{\rm NN}} = 17.3$~GeV (left) and 10.6 GeV
  (right) with a vertex detector based on MAPS. The yield per event is
  estimated assuming $\sigma_{\rm c\bar{c}}=5$ $\mu$b at
  $\sqrt{s_{\rm NN}} = 17.3$~GeV (based on
  ~\cite{Lourenco:2006vw,Vogt:2001nh} and POWHEG) and
  $\sigma_{\rm c\bar{c}}=0.5$ $\mu$b at
  $\sqrt{s_{\rm NN}} = 10.6$~GeV.}
\label{fig:D0invmass}
\end{center}
\end{figure}

In addition to charmonium measurements, also the detection of open
charm mesons and baryons represents an important observable in the low
SPS energy range. Their hadronic decay products can be measured in the
vertex spectrometer, see figure~\ref{fig:D0invmass} for the expected
performance of the $D^0$ measurement.  An indirect measurement via
muon pairs from simultaneous semileptonic decays of meson or baryon
pairs is also feasible, see figure~\ref{fig:D0invmass}. Indeed the
charm diffusion coefficient is predicted to be larger in the hadronic
phase at temperatures approaching the critical temperature $T_{\rm c}$
from below than in the QGP phase at temperatures larger than
$T_{\rm c}$, and the system is expected to spend a relatively longer
time in the hadronic phase when the collision energy becomes lower.
For what concerns hadronisation mechanisms, recombination effects
could lead to a large enhancement of the $\rm \Lambda_c/D$ ratio. The
enhancement could be larger at low SPS energy than at RHIC and LHC
energies, because of the larger baryon content of the system.
%
%

%
%

\paragraph{Detector.} This rich physics program can be addressed by an
experiment which includes a high-resolution vertex spectrometer,
consisting of 5 silicon pixel tracking planes immersed in a 1.2 Tm
dipole field. The tracker will be based on a new generation of CMOS
Monolithic Active Pixel Sensors (MAPS), developed in close synergy
with the ALICE experiment, immersed in a dipole magnetic field. The
goal is to obtain very large area sensors up to $\sim$14x14 cm$^2$,
retaining a thickness of 50 $\mu$m or less.  The vertex spectrometer
will be followed by a thick absorber (mainly graphite) that will
filter out hadrons and finally by a muon spectrometer, with a toroidal
magnet and several large planes of tracking and triggering
detectors. In particular, for the 4 tracking planes, the use of GEM
detectors providing spatial resolution $<100$ $\mu$m, time resolution
$<10$ ns and sustaining high particle rates (the current requirement
is $\sim$10 kHz/cm$^2$) is currently foreseen, with a total surface of
the order of 100 m$^2$. Finally, for the muon triggering detectors,
which will be placed behind a further hadron absorber, the default
choice is the use of RPCs, which have been shown to provide good
resistance to ageing and can easily sustain the expected maximum rate
of $\sim$100 Hz/cm$^2$.

\paragraph{Uniqueness.} This rich physics programme can only be
pursued at the CERN SPS. Although other facilities for the study of
ultra-relativistic heavy-ion collisions in the high-$\mu_{\rm B}$
domain are being built (FAIR, NICA) they are either lacking the
necessary interaction rate (NICA) or reach too low energy (SIS100
@FAIR) to address all the physics topics detailed above. Similarly,
the beam energy scan program at RHIC (BES-II) covers the same energy
range as the CERN SPS, but with interaction rates lower by orders of
magnitude.

\paragraph{Location.} In order to access all the observables detailed
above, a Pb ion beam intensity $\sim 10^7$s$^{-1}$ impinging on a
$\sim 4$ mm Pb target is required, leading to interaction rates of the
order of 1 MHz. This choice is dictated in particular by the need of a
high precision for the temperature measurement and by the relatively
low charmonium production cross section at SPS energy. Such an
intensity can be provided by using the existing ECN3 underground hall
in the CERN North Area.

\paragraph{Collaboration, estimated timeline and cost.}

At present a proto-collaboration is established with about 60
physicists from the following institutions:

University of Lyon, CNRS/IN2P3, France,
University of Frankfurt, Germany,
University  of Heidelberg, Germany,
Technical University Munich, Germany,
Saha Institute of Nuclear Physics, India,
INFN Cagliari, Italy,
University of Cagliari, Italy,
INFN Padova, Italy,
University of Padova, Italy,
INFN Torino, Italy,
Politecnico di Torino, Italy,
University of Torino, Italy,
University of Piemonte Orientale, Italy,
Tohoku University, Japan,
CERN, Switzerland,
Rice University, USA,
Stony Brook University, USA.

The estimated cost is in the 15-25 MCHF range. As indicated in
figure~\ref{fig:timelines} for the timeline, the experiment might take
data during LHC run 4 and run 5."

\subsection{NA61++}
\label{sec:NA61+}

\paragraph{Physics motivation.}

The continuation of the \NASixtyOne programme concerns measurements of hadron and nuclear
fragment production properties in reactions induced by hadron and ion
beams after the Long Shutdown 2.  The measurements are requested by
heavy ion, cosmic ray and neutrino communities and they will include:
\begin{itemize}
\item[(i)] measurements of charm hadron production in Pb+Pb collisions for heavy ion physics,
\item[(ii)] measurements of nuclear fragmentation cross section for cosmic ray physics,
\item[(iii)] measurements of hadron production induced by proton and kaon beams for neutrino physics.
\NASixtyOne is the only experiment which will conduct such measurements in the near future.
\end{itemize}

\begin{figure}[t]
\begin{center}
  \includegraphics*[width=110mm,angle=0.]{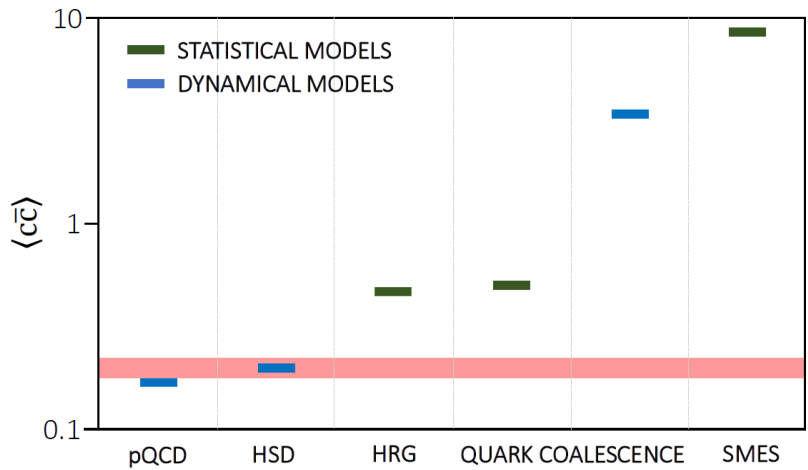}
\end{center}
\caption[Mean multiplicity of charm quark pairs produced in central
  Pb+Pb collisions at 158$A\,$GeV/$c$ calculated within dynamical
  models and statistical models, compared to the foreseen accuracy
  of the NA61/SHINE 2020+ result.]{Mean multiplicity of charm quark pairs produced in central
  Pb+Pb collisions at 158$A\,$GeV/$c$ calculated within dynamical
  models (blue bars): pQCD-inspired
  \cite{Gavai:1994gb,BraunMunzinger:2000px}, HSD \cite{Linnyk:2008hp},
  and Dynamical Quark Coalescence \cite{Levai:2000ne}, as well as
  statistical models (green bars): HRG [48], Statistical Quark
  Coalescence \cite{Kostyuk:2001zd}, and SMES
  \cite{Gazdzicki:1998vd}. The width of the red band, at the
  location assuming HSD predictions, shows the foreseen accuracy
  of the NA61++ result. (Figure taken from
  \cite{Aduszkiewicz:2309890}.)}
\label{fig:ModelsNewImpact}
\end{figure}

\begin{figure}[t]
\begin{center}
\includegraphics*[width=95mm,angle=0.]{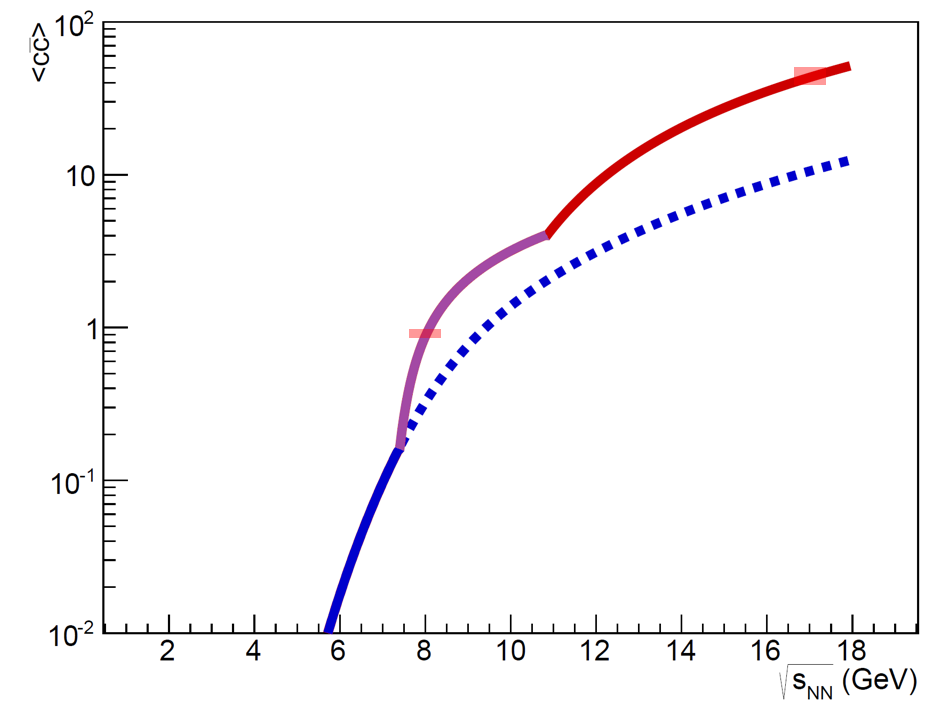}
\end{center}
\caption[Energy dependence of $\langle c\bar{c}\rangle$ in central
Pb+Pb collisions calculated within the SMES model in comparison to the
foreseen accuracy of the NA61/SHINE results for
$\sqrt{s_{NN}} = 8.6$\,GeV and $ \sqrt{s_{NN}} = 16.7$\, GeV.]{Energy
  dependence of $\langle c\bar{c}\rangle$ in central Pb+Pb collisions
  calculated within the SMES model. The red bars show the foreseen
  accuracy of the NA61/SHINE results for two energies: 40A
  GeV$\, (\sqrt{s_{NN}} = 8.6$\,GeV) and 150A
  GeV$\, (\sqrt{s_{NN}} = 16.7$\, GeV), assuming the SMES model yields
  \cite{Gazdzicki:1998vd}. (Figure taken from
  \cite{Aduszkiewicz:2309890}.)}
\label{fig:charm_resolution3}
\end{figure}

\begin{figure}[t]
\begin{center}
\includegraphics*[width=95mm,angle=0.]{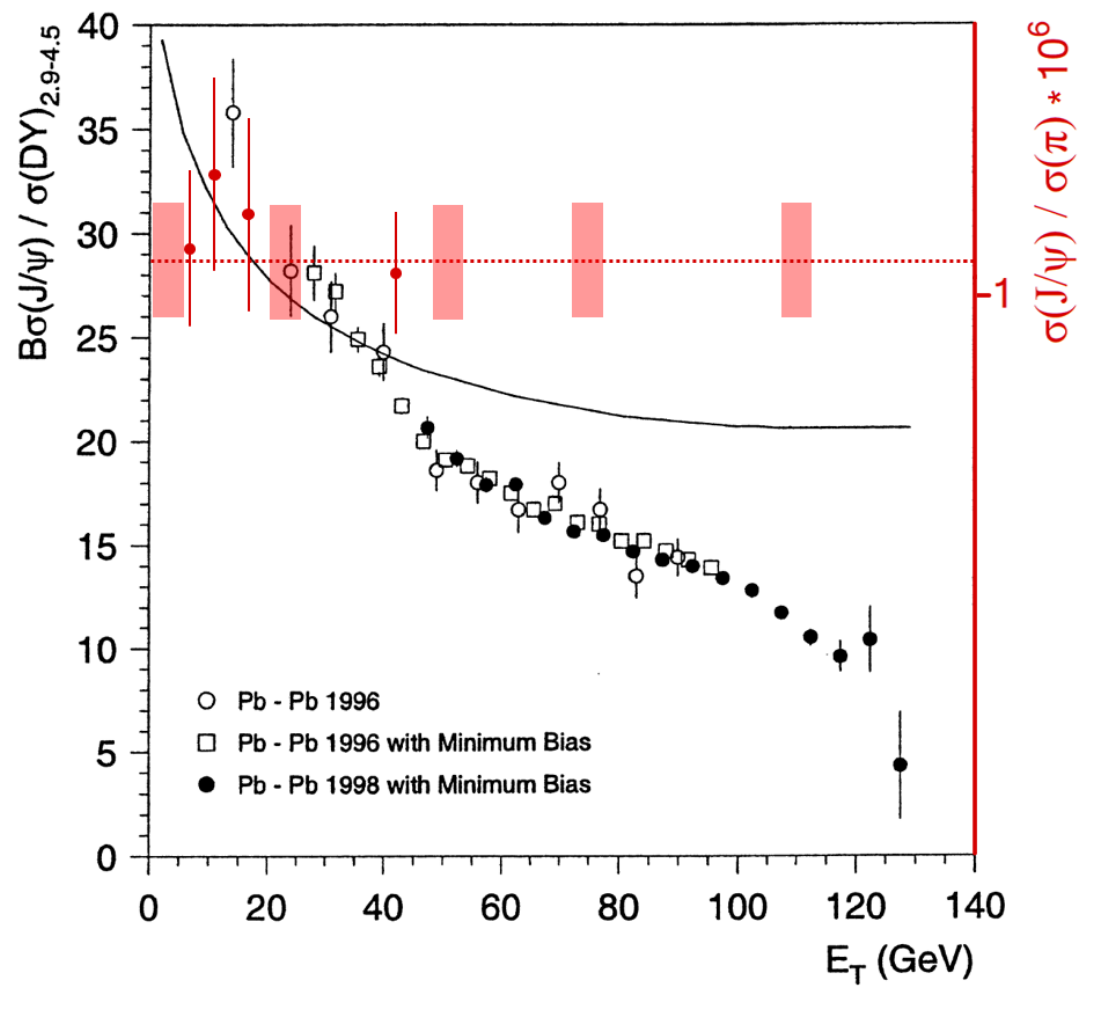}
\end{center}
\caption[The ratios $\sigma_{J/\psi}/\sigma_{\textrm{DY}}$
  and $\sigma_{J/\psi}/\sigma_\pi$ as functions of transverse
  energy in Pb+Pb collisions at 158$A$\, GeV,
  in comparison to the expected accuracy of
  $\sigma_{J/\psi}/\sigma_{c\bar c}$ from NA61/SHINE 2020+.]
  {The ratio of $\sigma_{J/\psi}/\sigma_{\textrm{DY}}$ (left)
  and $\sigma_{J/\psi}/\sigma_\pi$ (right) as a function of transverse
  energy in Pb+Pb collisions at 158$A$\, GeV. The
  $\sigma_{J/\psi}/\sigma_{\textrm{DY}}$ ratio was measured by NA50
  \cite{Abreu:2000ni}, and was used to calculate the
  $\sigma_{J/\psi}/\sigma_\pi$ ratio in \cite{Gazdzicki:1998jx}. Red
  bars mark the expected accuracy of the
  $\sigma_{J/\psi}/\sigma_{c\bar c}$ result of NA61/SHINE 2020+
  assuming $\sigma_{c\bar c} \propto \sigma_\pi$ and scaled to the
  $\sigma_{J/\psi}/\sigma_{\textrm{DY}}$ ratio in peripheral
  collisions. (Figure taken from \cite{Aduszkiewicz:2309890}.)}
\label{fig:JpsiDY4}
\end{figure}

\noindent The objective of {\bf charm hadron production measurements} in Pb+Pb
collisions is to obtain the first data on mean number of $c\bar{c}$
pairs produced in the full phase space in heavy ion collisions. Moreover, first results on the
collision energy and system size dependence will be provided.  This,
in particular, should significantly help to answer the questions:
\begin{itemize}
\item[ (i)]	What is the mechanism of open charm production? \\[1ex]
  The mechanism for open charm production is either of dynamical or of
  predominantly statistical nature. NA61++ offers the unique
  possibility to single out the production mechanism, see
  figure~\ref{fig:ModelsNewImpact}.
\item[(ii)] How does the onset of deconfinement impact open charm production?\\[1ex]
  The charm production in deconfined matter (dominantly $\bar c c$
  pairs) is expected to be more abundant than in the confined phase
  (dominantly $\textrm{D} \bar{\textrm{D}}$-pairs) due to the higher production energy
  (about 1 GeV) of the latter. A model computation for the energy
  dependence is shown in figure~\ref{fig:charm_resolution3}.
\item[(iii)] How does the formation of quark-gluon plasma impact
  $J/\psi$ production?\\[1ex]
  $J/\psi$-suppression has been used as an important indication for
  the formation of the quark-gluon plasma. So far it has been accessed
  by an assumed proportionality of the mean multiplicity of
  $\bar c c $-pairs to that of Drell-Yan pairs,
  $\langle c\bar c\rangle \sim \langle \textrm{DY}\rangle $. This
  assumption is challanged by the fact that the ratio
  $\sigma_{J/\psi}/\sigma_\pi$ is independent of centrality and close
  to the prediction of the hadron-resonance gas, 
  see figure~\ref{fig:JpsiDY4}.
\end{itemize}

\noindent The objective of {\bf nuclear fragmentation cross section
  measurements} is to provide high-precision data needed for the
interpretation of results from current-generation cosmic ray
experiments.  The proposed measurements are of paramount importance to
extract the characteristics of the diffuse propagation of cosmic rays
in the Galaxy. A better understanding of the cosmic-ray propagation is needed to
\begin{itemize}
\item[ (i)] study the origin of Galactic cosmic rays and
\item[ (ii)] evaluate the cosmic-ray background for signatures of
  astrophysical dark matter.
\end{itemize}
The objectives of {\bf new hadron production measurements for neutrino
  physics} are
\begin{itemize}
\item[ (i)] to improve further the precision of hadron production
  measurements for the currently used T2K replica target, paying
  special attention to the extrapolation of produced particles to the
  target surface,
\item[ (ii)] to perform measurements for a new target material (super-sialon),
both in thin target and replica target configurations, for T2K-II and
Hyper-Kamiokande,
\item[ (iii)] to study the possibility of measurements at low incoming beam
momenta (below 12~GeV) relevant for improved predictions
of both atmospheric and accelerator neutrino fluxes,
\item[ (iv)] to ultimately perform hadron production measurements with
prototypes
of Hyper-Kamiokande and DUNE targets.
\end{itemize}

\paragraph{Detector.}
The new measurements require upgrades of the \NASixtyOne detector that
shall increase the data taking rate to about 1~kHz. These are:
\begin{itemize}
\item[ (i)] construction of a new Vertex Detector,
\item[ (ii)] replacement of the TPC read-out electronics,
\item[ (iii)]  construction of a new trigger and data acquisition system,
\item[ (iv)] upgrade of the Projectile Spectator Detector.
\end{itemize}
Furthermore, the construction of new Time-of-Flight detectors would be
highly desirable for potential future measurements of hadron
production in C+C and Mg+Mg collisions which are expected to be needed
to understand the onset of fireball phenomenon.  The detector upgrade
is planned to be executed during the LS2 and the measurements are
scheduled in the period 2021-2024.  The total upgrade cost is
estimated to be about 2M~CHF.

\paragraph{Collaboration, estimated timeline and cost.}

The collaboration consists of about 150 physicists from 30
institutions and 14 countries. Detailed description can be found in
Ref.~\cite{Aduszkiewicz:2309890}.

\subsection{DIRAC++}
\label{sec:DIRAC}


The DIRAC++ proposal aims at performing the first high-precision
measurement of the scattering lengths in the $\pi K$ system, and at
further improving the accuracy of the corresponding results in the
$\pi \pi$ sector.  As mentioned in section~\ref{sec:motivation}, these
are key quantities for the study of QCD in the low-energy limit, where
much of the dynamics follows from chiral symmetry breaking and can be
described within chiral perturbation theory ($\chi$PT).

\begin{figure}[b]
\begin{center}
\includegraphics[width=0.85\textwidth,trim=115 0 120 0]{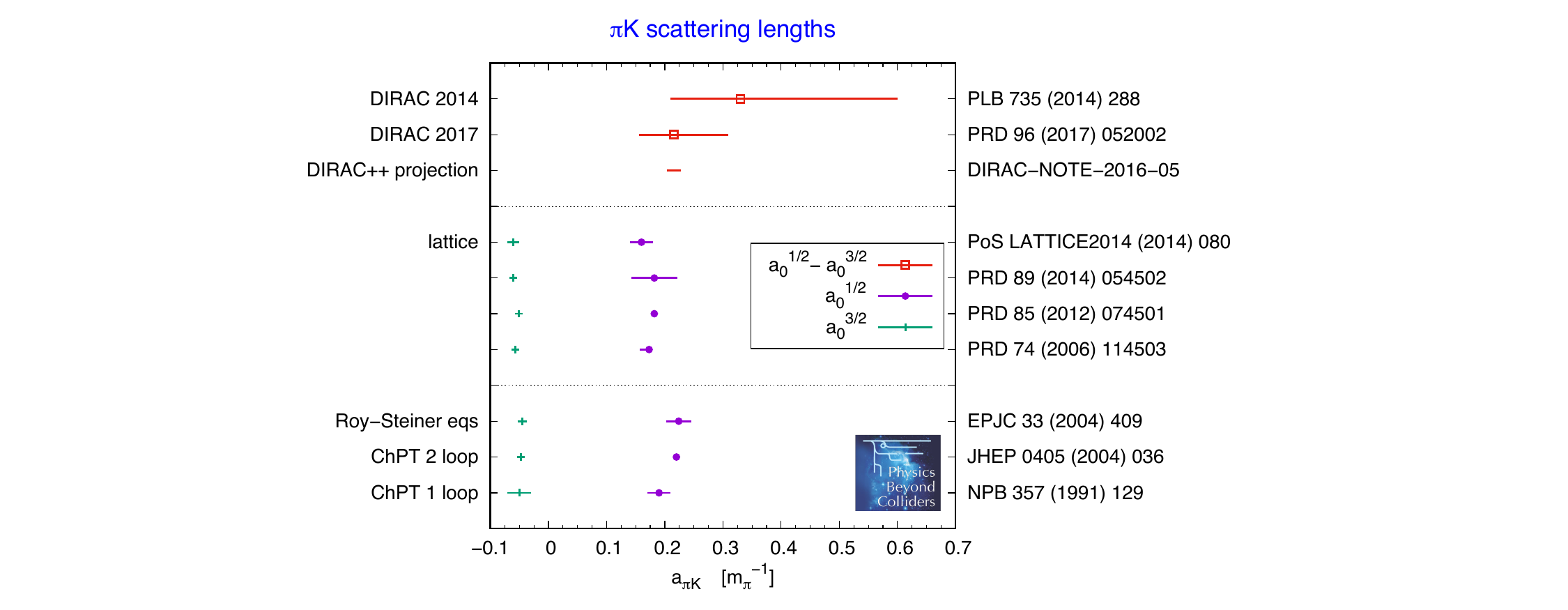}
\caption{\label{fig:a-Kpi-overview} Theoretical predictions for the
  $\pi K$ scattering lengths in the isospin $1/2$ and $3/2$ channels,
  together with the DIRAC measurements of the difference
  $a_0^{1/2} - a_0^{3/2}$ and the projected uncertainty of a
  measurement with DIRAC++.}
\end{center}
\end{figure}

\paragraph{Precise measurements in the $\pi K$ sector.}
The $S$ wave $\pi K$ scattering lengths $a_{1/2}$ and $a_{3/2}$ (where
the subscripts denote the isospin channel) are zero in the limit of
chiral symmetry and hence particularly sensitive to chiral symmetry
breaking.  An overview of theoretical predictions is shown in
figure~\ref{fig:a-Kpi-overview}.  Experimentally, the combination
$|a_{1/2} - a_{3/2}|$ can be extracted from the lifetime of the
short-lived $\pi K$ atoms $A_{\pi^+ K^-}$ and $A_{\pi^- K^+}$ in the
ground state; the relation between the two quantities has a
theoretical precision of 1\%.  Such an extraction has been achieved by
the DIRAC experiment at the CERN PS with a proton beam momentum of 24
GeV/$c$: the DIRAC collaboration observed $349 \pm 62$ $\pi K$ atomic
pairs after breakup of the produced $\pi K$ atoms, measured the
$\pi K$ atom lifetime, and extracted $|a_{1/2} - a_{3/2}|$ with a
precision of 34\% \cite{Adeva:2017oco}.

As estimated in \cite{Gorchakov:2016thm}, the number of
$A_{\pi^+ K^-}$ and $A_{\pi^- K^+}$ produced per time unit at SPS CERN
at $E_p = 450 \gev$ would be 53$\pm$11 and 24$\pm$5 times higher than
in the DIRAC experiment. For an experimental setup with the same
parameters as in the DIRAC experiment and 5 months running time, this
results in an expected number $n_{A}$ =13000 of $\pi K$ atomic pairs.
The statistical (systematic) precision of the $\pi K$ scattering
length $|a_{1/2} - a_{3/2}|$ is estimated to be around 5\% (2\%)
\cite{Yazkov:2207227}.  As seen in figure~\ref{fig:a-Kpi-overview},
such a precision is highly competitive at the scale of theoretical
predictions, concerning both their spread and quoted errors.

\begin{figure}
\begin{center}
\includegraphics[width=0.47\textwidth]{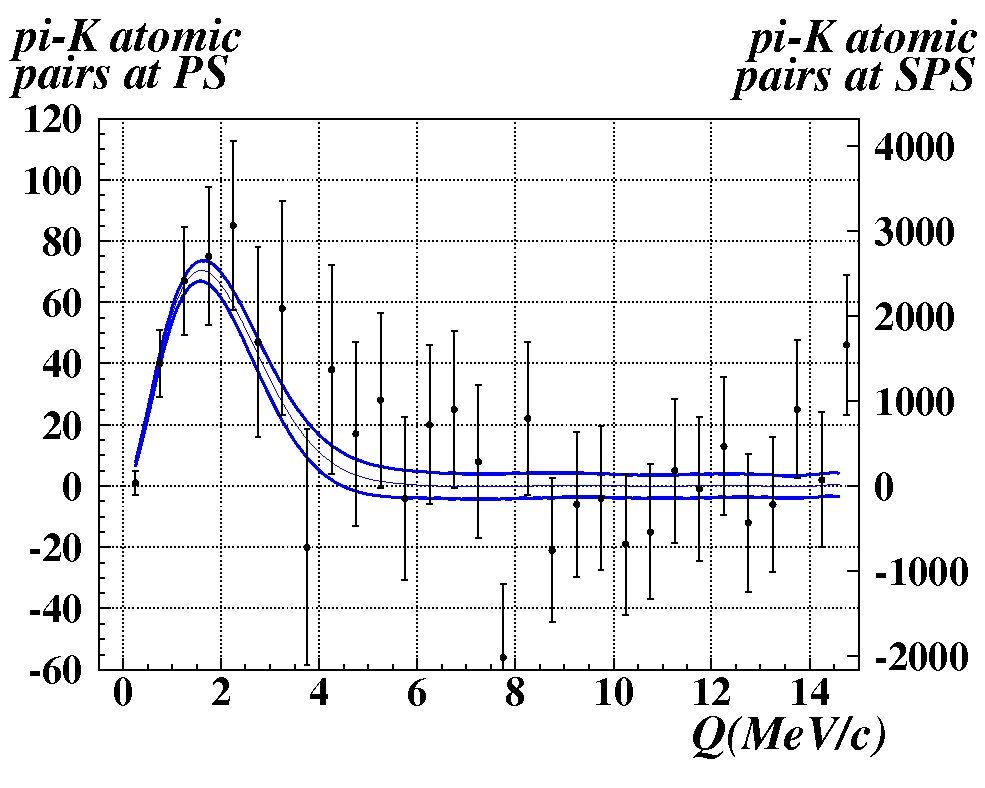}
\caption[Change of statistical uncertainties on atomic $\pi K$ pairs
between the PS (DIRAC) and the SPS
(DIRAC++).]{\label{fig:DIRAC-Q-spec} Change of statistical
  uncertainties on atomic $\pi K$ pairs between the PS (DIRAC) and the
  SPS (DIRAC++), as explained in the text.}
\end{center}
\end{figure}

Figure~\ref{fig:DIRAC-Q-spec} illustrates the gain in statistics
between the PS and the SPS.  Shown is the distribution of atomic
$\pi K$ pairs in their relative momentum $Q$, from which the number of
produced atoms extracted.  The points with errors and the left scale
refer to the DIRAC measurement, where $\pi^- K^+$ and $\pi^+ K^-$
349$\pm$62 atomic pairs were observed.  The blue error band and the
scale on the right correspond to the projection of 13000 atomic pairs
produced at DIRAC++ during 5 months of data taking.  The PS and SPS
distributions are normalised to the same area in the interval of $Q$
from 0 to 1.5 MeV/$c$.

\paragraph{Improvements in the $\pi \pi$ sector.}
In the same experiment, a large number of $\pi^+ \pi^-$ atoms would be
observed.  Theoretical calculations of the $S$-wave $\pi \pi$
scattering lengths $a_{0}$ and $a_{2}$ have reached a high degree of
accuracy: lattice calculations quote a relative precision of 4-10\%
and 1\% for $a_{0}$ and $a_{2}$, respectively
\cite{Beane:2007xs,Feng:2009ij,Fu:2013ffa,Sasaki:2013vxa}. Chiral
Perturbation Theory predicts $a_{0}$ and $a_{2}$ with a precision of
2.3\% and $|a_{0} - a_{2}|$ with an accuracy of 1.5\%
\cite{Colangelo:2001df}. For the short-lived $\pi^+ \pi^-$ atom
($A_{2\pi}$) in the ground state, the lifetime is related to the
scattering length combination $|a_{0} - a_{2}|$. The most accurate
experimental measurements of $a_{0}$ and $a_{2}$ have a precision of
6\% and 22\% respectively and were done by NA48 using kaon decays
\cite{Batley:2000zz,Batley:2010zza}).  The combination
$|a_{0} - a_{2}|$ can also be extracted from the lifetime of the
short-lived $\pi^+ \pi^-$ atom ($A_{2\pi}$) in the ground state.  This
was done by DIRAC \cite{Adeva:2011tc} with an accuracy around 4\% and
a value consistent with the NA48 results.

With a proton beam momentum 450 GeV/$c$, the number of $A_{2\pi}$
generated per time unit would be \revised{$12 \pm 2$} times higher
than in the DIRAC experiment. The expected number of $\pi \pi$ atomic
pairs detected simultaneously with $\pi K$ atomic pairs would be
$n_A =4 \times 10^{5}$.  The statistical (systematic) precision of
$\pi \pi$ scattering lengths $|a_{0} - a_{2}|$ is estimated to be
around 0.7\% (2\%), thus improving the current precision by a factor
of two.

One may wonder whether the $\pi K$ scattering lengths could be
extracted from charmed meson decays, in analogy to the method used by
NA48 for $\pi \pi$.  However, an analysis in the strange sector is
expected to be more delicate because of prominent resonances on top of
the $\pi K$ continuum, starting with $K^*(892)$ not far away from the
threshold at 633 MeV.

\paragraph{Long-lived atoms.}
A sample of $436^{+157}_{-61}$ atomic pairs from the breakup of
long-lived $A_{2\pi}$ atoms was observed by DIRAC
\cite{Adeva:2015vra}.  \revised{From this sample, the collaboration
  extracted the lifetime of the $2p$ state, which is three orders of
  magnitude larger than the one of $A_{2\pi}$ in the ground state
  ($\tau_{2p} = 0.45^{+ 1.08}_{-0.30} \times 10^{-11}$ s vs
  $\tau_{1s} = 3.15^{+0.28}_{-0.26} \times 10^{-15}$ s)
  \cite{Adeva:2018fwr}.  At the SPS, the 450 GeV/$c$ proton beam and
  experimental adaptations envisaged for DIRAC++ (see below) open up
  new possibilities for study of long-lived $\pi\pi$ and $\pi K$ atoms
  \cite{Gorchakov:2016thm}.  The number of $A_{2\pi}$, $A_{\pi^+ K^-}$
  and $A_{\pi^- K^+}$ detected per unit of time would be increased by
  factors of 609, 26553 and 12024, respectively and the background
  would be decreased by about two orders of magnitude compared to the
  DIRAC experiment.} This would allow one to apply a resonance method
for measuring only one parameter, the Lamb shift, and evaluating from
this parameter the linear combination $2 a_0 + a_2$
\cite{Nemenov:2002wz}.  Combined with the measurement of
$|a_0 - a_2|$, this would allow for the experimental separation of
both scattering lengths from measurements with $\pi\pi$ atoms.


\paragraph{Required upgrades and additions to experimental equipment.}
For an experimental setup at the SPS, parts of the original DIRAC
detector can be re-used.  A scheme currently under investigation is
the use of two magnets, a new small one, and the DIRAC magnet as
second one.  The use of modern coordinate detectors is foreseen.  This
modification will enlarge the setup acceptance and will allow the
experiment to work with a higher proton beam intensity than DIRAC,
thus improving the statistical precision significantly.  A rough
estimate of the cost of such a new set up is 3 million CHF.

\paragraph{Collaboration.}
Based on the experience with the original DIRAC experiment, it is
hoped that a collaboration of adequate strength to build and operate
DIRAC++ would form if a Letter of Intent were positively reviewed by the SPS
Committee.


\begin{figure}[t]
\begin{center}
  \includegraphics*[width=90mm]{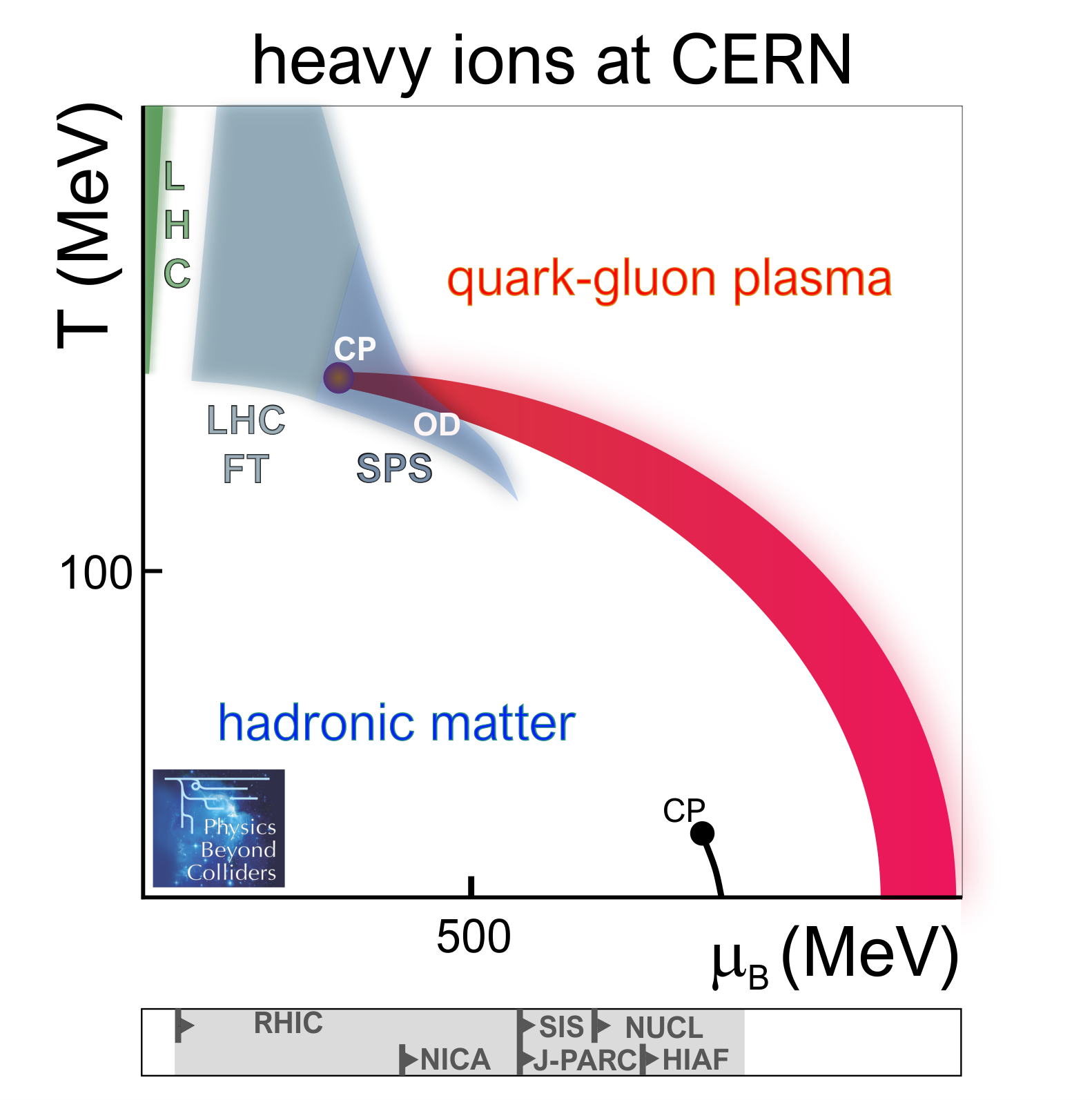}
\end{center}
\caption[Phase diagram of QCD and density range of running or planned
  experiments.]{Phase diagram of QCD and density range of running or planned
  experiments. The density range of LHC, LHC-FT and SPS experiments is
  indicated with the shaded areas in the figure. The lower boundary of
  the grey and blue shaded area follows the chemical freeze-out. The upper boundary
  relates to the parameters at the early stage of the collisions. The
  potential critical end point is labelled with CP, the onset of
  deconfinement with OD.  The back line at small temperatures and high densities
  signifies the nuclear liquid-gas transition, also ending in a
  critical end point CP. The density range of other experiments is
  indicated in the bar below the figure. This includes RHIC at BNL,
  NICA at JINR, SIS100 at FAIR, J-PARC-HI at J-PARC, the Nuclotron at
  JINR (NUCL), and HIAF at HIRFL. }
\label{fig:PhaseStructure}
\end{figure}
\section{Summary of heavy-ion measurements}
\label{sec:HIC-Summary}

\begin{figure}[t]
\begin{center}
  \includegraphics*[width=110mm]{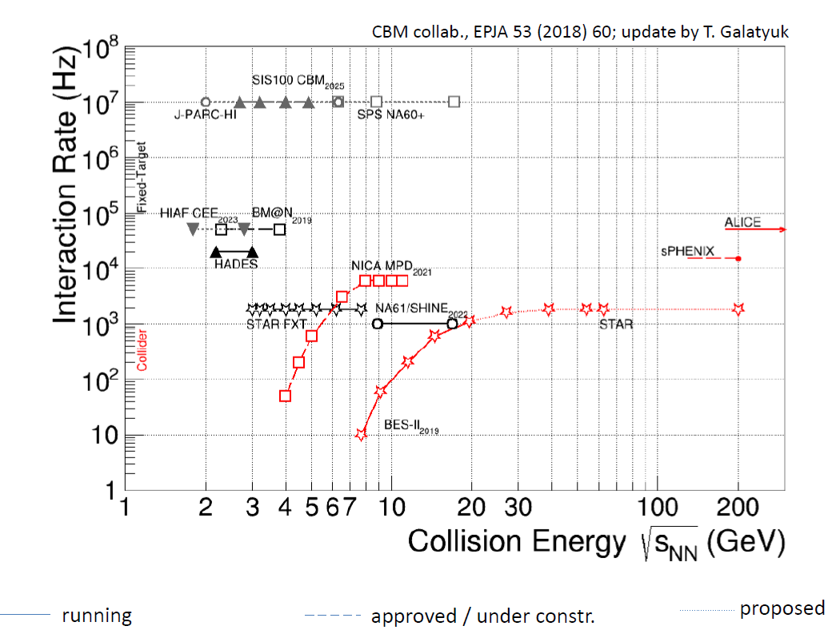}
\end{center}
\caption[Interaction rate as a function of collision energy of running
or planned experiments.]{Interaction rate of running or planned
  experiments (figure taken from \cite{Ablyazimov:2017guv}, update by
  T.~Galatyuk).  The fixed target experiments at CERN have large
  interaction rates relevant for the accuracy required for the physics
  questions at hand. They cover a very interesting regime in the
  collision energy $\sqrt{s_\textrm{NN}}$, see
  figure~\ref{fig:PhaseStructure}, complementary to other running or
  planned experiments. }
\label{fig:InteractionRates}
\end{figure}

The currently running and planned heavy ion experiments sweep over a
large temperature and density regime in the phase structure, scanning
$\sqrt{s_\textrm{NN}}$ from about $2$\, GeV to LHC energies. Combined,
the experiments will unravel the phase structure of QCD, see
figures~\ref{fig:PhaseStructure} and \ref{fig:InteractionRates}.

In combination, the heavy ion experiments at CERN already offer a
unique opportunity with their positioning from the very high energy
regime covered by LHC, and the connected regime covered by the
fixed target experiments at SPS and LHC.

Moreover, the SPS experiments cover a specifically interesting energy
regime. In this regime several anomalies in the energy dependence of
hadron production properties have been observed in central lead-lead
collisions. They are interpreted as the beginning of the creation of
QGP (the onset of deconfinement, OD), see
figure~\ref{fig:PhaseStructure}.  This regime possibly also include
the critical end point in the QCD phase diagram (CP), see
figure~\ref{fig:PhaseStructure}. They are complementary to the other
experiments in this regime, RHIC BES-II (running (2019), NICA (starting
2020). They are unique in their access to open charm (NA61++, NA60++)
and the precise determination of the caloric curve (NA60++), for more
details see the chapters~\ref{sec:NA60+}, \ref{sec:NA61+}, as well as
connecting to CBM, FAIR and J-PARC-HI with high interaction rates at
lower collision energies. In summary this offers the opportunity at
SPS CERN for significantly contributing to the resolution of the
density region potentially hosting the critical end point of QCD.

The FT experiments at LHC cover an energy regime overlapping with that
of RHIC, but allow for different and partially more accurate
measurements.  For example, the large $pA$ rates are orders of
magnitudes larger than those reached at RHIC in collider mode. The
high statistics data for heavy flavour azimuthal correlations
($D-D, J/\psi-D$) together with measurements of the D-meson elliptic
flow and the nuclear modification factor allow for a study of the QGP
transport properties ($\hat q$ and charm quark diffusion
coefficients), for more details see
\cite{Hadjidakis:2018ifr}.

\section{Measurements for cosmic-ray physics and for neutrino experiments}
\label{sec:cosmics}

Precise measurements of cosmic-ray fluxes up to very high energies,
ranging from about tens of MeV up to hundreds of TeV, have become
available for many particle species.  High-precision data on the
interaction (production) cross sections of typical cosmic-ray
particles with (from) nuclei most abundant in the universe are needed
for the interpretation of results from current- and future-generation
cosmic-ray experiments.  In this respect a better understanding of the
cosmic-ray propagation is needed to study the origins of galactic
cosmic rays and to evaluate the cosmic-ray background for signatures
of astrophysical dark matter cosmic-ray acceleration.

\begin{figure}
	\centering
	\includegraphics[width=0.50 \textwidth]{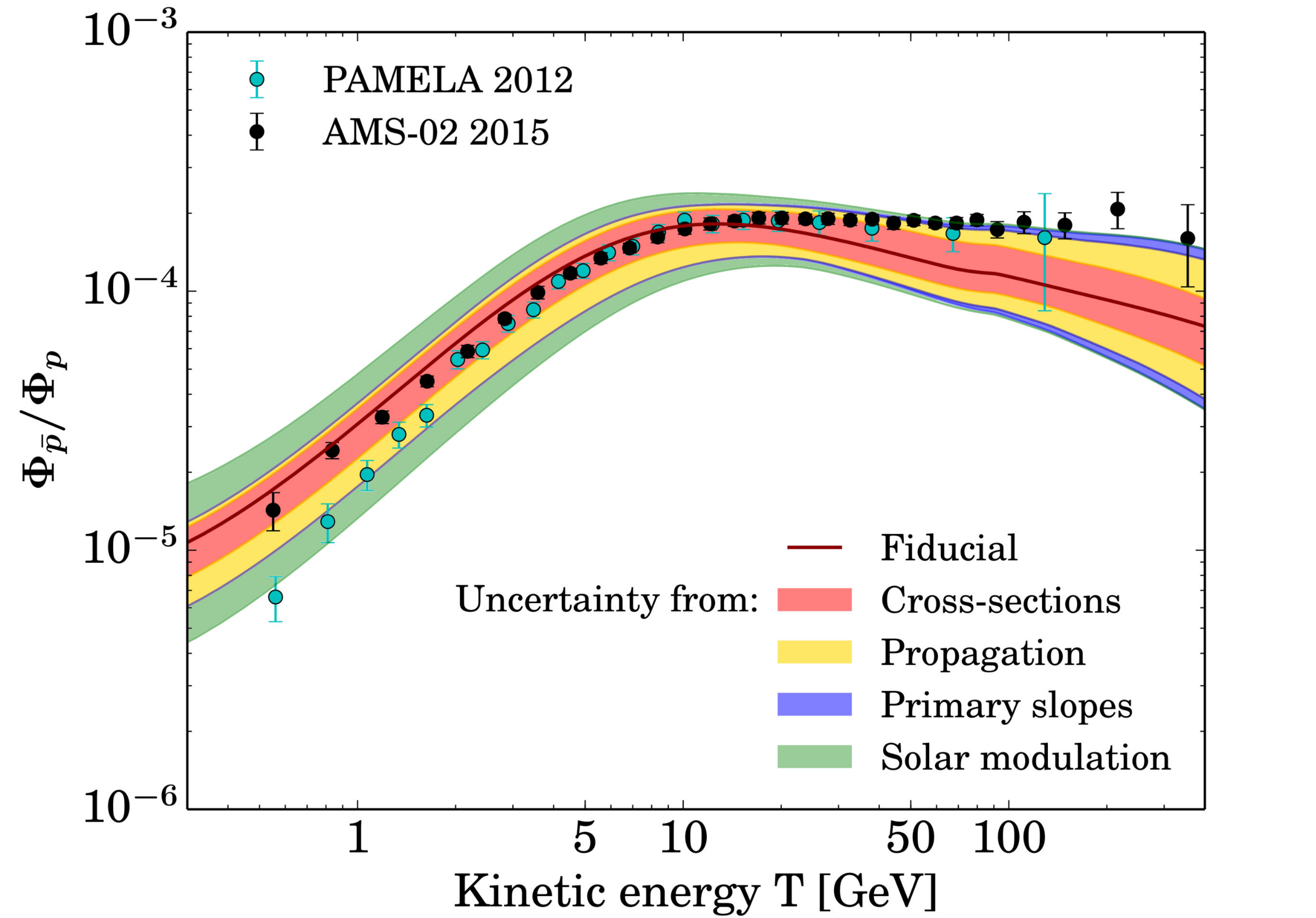}
	\includegraphics[width=0.48 \textwidth]{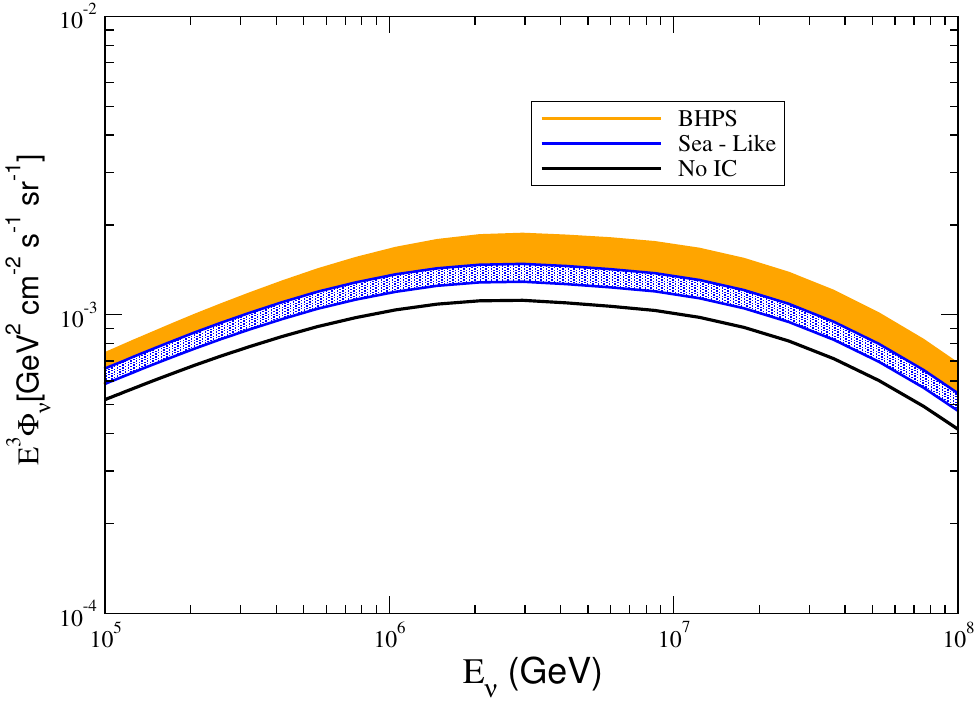}
	\caption[Examples of QCD-related limitations in flux
        calculations for cosmic-ray physics.]{Examples of QCD-related
          limitations in flux calculations for cosmic-ray physics.
          (left) The combined total uncertainty on the predicted
          secondary $\bar{p}/p$ ratio, superimposed to
          PAMELA~\cite{Adriani:2012paa} and AMS~\cite{Aguilar:2016kjl}
          data.  (right) Predictions for high-energy atmospheric
          neutrinos from collisions of cosmic rays with atmospheric
          nuclei. The different curves correspond to different
          assumptions on the high-\(x\) charm content of the
          proton~\cite{Giannini:2018utr,Hadjidakis:2018ifr}.  (Figures
          taken from~\cite{Giesen:2015ufa,Hadjidakis:2018ifr}.)  }
	\label{fig:AMSandCharm}
\end{figure}

\begin{figure}
  \centering
  \includegraphics[width=1.02\textwidth]{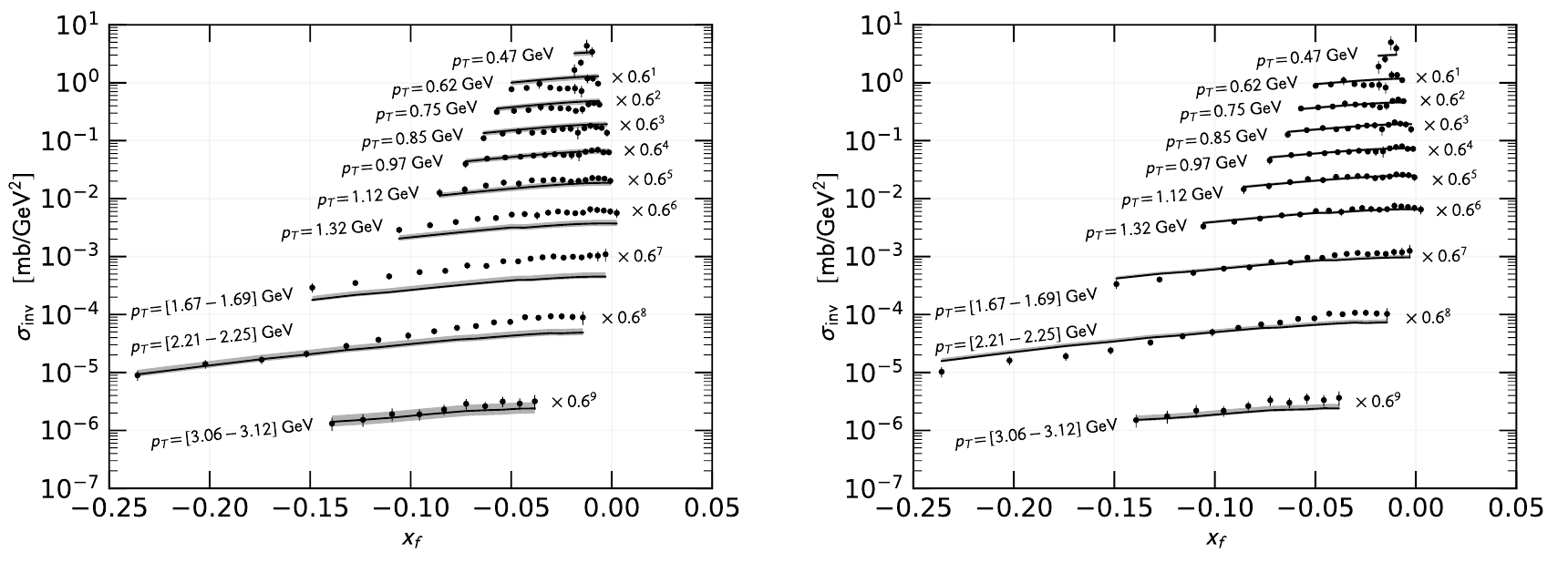}
  \caption[Comparison of parameterisations for the $p$He cross section
  with the LHCb data.]{Comparison of two
    parameterisations~\cite{Korsmeier:2018gcy} for the $p$He cross
    section with the LHCb data~\cite{Aaij:2018svt}.  (Figure taken
    from~\cite{Korsmeier:2018gcy}.)  }
	\label{fig:pHe-pbar}
\end{figure}

Anti-protons are of special interest as expected to be a result of the
diffuse propagation of primary cosmic rays in the galaxy.  Recently,
the AMS experiment reported anti-proton to proton flux ratios for
1--400 GeV~\cite{Aguilar:2016kjl} with much higher precision than
earlier PAMELA data~\cite{Adriani:2012paa} (see
figure~\ref{fig:AMSandCharm}).  Precision knowledge of anti-proton
production cross sections over a wide range of energy and
initial-state nuclei opens the portal to indirect detection of dark
matter or unknown astrophysical mechanisms of cosmic-ray acceleration.
Nuclei that are abundant in the galaxy play a particular role, as
prime candidates for anti-proton sources.  Data from
NA61/SHINE~\cite{Aduszkiewicz:2017sei} and from
LHCb~\cite{Aaij:2018svt} with the current SMOG fixed-target system
have already been used in analyses and demonstrated the usefulness of
such measurements~\cite{Korsmeier:2018gcy}.  This is illustrated in
figure~\ref{fig:pHe-pbar} in which different parameterisations of the
\(\bar{p}\) production cross section in proton-He collision is
compared to the LHCb SMOG data, clearly favouring one over the other
parameterisation~\cite{Korsmeier:2018gcy}.

In the case of cosmic neutrinos, atmospheric neutrinos are sources of
background in terrestrial studies of galactic neutrino generation and
propagation.  Heavy-quark decay constitutes a major uncertainty that
can be addressed, e.g., by charm production cross sections in $pA$ and
$AA$ collisions.  In particular, charm production at large-$x_{F}$,
e.g., very forward and at high-energies, strongly depends on the charm
content of the nucleon, illustrated in figure~\ref{fig:AMSandCharm}
for three models for charm in the projectile
proton~\cite{Giannini:2018utr,Hadjidakis:2018ifr}.  \revised{The
  capability to measure charm production in fixed-target mode at the
  LHC has been already demonstrated by LHCb using few nb$^{-1}$ of
  data collected with the SMOG target with helium and
  argon~\cite{Aaij:2018ogq}. The use of a hydrogen target and the
  large increase in luminosity allowed by the SMOG2 target upgrade are
  expected to provide precise measurements on the amount of intrinsic
  charm in nucleons.}

Last but not least accelerator-based neutrino experiments
(cf.~Refs.~\cite{Ahn:2006zza,Adamson:2007gu,Abe:2011ks,Acciarri:2015uup})
rely on a precise description of particle production in primary and
follow-up interactions of beam and secondary hadrons in the often
extended nuclear targets. It is, however, far from trivial to predict
those precisely from standard Monte Carlo simulations due to the
complexity and ranges of different interactions involved. Measurements
of particle fluxes using replica targets or thin targets of the same
material under similar conditions can provide important input to these
simulations.  For example, the NA61/SHINE collaboration has performed
several of such measurements that have been exploited in flux
calculations for the T2K long-baseline neutrino
experiment~\cite{Abgrall:2011ae,Abgrall:2011ts,Abgrall:2013wda,Abgrall:2016jif}.
NA61/SHINE plans to continue this program including measurements for
present and future J-PARC long-baseline experiments (T2K, T2K-II,
Hyper-K), and the future Long-Baseline Neutrino Facility (LBNF) beam
for DUNE~\cite{Acciarri:2015uup}, strongly endorsed by those collaborations~\cite{Aduszkiewicz:2309890}.\\

\begin{figure}
	\centering
	\includegraphics[width=0.84 \textwidth]{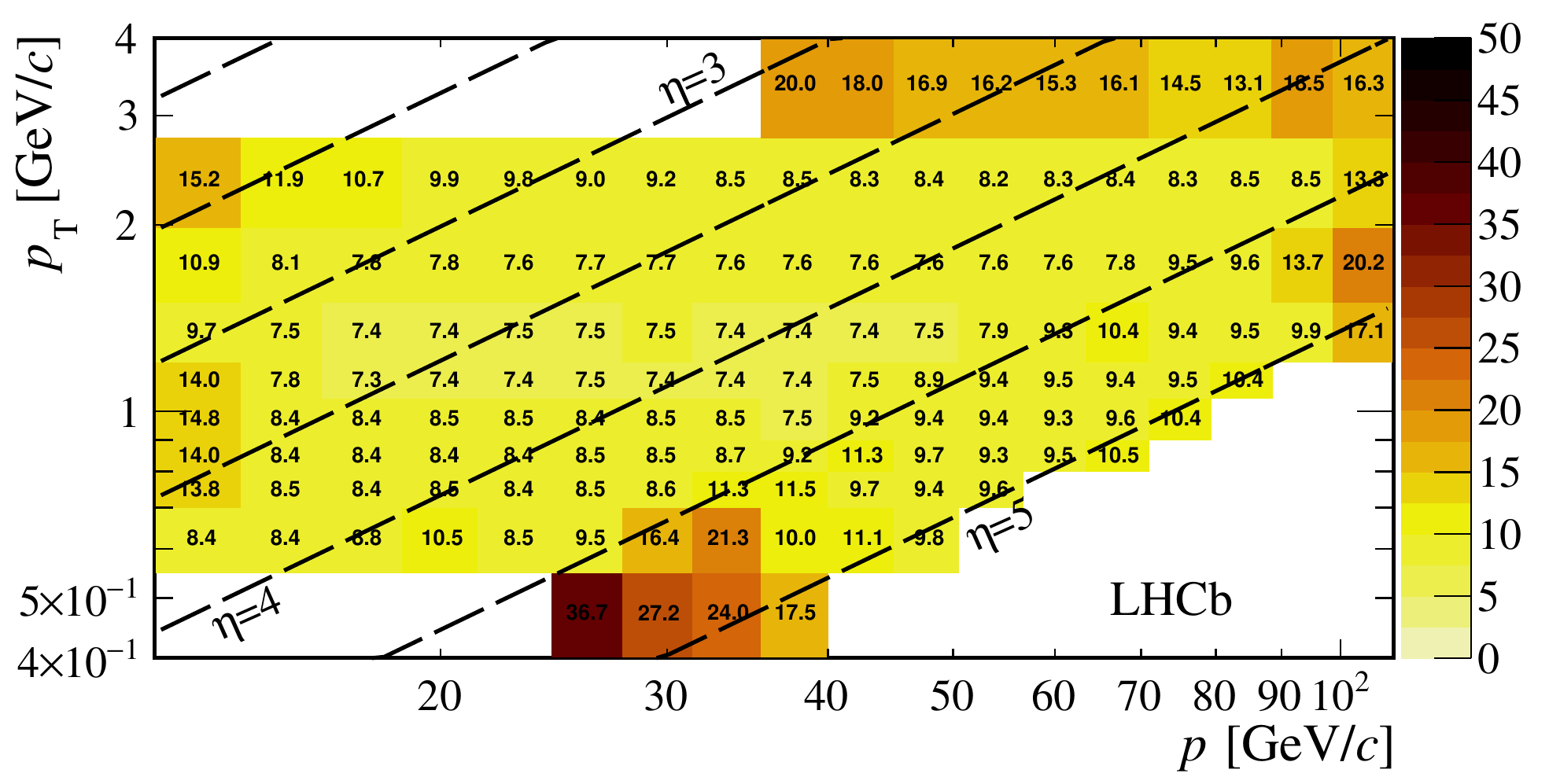}
	\caption[Relative precision of LHCb results on
		\(\bar{p}\) production in $p$He collision at $\sqrt{s}$=110~GeV.]
		{Relative precision (in \%) of recent LHCb results~\cite{Aaij:2018svt} on
		\(\bar{p}\) production in $p$He collision at $\sqrt{s}$=110~GeV. The integrated luminosity used in this
		measurement amounts to 0.5~nb$^{-1}$. With SMOG2 it should be possible to collect about
		100~nb$^{-1}$ in an hour beam time.
		\rev{Moreover, SMOG2 will significantly reduce the systematic uncertainty on the luminosity,}
		\rev{the dominating uncertainty of the present measurement.}
			}
	\label{fig:LHCb-pHe}
\end{figure}

\begin{figure}
	\centering
	\includegraphics[width=0.54 \textwidth]{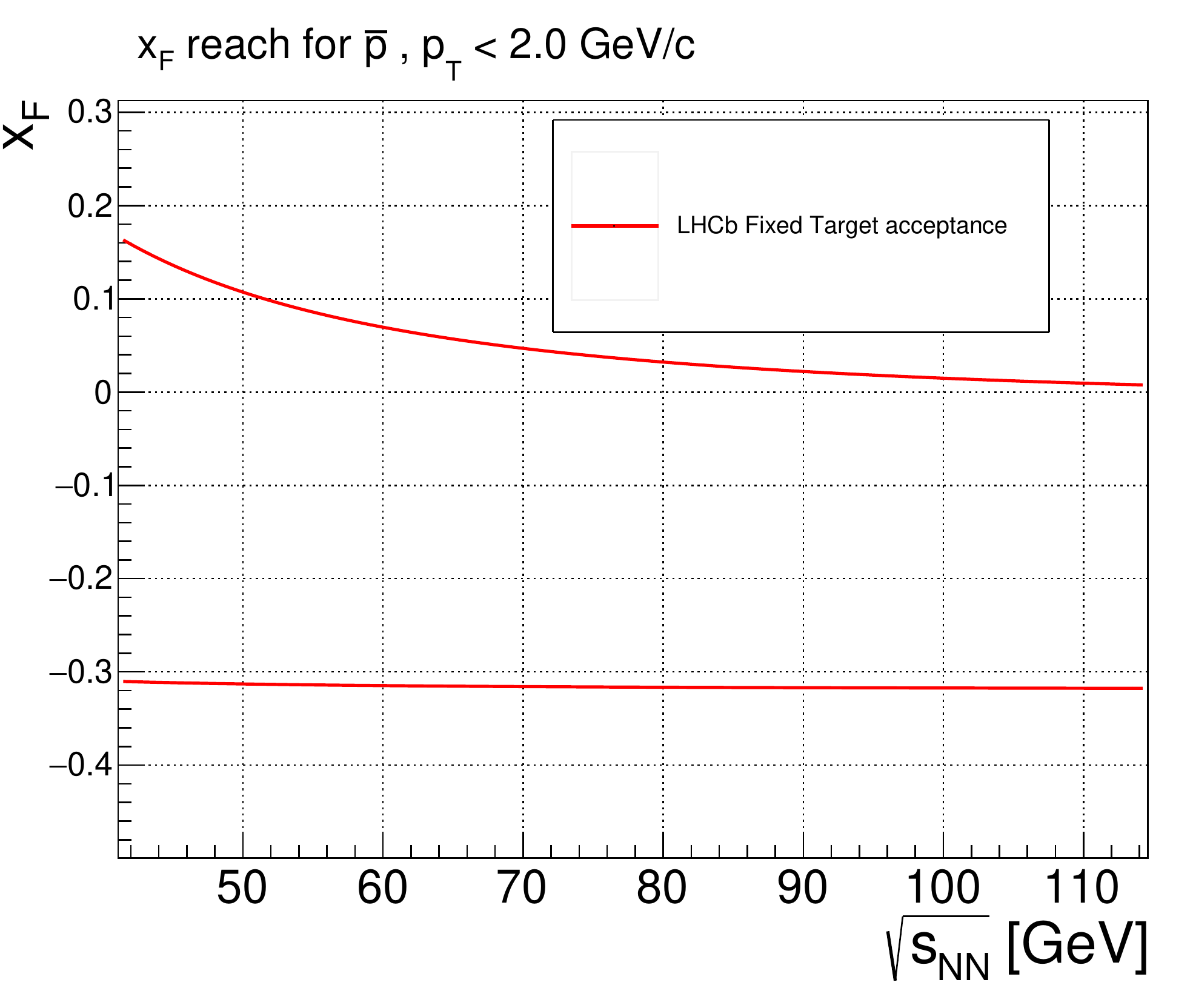}
	\caption[Acceptance in Feynman-$x$ of LHCb-FT (SMOG2) for \(\bar{p}\)  production for $p_{T}$ up to 2 GeV.]{
		Acceptance in Feynman-$x$ of LHCb-FT (SMOG2) for $p_{T}$ up to 2 GeV in \(\bar{p}\) production.
		Note that the majority of the production is in the range $|x_F|<0.1$.
			}
	\label{fig:LHCb-accept}
\end{figure}

Among the various PBC-QCD proposals, COMPASS++, LHC-FT (AFTER@LHC,
LHCb-FT, and ALICE-FT), as well as NA61++ have cosmic-ray and
neutrino-flux related measurements as part of their program. They are
discussed below.

\begin{figure}
	\centering
	\includegraphics[width=0.55 \textwidth]{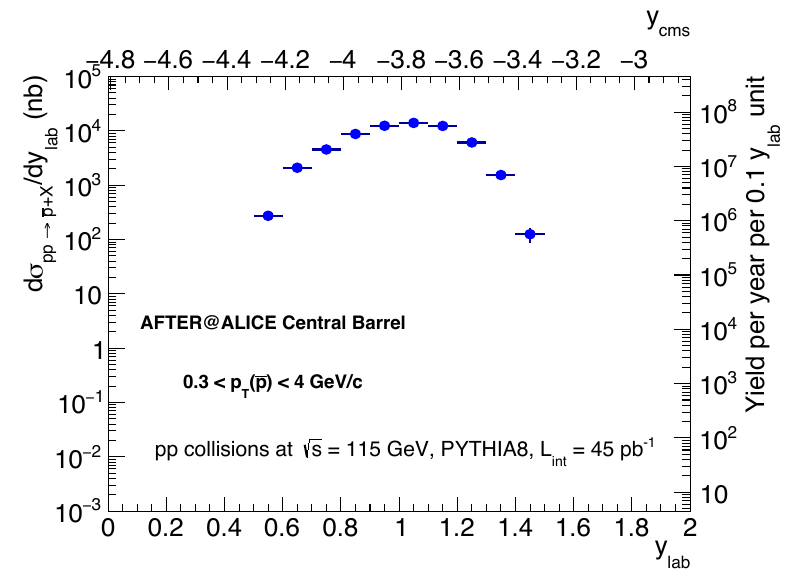}
	\caption[Expected kinematic coverage of \(\bar{p}\) production
        for ALICE-FT.]{Expected kinematic coverage of \(\bar{p}\)
          production for ALICE-FT (for details,
          see~\cite{Hadjidakis:2018ifr}; figure taken
          from~\cite{Hadjidakis:2018ifr}).  }
	\label{fig:ALICE-pbar}
\end{figure}

\begin{figure}
	\centering
	\includegraphics[width=0.75 \textwidth]{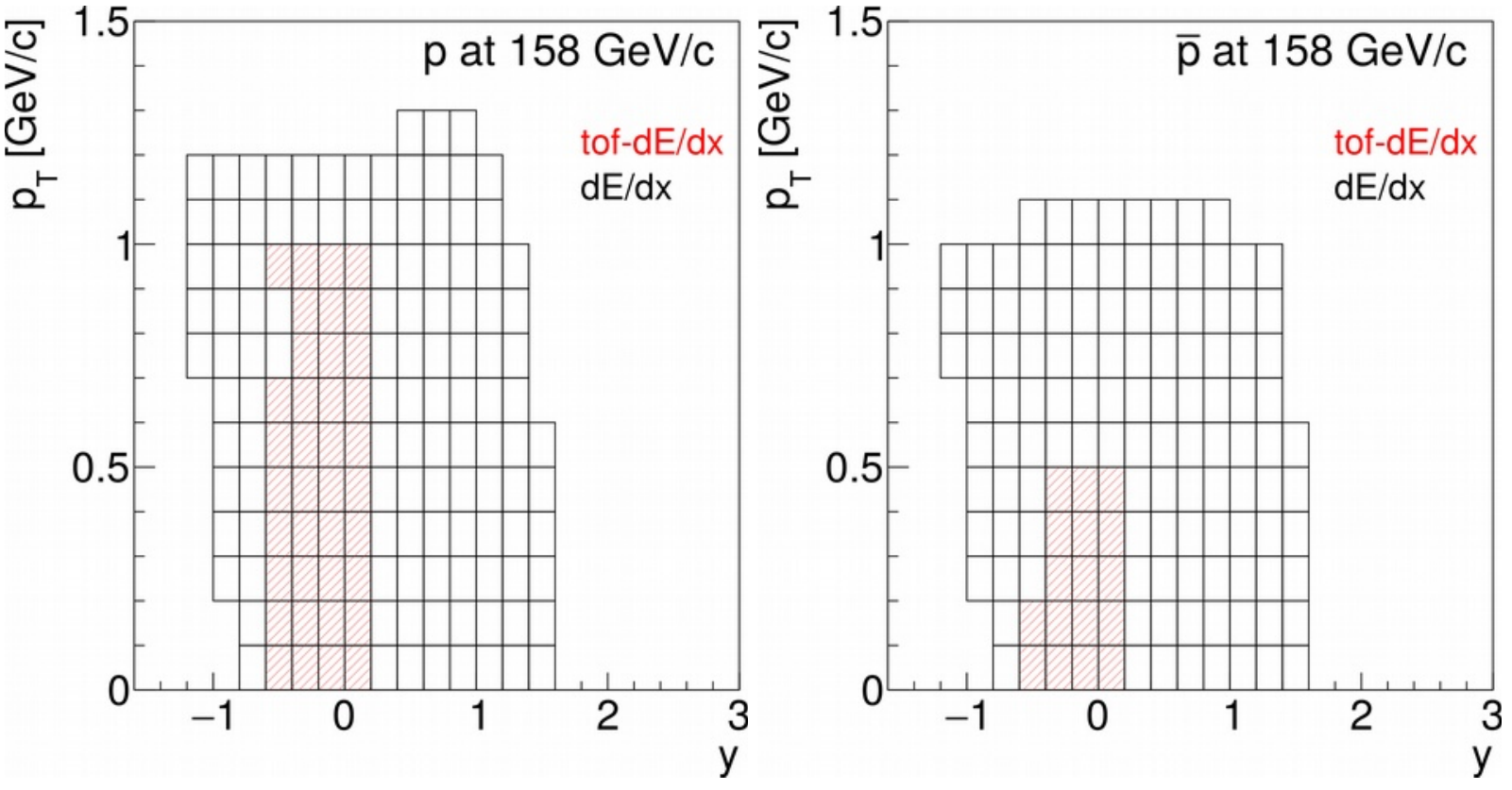}
	\caption[Accepted range in \(p_{T}\) and rapidity \(y\)
        (calculated in the collision centre-of-mass system) for \(p\)
        and \(\bar{p}\) production in $pp$ collision at
        \(\sqrt{s}=158\) GeV.]{Accepted range in \(p_{T}\) and
          rapidity \(y\) (calculated in the collision centre-of-mass
          system) for proton and anti-proton production in $pp$
          collision at \(\sqrt{s}=158\) GeV.  (Figures taken
          from~\cite{Aduszkiewicz:2017sei}.)  }
	\label{fig:NA61-pbar}
\end{figure}

\paragraph{Anti-proton production and nuclear fragmentation:}
\begin{itemize}
\item {\bf LHCb-FT} (SMOG2
  \cite{Bursche:2649878,CERN-PBC-Notes-2018-007}) plans to perform
  similar measurements as those mentioned above using a He
  target~\cite{Aaij:2018svt}, but with hydrogen and deuterium under
  the same kinematic condition to provide the production in $pp$
  collision and to constrain isospin violations, which affect the
  prediction for the anti-neutron contribution to the antiproton flux.
  The better control over the gas density provided by SMOG2
    will reduce the systematic uncertainty on the luminosity, which is
    one of the largest limitations to the precision of the current
  measurement with He, which is depicted in
  figure~\ref{fig:LHCb-pHe}, with the reach in \(x_{F}\) for
  $p_{T}<2$~GeV of the \(\bar{p}\) shown in
  figure~\ref{fig:LHCb-accept}. Measurements at lower $\sqrt{s}$ are
  also foreseen.
\item {\bf ALICE-FT} could employ the central barrel to measure very
  slow \(\bar{p}\) with almost zero momentum. The production of such
  slow \(\bar{p}\) with the LHC proton beam correspond to the highest
  possible energies in the inverse kinematics, where the nuclear
  target (C, N, O, He) travels at TeV energies, hit an interstellar
  \(p\) at rest and produces a \(\bar{p}\) in the limit of
  \(x_{F} \to 1\)~\cite{Hadjidakis:2018ifr}. Using, e.g., a $^{4}$He
  target would then correspond to the high-energy tail of the
  $^{4}\text{He } + p \to \bar{p} + X $ process ($^{4}$He as
  projectile), which is one of the leading process in the cosmic
  \(\bar{p}\) spectrum. Similarly, using C, N or O targets, one can
  study the high-energy \(\bar{p}\) tail for
  (C,N,O)$ +p \to \bar{p} + X$. As an example, the yield of
  low-momentum \(\bar{p}\) in the range
  \( 0.3~\text{GeV} < p_{T}^{\bar{p}}<4~\text{GeV}\) is shown in
  figure~\ref{fig:ALICE-pbar} for 45~pb$^{-1}$.
\item {\bf NA61++} proposes to measure a range of nuclear
  fragmentation cross sections relevant for the production of Li, Be,
  B, C and N nuclei. In addition, (anti)proton and (anti)deuteron
  cross sections will be studied under the same kinematic conditions
  allowing for determinations of $d/p$ and $\bar{d}/\bar{p}$
  ratios. Details are to be worked out and will be subject to a future
  addendum to the NA61/SHINE proposal. For guidance, from 60M $pp$
  collisions recorded between 2009 and 2011, 13k $\bar{p}$ and 10
  $\bar{d}$ are expected based on
  simulations~\cite{Aduszkiewicz:2309890}. 600M events each are
  foreseen to be taken with a liquid-hydrogen target at beam energies
  of 40 GeV, 180 GeV, and 350 GeV, respectively, running each 100 days
  at a rate of 6M \(pp\)-events/day.  Yields of \(\bar{p}\) and
  \(\bar{d}\) after selection and identification requirements are
  estimated based on the preliminary \(p\) and \(d\) analyses of
  already existing NA61/SHINE 158GeV $pp$
  data~\cite{Aduszkiewicz:2287091} as well as an analysis of other
  statistically limited \(\bar{d}\) production
  data~\cite{Gomez-Coral:2018yuk}, and are summarised in
  Table~\ref{tab:NA61antipd}, \revised{taking into account the about 1
    kHz rate limit of NA61++ read-out (a rate that also saturates the
    available radiation budget in the case of PbPb collision)}. The
  acceptance, estimated based on the published analysis of $pp$ data
  from 2009~\cite{Aduszkiewicz:2017sei}, is depicted for protons and
  anti-protons in figure~\ref{fig:NA61-pbar}.
\item {\bf COMPASS++} proposes, similarly as NA61++, to measure
  \(\bar{p}\) production with a liquid-hydrogen target, but also with
  a $^{4}$He target. \(\bar{p}\) will be accepted in a momentum range
  of 10--45~GeV (where an absence of a RICH signal is used for
  particle-ID in the range of 10--18~GeV) and 2.4--8 in
  pseudo-rapidity. A 1\% relative statistical precision in each of
  20$\times$20 bins in momentum and pseudo-rapidity is envisaged using
  a beam intensity of 10$^{5}~p$/s on target, corresponding to a
  trigger rate of 25kHz. With a total of 10s beam per minute,
  2.5$\times$10$^{5}$ collision events will be collected. An estimated
  \(\bar{p}\) identification efficiency of 70\%, 4 hours (6 hours
  incl. contingency) will be required for each combination of target
  and beam setting.
\end{itemize}

\begin{table}
  \caption{\label{tab:NA61antipd} Estimated \(\bar{p}\) and
    \(\bar{d}\) event yields for 100 days running each at 40 GeV, 180
    GeV, and 350 GeV beam momentum and a liquid-hydrogen target at
    NA61++ \rev{and a 1 kHz read-out rate}.}
\centering
 \vspace*{2mm}
\begin{tabular}{lccc}
\hline
beam momentum~~~~~~	&	~~~~40~GeV~~~~	&	~~~~180~GeV~~~~ 	&	~~~~350~GeV~~~~	 	\\
\hline
$\sqrt{s}$			&	8.8~GeV	&	18.4~GeV	&	25.7~GeV 	\\
beam time			&	100~d	&	100~d	&	100~d 		\\
\(pp\) collisions		&	\(6\times10^{8}\)	&	\(6\times10^{8}\)	&	\(6\times10^{8}\) 	\\
estimated \(\bar{p}\) events	&	\(1.3 \times 10^{4}\) 	&  \(1.4 \times 10^{5}\) &  \(2.3 \times 10^{5}\) 	\\
estimated \(\bar{d}\) events	&	10 	& 110 &  180 	\\
\hline
\end{tabular}
\end{table}

\paragraph{Charm:}
{\bf LHC-FT} offers a unique opportunity to study high-$x$ parton
distributions, as discussed already in Sec.~\ref{sec:LHC-FT}. LHCb-FT,
in particular, has excellent acceptance and identification
capabilities for various charm-sensitive channels. In addition to the
discussion in Sec.~\ref{sec:LHC-FT}, example prospects for measuring
charm at LHCb with SMOG2 are given in Table~\ref{tab:SMOG2charm}.
Please note that
\begin{itemize}
\item the list is far from being exhaustive, e.g., different target gases are possible;
\item extrapolations are currently crude estimates;
\item assuming quarkonium absorption by the nuclear target leads to
  a decrease of its cross section by a factor 0.75 (0.6) in $p$Ne and
  0.5 (0.4) in $p$Ar for $J/\psi$ ($\psi'$) with respect to $p$He;
\item the smaller systematic uncertainty with SMOG2 is expected from
  the reduction of the dominant uncertainty on the luminosity (6\%)
  for SMOG data.
\end{itemize}
\vspace{3mm}

\begin{table}
  \caption{\label{tab:SMOG2charm} Current and estimated yields in
    charm events as well as for $\Upsilon(1S)$ production and \DY for
    SMOG and SMOG2 at LHCb-FT.}
 \centering
 \vspace*{2mm}
 \begin{tabular}{lccc}
 \hline
                              		&     SMOG   			&  SMOG 				&    SMOG2   \\
					&	published 		& largest sample		& 	example \\
                               		&     pHe@87 GeV   		&  pNe@68 GeV            	&  pAr@115 GeV   \\ \hline
Integrated luminosity 	&   7.6 nb$^{-1}$          	& $\sim$100 nb$^{-1}$ 	&    $\sim$45 pb$^{-1}$  \\
Systematic unc. on $J/\psi$ cross section &      7\%                 	&      6--7\%            		&      2--3  \%     \\
$J/\psi$ yield                 	&     400                  		&      15k               		&      15M          \\
$D^0$ yield                    	&     2000                 		&      100k              		&      150M         \\
$\Lambda_c^{+}$ yield          	&      20                  		&       1k               		&      1.5M         \\
$\psi(2S)$ yield                  	&     negl.                		&       150              		&      150k         \\
$\Upsilon(1S)$ yield  	&     negl.                		&       4               		&       7k         \\
DY $\mu^+\mu^-$ yield  (5 GeV$<M<9$ GeV)	&     negl.                		&       5               		&       9k         \\
\hline
 \end{tabular}
\end{table}

\paragraph{Hadron production measurements for accelerator-based neutrino experiments:}
{\bf NA61++} proposes a series of measurements for J-PARC and LBNF in order to
\begin{itemize}
\item improve further the precision of hadron production measurements
  for the currently used T2K replica target, paying special attention
  to the extrapolation of produced particles to the target surface,
\item perform measurements for a new target material (super-sialon),
  both in thin target and replica target configurations, for T2K-II
  and Hyper-Kamiokande,
\item study the possibility of measurements at low incoming-beam
  momenta (below 12 GeV) relevant for improved predictions of both
  atmospheric and accelerator neutrino fluxes,
\item ultimately perform hadron production measurements with prototypes of Hyper-Kamio\-kande and DUNE targets.
\end{itemize}

\section{Compatibility of COMPASS++ and MUonE at the M2 beam line}
\label{sec:compass-muone}

The proposed measurements of elastic $\mu p$ scattering by COMPASS++
and of elastic $\mu e$ scattering by MUonE are both envisaged to take
place (or at least start) during Run 3 of the LHC, given the
scientific urgency of the physics problems they address and the
activity of related experiments elsewhere in the world.  The question
therefore arises whether and under which condition a concurrent
running of both experiments might be possible.  This initiated a study
within the PBC QCD working group, whose status is presented in this
section.  Preliminary answers were already provided at the 2018 June
PBC workshop.\footnote{See the presentation by A. Magnon at
  \url{https://indico.cern.ch/event/706741/timetable/\#20180614.detailed}.}

\paragraph{Compatible locations for $\mu$-$e$ and the $\mu$-proton-radius foreseen setups.}

\begin{figure}
\begin{center}
\includegraphics*[width=0.9\textwidth]{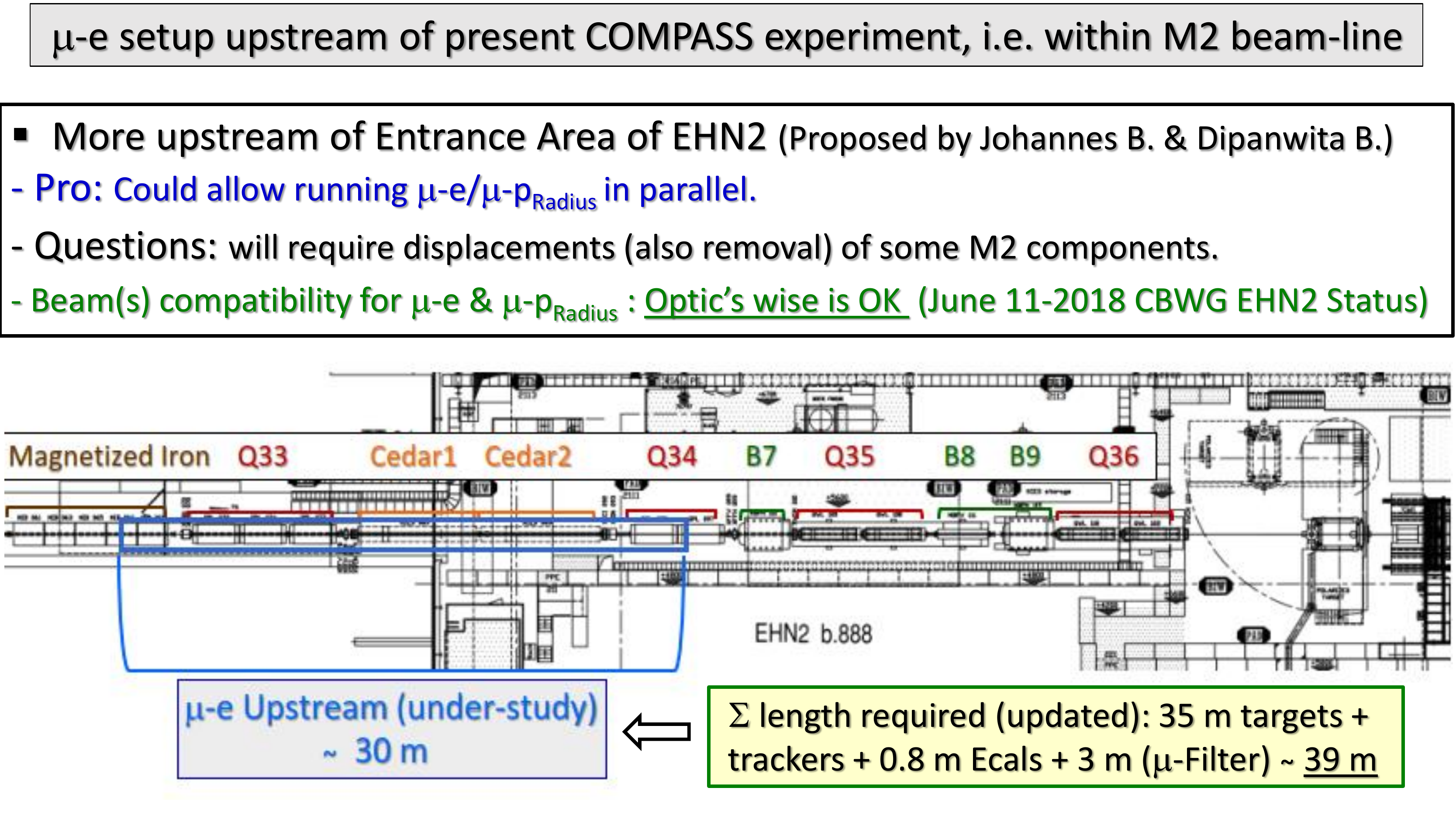}
\\[3em]
\includegraphics*[width=0.98\textwidth]{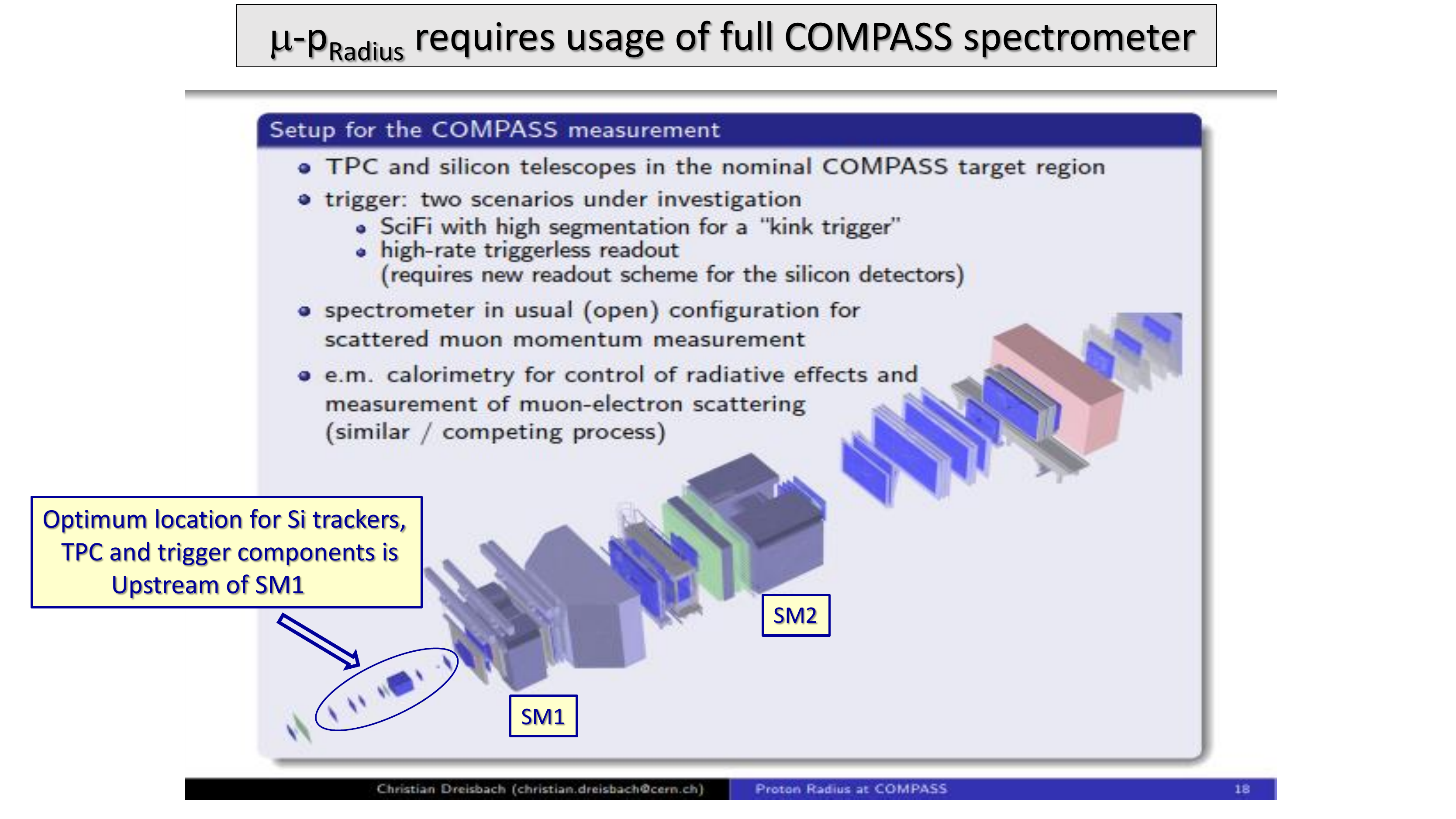}
\end{center}
\caption[Foreseen experimental setups for the $\mu$-$e$ and the
$\mu$-proton-radius measurements.] {Foreseen setups: (Top) $\mu$-$e$
  and (Bottom) $\mu$-proton-radius setups.}
\label{fig:M2-setups}
\end{figure}

Concerning the setup, several options were studied. The $\mu$-$e$ is a
long setup and among the two options studied (upstream or downstream
of EHN2 Hall) the retained option is presently upstream
(figure~\ref{fig:M2-setups}~top), within a possible available space in
the M2 beam-line. The $\mu$-proton-radius requires the usage of
several components of the COMPASS~spectrometer, therefore it should be
installed, as shown in figure~\ref{fig:M2-setups}~(bottom). Dedicated
studies were performed~\footnote{This was presented by L. Gatignon at
  the Conventional Beams PBC General Meeting on 13 June 2018.  An ATS
  Note by D. Banerjee and J. Bernhard is in preparation.}
showing that an adequate $\mu$-beam tuning can be achieved
for the two experiments running in parallel.

\begin{figure}
\begin{center}
\includegraphics*[width=0.8\textwidth]{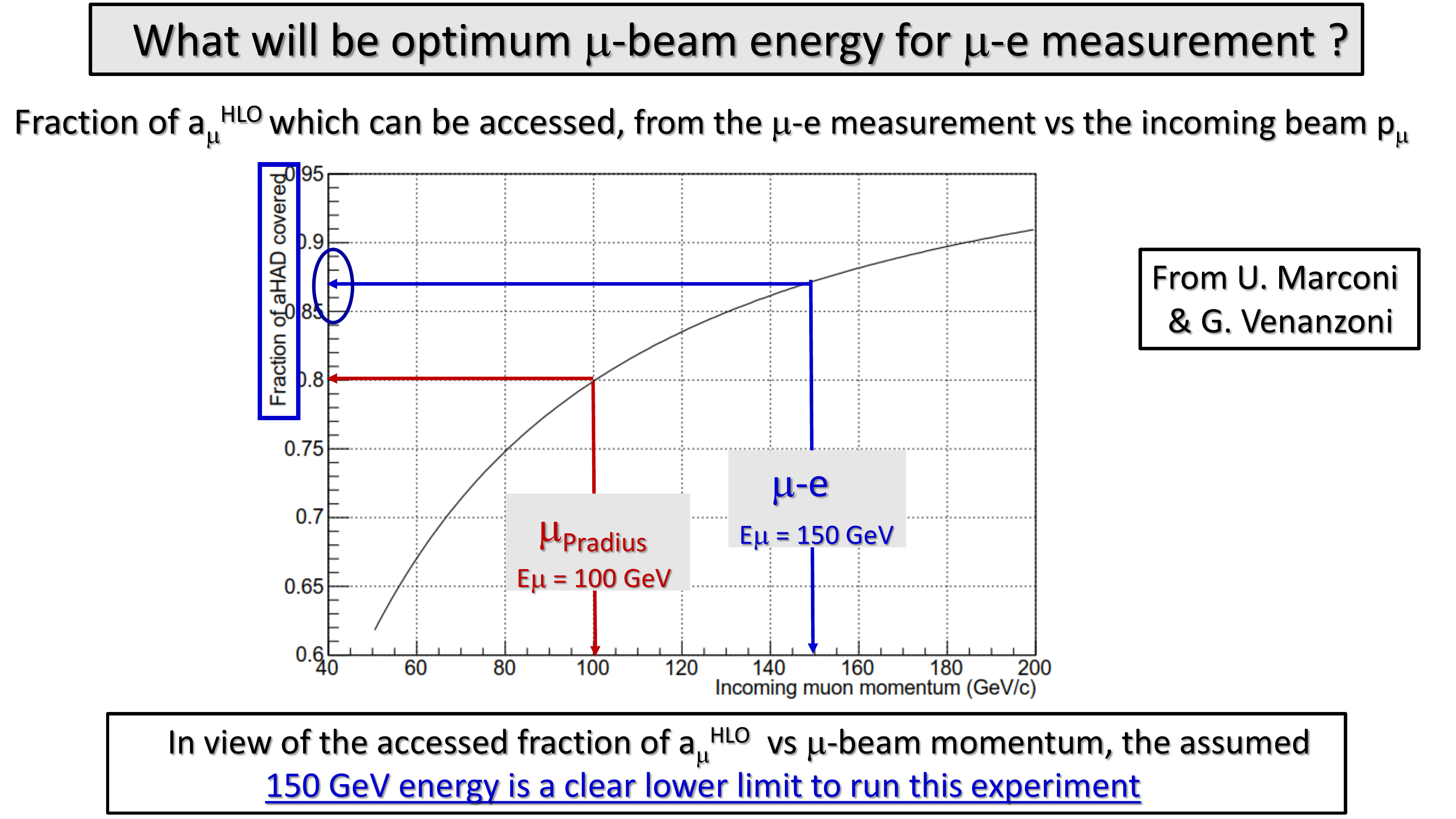}
\\[2em]
\includegraphics*[width=0.8\textwidth]{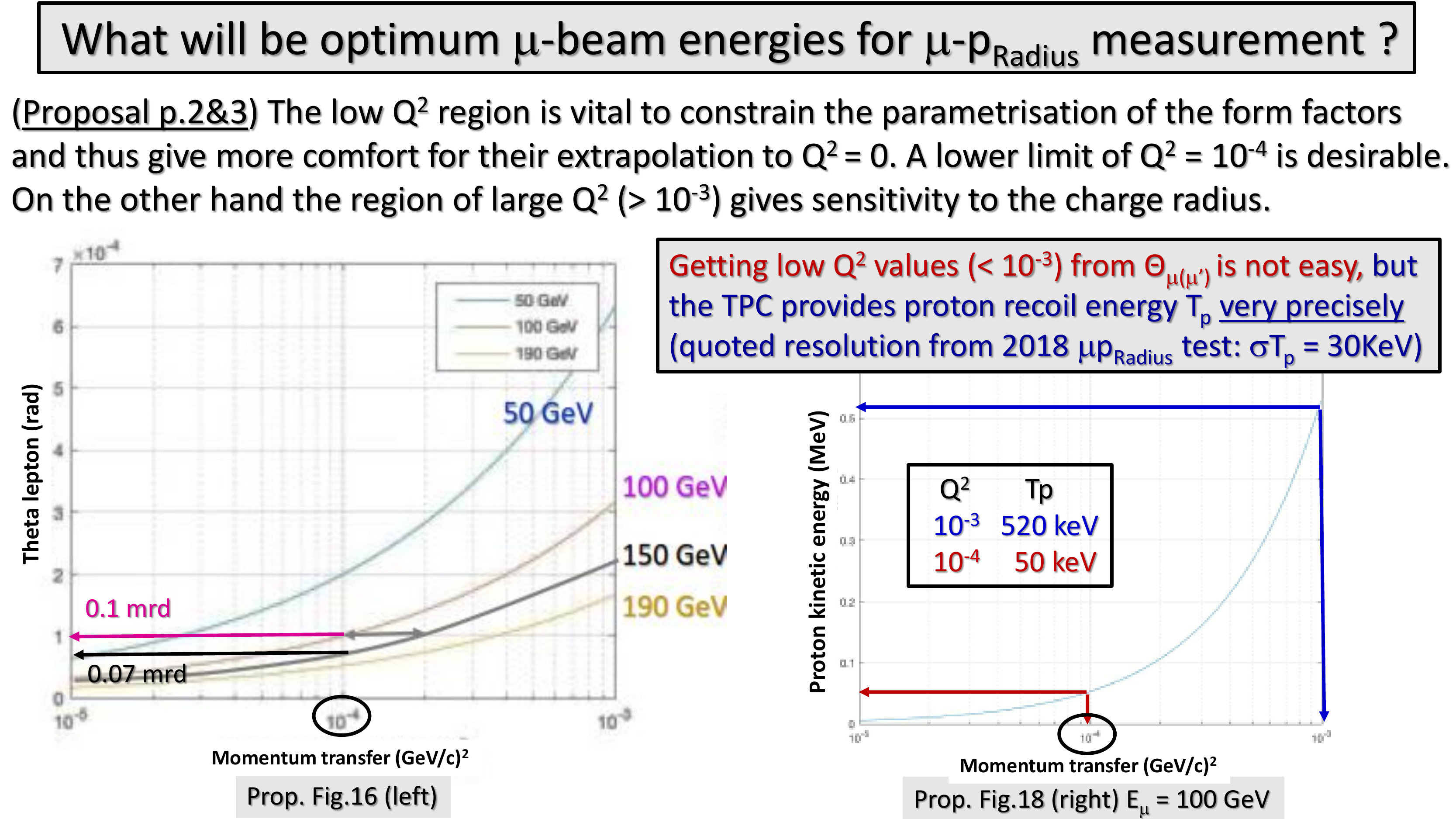}
\end{center}
\caption[Quoted optimum energies for the $\mu$-$e$ and the
$\mu$-proton-radius measurements.]{Quoted optimum energies: (Top) at
  least 150~GeV for $\mu$-$e$ and (Bottom) 100~GeV and below for
  $\mu$-proton-radius.}
\label{fig:energies}
\end{figure}

\paragraph{Optimum $\mu$-beam energies for $\mu$-$e$ and the $\mu$-proton-radius.}
Presently, the optimum beam energies for the two measurements are
different. Quoted is maximum 100~GeV for $\mu$-proton-radius and (at
least) 150~GeV for $\mu$-$e$. Figures~\ref{fig:energies}~(top and
bottom) illustrate this issue.  The $\mu$-proton-radius will also need
data taken $< 100$~GeV to get an accurate measurement at the lowest
foreseen $Q^2 = 10^{-4}$, in order to check the quality of the
extrapolation to $Q^2 = 0$. One should definitely \emph{study the
  option of running $\mu$-proton-radius} \emph{at 150 GeV}.

\begin{figure}
\begin{center}
\includegraphics*[width=0.75\textwidth]{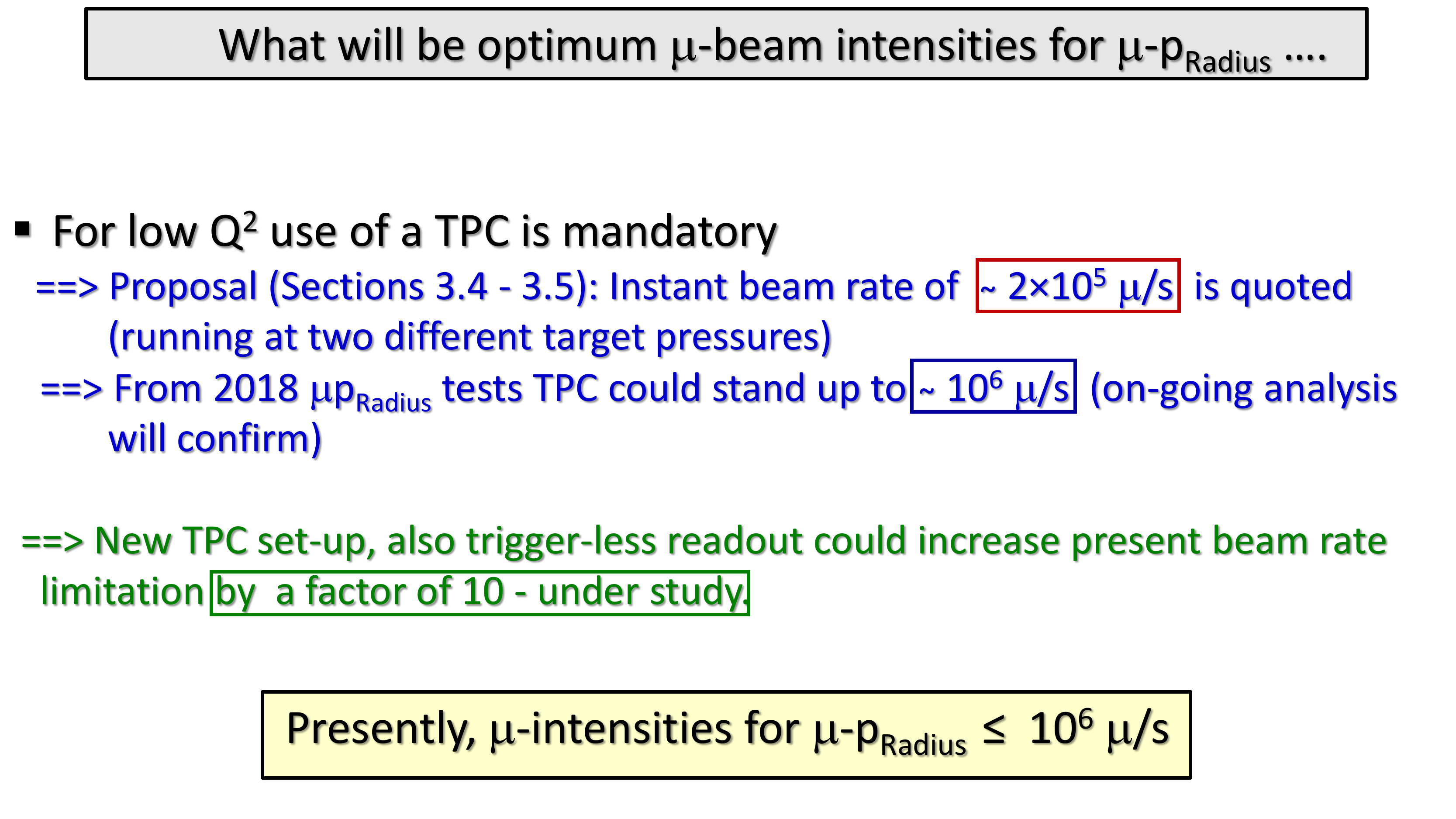}
\\[2em]
\includegraphics*[width=0.75\textwidth]{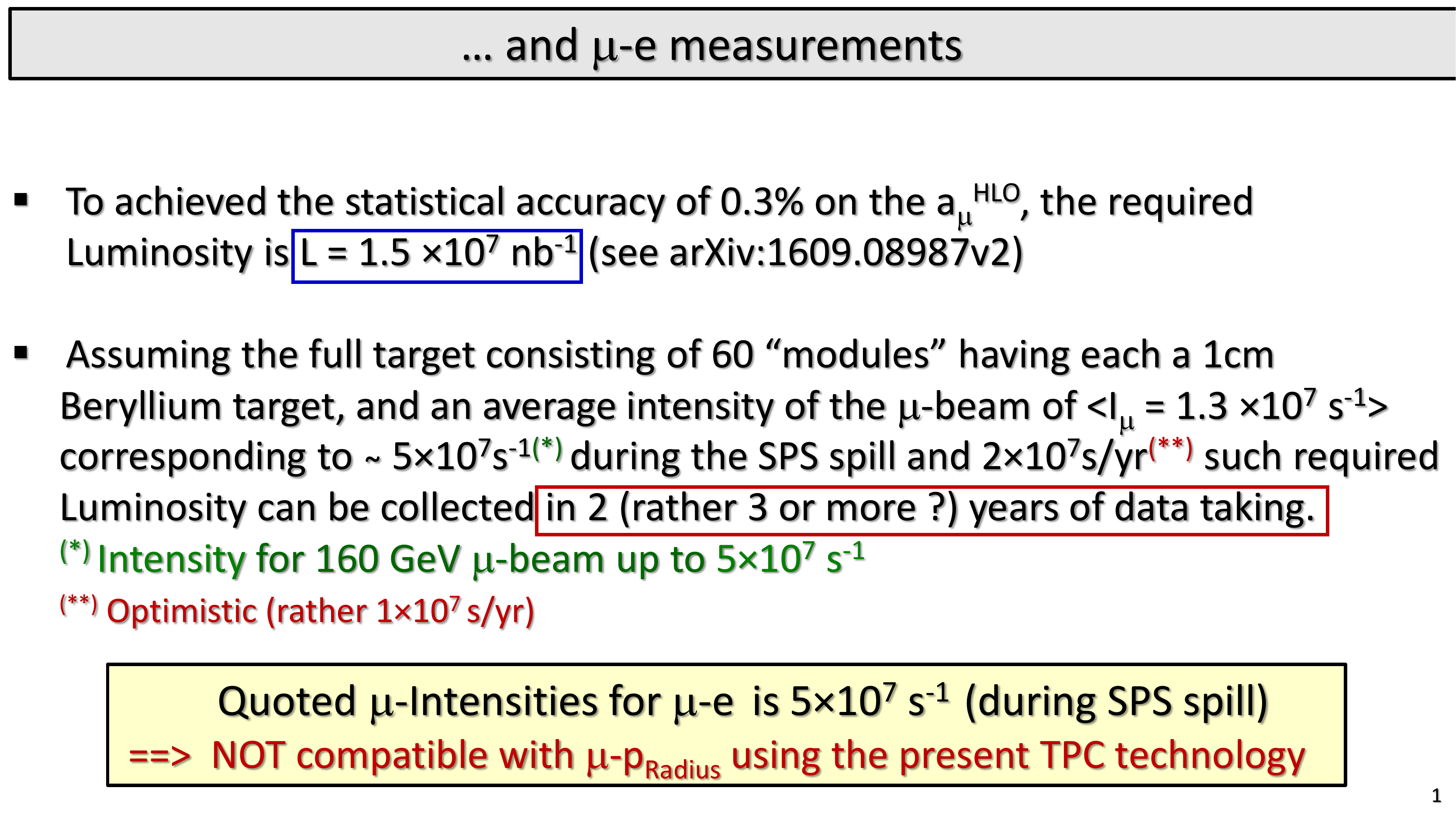}
\end{center}
\caption[Quoted optimum intensities for the $\mu$-$e$ and the
$\mu$-proton-radius measurements.]{Quoted expected optimum
  intensities: (Top) $\mu$-proton-radius and (Bottom) $\mu$-$e$.}
\label{fig:intensities}
\end{figure}

\paragraph{Optimum $\mu$-beam intensities for $\mu$-$e$ and the $\mu$-proton-radius.}
The values of intensities quoted in the $\mu$-proton-radius proposal
\cite{Friedrich:2286954} and in the $\mu$-$e$ proposal
\cite{Abbiendi:2016xup} (and subsequent presentations) are quite
different. Even though they cannot be taken yet as finalised
parameters, they differ by at least one order of magnitude. Therefore
raising a critical issue, concerning compatibility, see illustration
in figure~\ref{fig:intensities}. To optimise the overall use of the
$\mu$-beam it would be of \emph{great interest for $\mu$-proton-radius
  to improve their setup}, with the goal of running at a higher beam
intensity.

\begin{figure}
\begin{center}
\includegraphics*[width=0.8\textwidth]{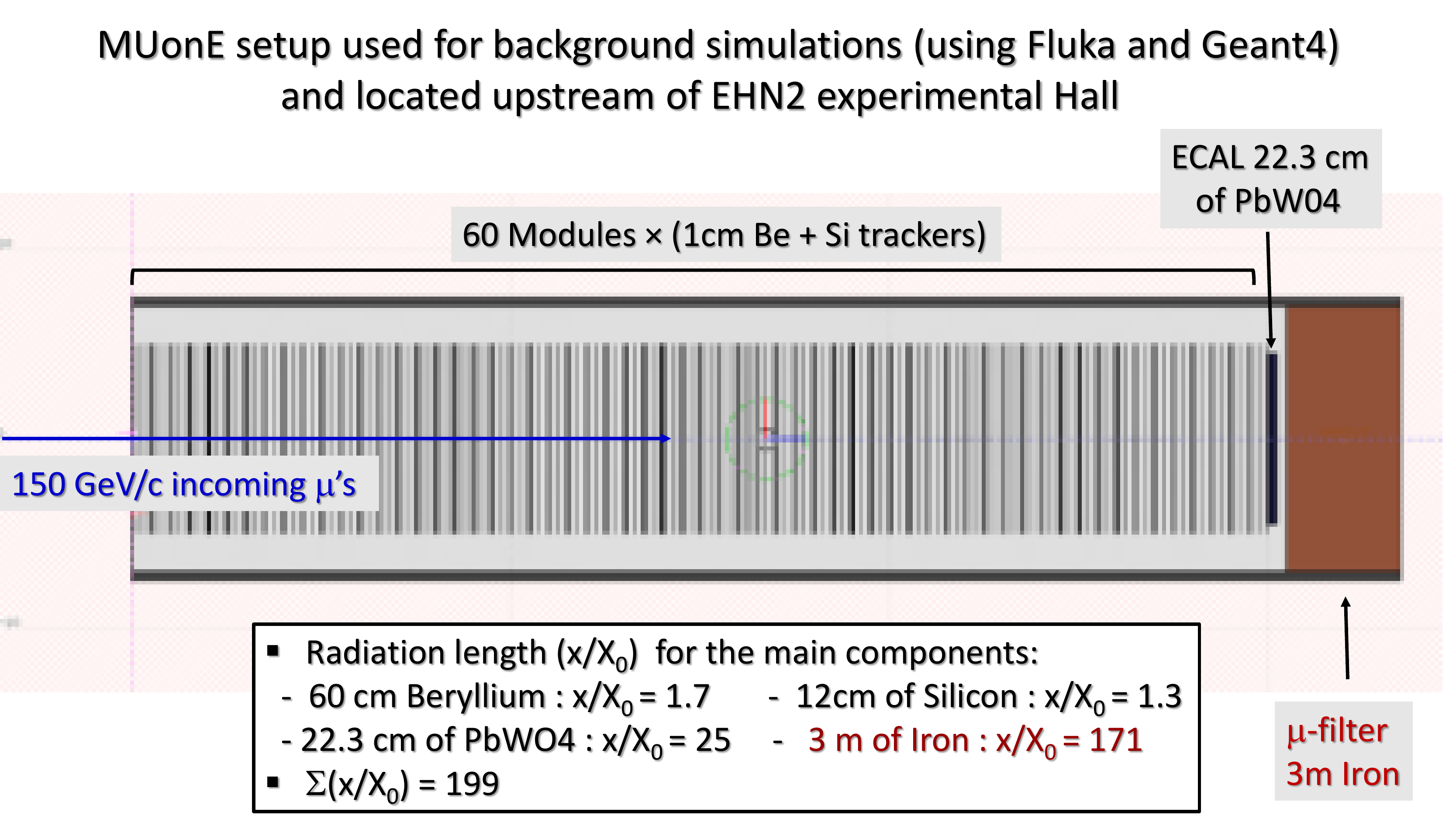}
\\[2em]
\includegraphics*[width=0.9\textwidth]{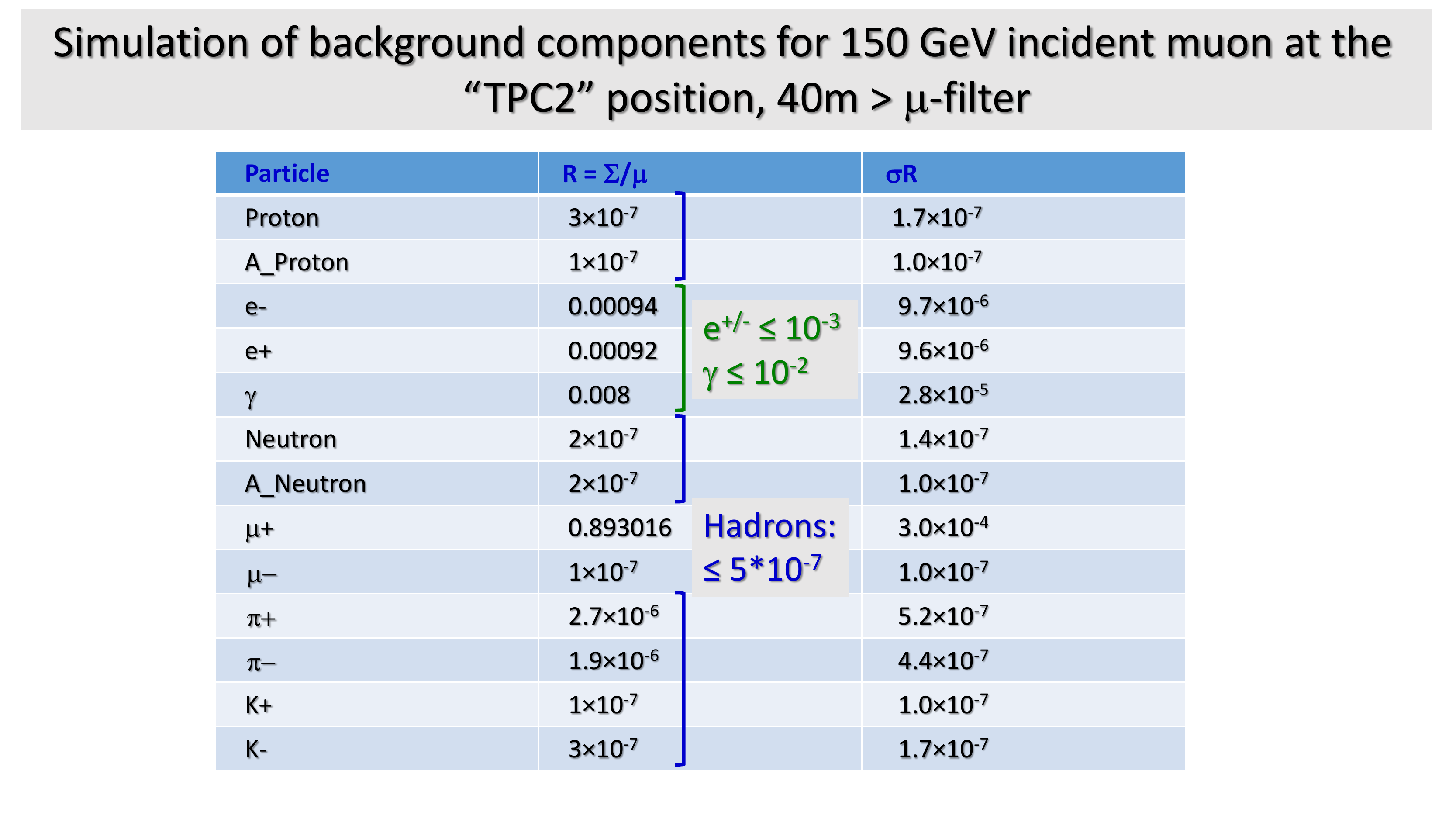}
\end{center}
\caption[Assumed $\mu$-$e$ setup composition and estimated radiations
length; fraction of background components entering the acceptance of
the COMPASS++ TPC used to detect the recoil low-energy proton and
located 40m downstream.]{(Top) Assumed $\mu$-$e$ setup composition and
  estimated radiations length.  (Bottom) Fraction (per-$\mu$) of
  background components entering the acceptance of the TPC used to
  detect the recoil low-energy proton and located 40m downstream, as
  shown in figure~\ref{fig:M2-setups}~(Bottom)}
\label{fig:pollution}
\end{figure}

\paragraph{Deterioration of the $\mu$-beam purity from interactions with the $\mu$-$e$ setup.}
Several remarks were expressed by the $\mu$-proton-radius proponents
concerning the option of the $\mu$-$e$ setup located upstream: "Such
hybrid solution {\it will not be compatible with the very strict
  requirements} of the proton radius measurement using high energy
$\mu$p elastic scattering within the COMPASS setup". One remark
concerned the generated background, resulting from interaction of the
muons with the $\mu$-$e$ setup. A quantification was
performed,\footnote{The simulation was done by D. Banerjee and
  M. V. Dick.}
using both Geant4 and Fluka softwares. A sketch of the assumed
material-input for $\mu$-$e$ is shown in figure~\ref{fig:pollution}
(top) and results from simulations are shown in
figure~\ref{fig:pollution} (bottom).

These $1^{st}$ estimates of the pollution from the $\mu$-$e$ setup
(upstream) are \emph{quite promising}. Note also that the two
different simulations (Geant4 and Fluka) provide \emph{very
  consistent} results.

\begin{figure}
\begin{center}
\includegraphics*[width=0.77\textwidth]{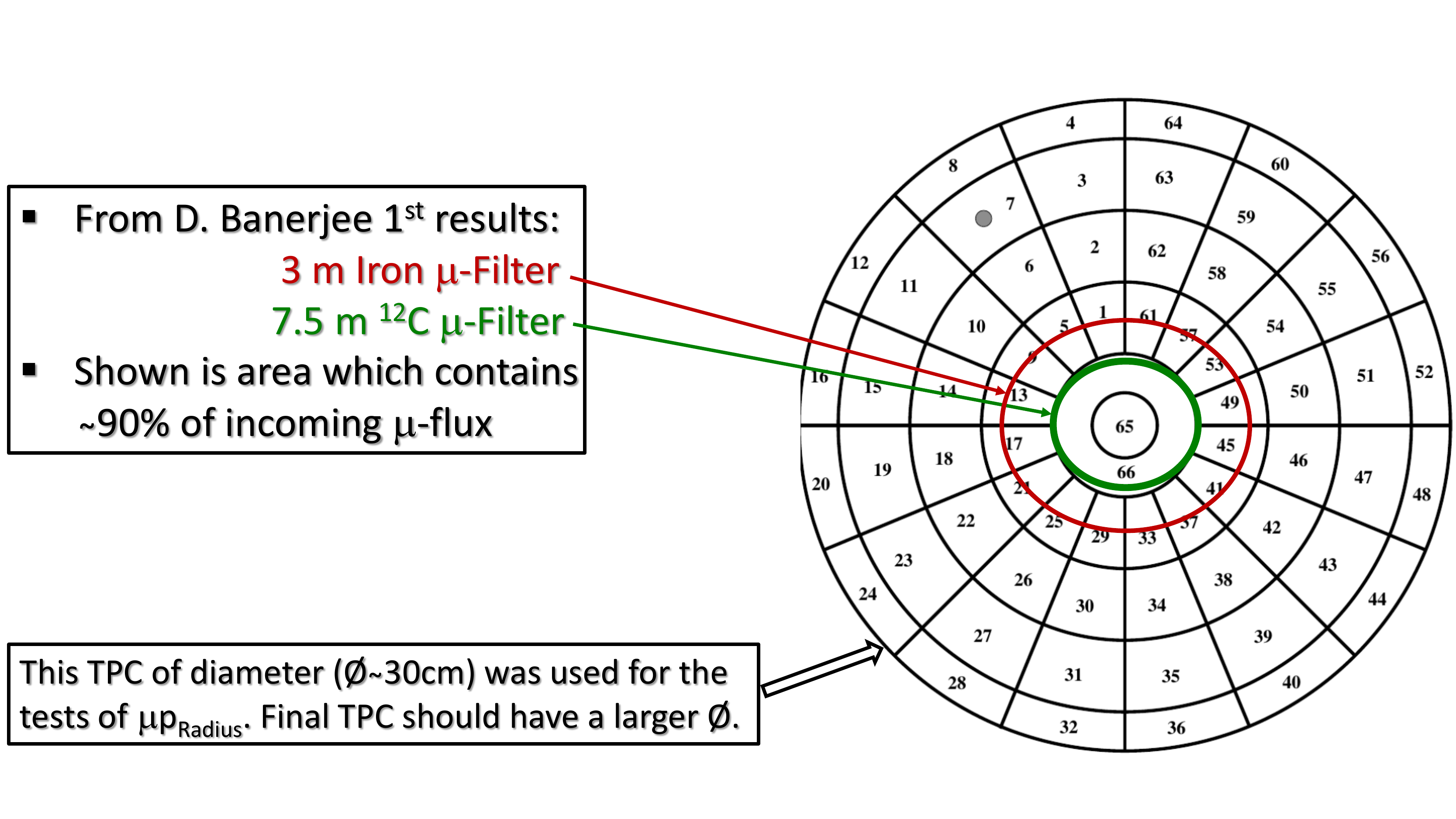}
\\[2em]
\includegraphics*[width=0.77\textwidth]{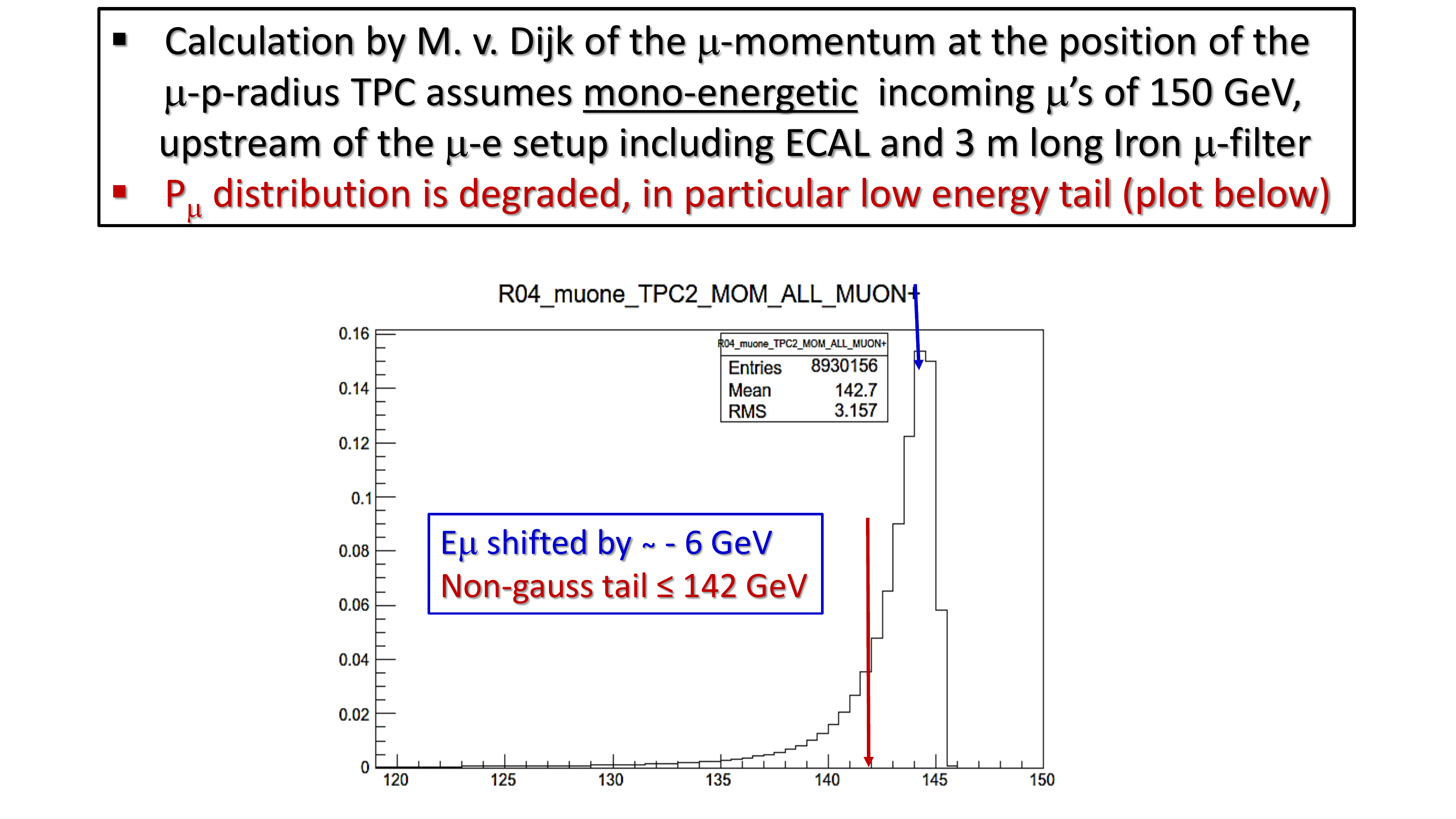}
\end{center}
\caption[Reduction of the $\mu$-beam size spreading from the MSC
angular divergence from a dedicated M2 $\mu$-beam tuning, optimised
for the $\mu$-proton-radius TPC position, and additional reduction by
replacing the Iron $\mu$-filter by a longer $\rm ^{12}C$
filter. Impact on the $\mu$ energy distribution of the energy losses
produced by the full $\mu$-$e$ setup.]{(Top) Reduction of the
  $\mu$-beam size spreading from the MSC angular divergence from a
  dedicated M2 $\mu$-beam tuning, optimised for the
  $\mu$-proton-radius TPC position (red). Also additional reduction by
  replacing the Iron $\mu$-filter by a longer $\rm ^{12}C$ filter
  (green). (Bottom) Impact on the $\mu$ energy distribution of the
  energy losses produced by the full $\mu$-$e$ setup.}
\label{fig:mscenergyloss}
\end{figure}

\paragraph{Deterioration of the $\mu$-beam angular divergence from MSC
  in the $\mu$-$e$ material, also momentum distribution spreading from
  energy losses.}
%
The large amount of radiation lengths of the $\mu$-$e$ setup quoted in
figure~\ref{fig:pollution}~(top), in particular downstream, generates
a significant beam divergence from Multiple SCattering (MSC). In
addition, the $\mu$'s loose energy which results in a shift and a
spreading of the energy distribution. Concerning the MSC a dedicated
optical beam tuning (suggested by J.Bernhard) has been proposed and
studied. Also suggested is the use of a $\rm ^{12}C$ $\mu$-filter
instead of Iron, which results in an additional reduction of the beam
spreading from MSC. The (preliminary) \emph{quite promising} results
are shown in figure~\ref{fig:mscenergyloss}~(top). However, the
effects from the $\mu$-beam energy losses in the $\mu$-$e$ material
which impact is shown on figure~\ref{fig:mscenergyloss}~(bottom)
cannot be reduced. We need a \emph{precise evaluation} of the
sensitivity of $\mu$-proton-radius measurement to such degradation of
the incoming $\mu$-beam energy (see below) and not ignore the
potential of the COMPASS~spectrometer to measure
$\rm E_{\mu'}(E_{\mu})$ precisely, and "tag" events issued from the
low energy tail.

\begin{figure}
\begin{center}
\includegraphics*[width=0.77\textwidth]{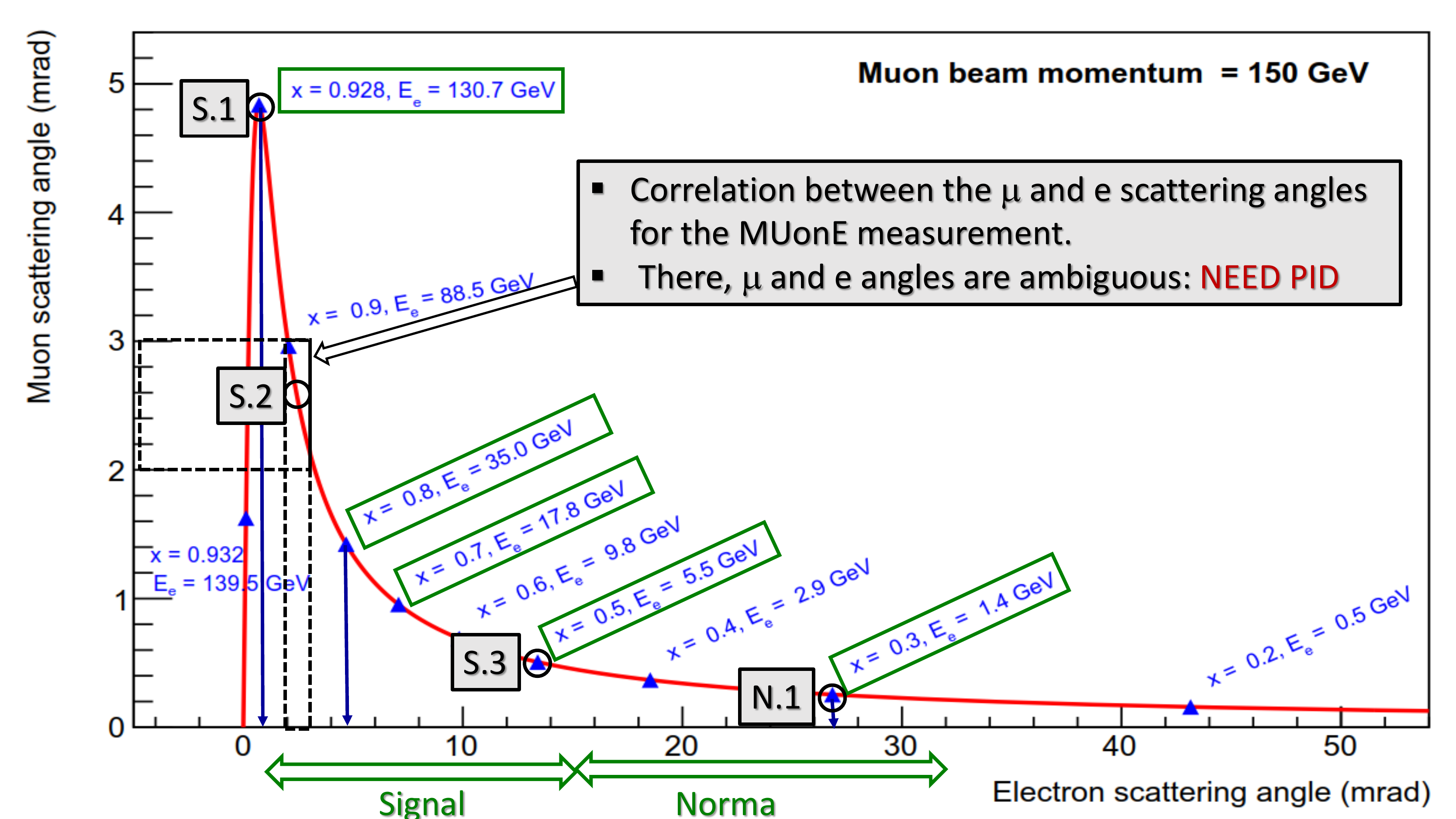}
\\[3em]
\includegraphics*[width=0.95\textwidth]{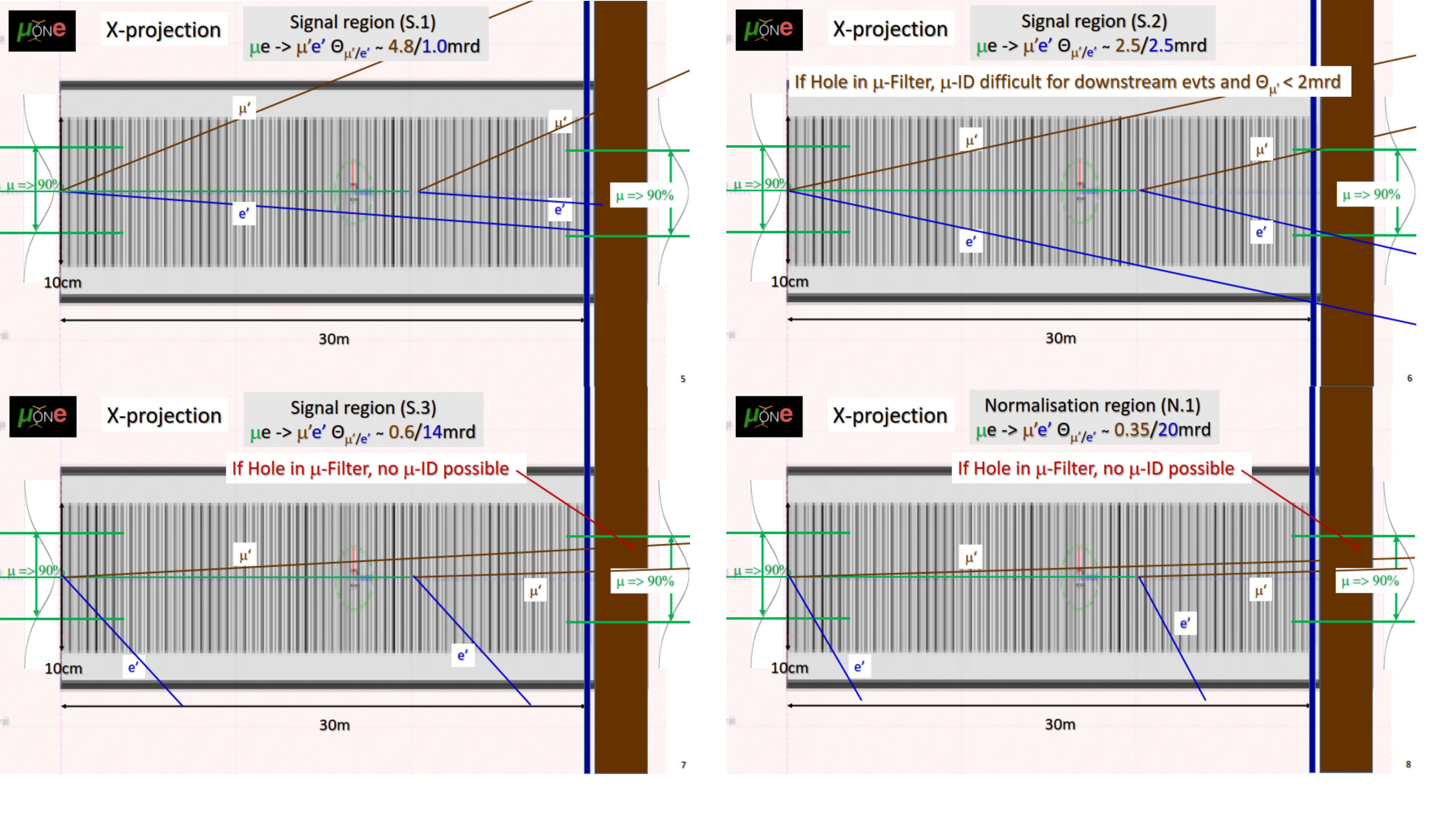}
\end{center}
\caption[Selected $\mu$-$e$ kinematics used to test the impact on the
PID of a hole in both ECAL and $\mu$-filter, and corresponding sample
of tracks for $\mu$-$e$ scattered events.]{(Top) Selected $\mu$-$e$
  kinematics used to test the impact on the PID of a hole in both ECAL
  and $\mu$-filter. (Bottom) corresponding sample of tracks for
  $\mu$-$e$ scattered events.}
\label{fig:holeinmufilter}
\end{figure}

\paragraph{Option of having a hole of appropriate size in the
  downstream $\mu$-$e$ heavy components.}
%
In order to reduce significantly the $\mu$-beam deterioration,
question was asked about the possibility to make a hole in the central
part of the ECAL and $\mu$-filter (heavy material) of the $\mu$-$e$
setup. Such hole should have an appropriate size to not interfere with
(at least 90\%) of the
$\mu$-beam. Figure~\ref{fig:holeinmufilter}~(top) shows a few
$\mu$-$e$ selected kinematics for which a sample of outgoing scattered
$\mu$ and $e$ tracks for events produced both upstream and downstream
of the setup are illustrated in
figure~\ref{fig:holeinmufilter}~(bottom). The conclusion is
  that such an option will seriously degrade the PID performances, in
particular for the Signal (S.2) region where the $\mu$ and $e$ angles
are ambiguous and PID mandatory.

\paragraph{Preliminary conclusions about $\mu$-$e$ and
  $\mu$-proton-radius compatibility.}
\begin{enumerate}
\item The option of having the $\mu$-$e$ installed upstream of EHN2
  looks quite promising. It guarantees, at least that the two projects
  can be setup together, with a priori, \emph{no interference}
  concerning their installation.  More technical input is expected
  from the EN/EA/LE group, in particular about the real available
  space and the required modifications/adjustments for some M2
  components.
\item Concerning a possible beam sharing: 1/- compatible beam energies
  is not granted. However, at this stage, we cannot exclude possible
  compromises which could allow a "partial" sharing, 2/- about beam
  intensities, compatibility appears to be a real problem, and given
  the present information(s) a \emph{true show-stopper}. More studies
  are requested from $\mu$-proton-radius to progress on this issue.
\item To proceed with further investigations, one will need more input
  about the overall issue of "feasibility", in particular for the very
  challenging $\mu$-$e$ project.
\item Given the "revised mandate" of the Physics Beyond Colliders
  Study group this should leave enough time for the future projects to
  provide the required additional input(s).
\end{enumerate}


\paragraph{Additional remarks concerning the $\mu$-proton-radius measurement.}

\begin{figure}
\begin{center}
\includegraphics*[width=0.98\textwidth]{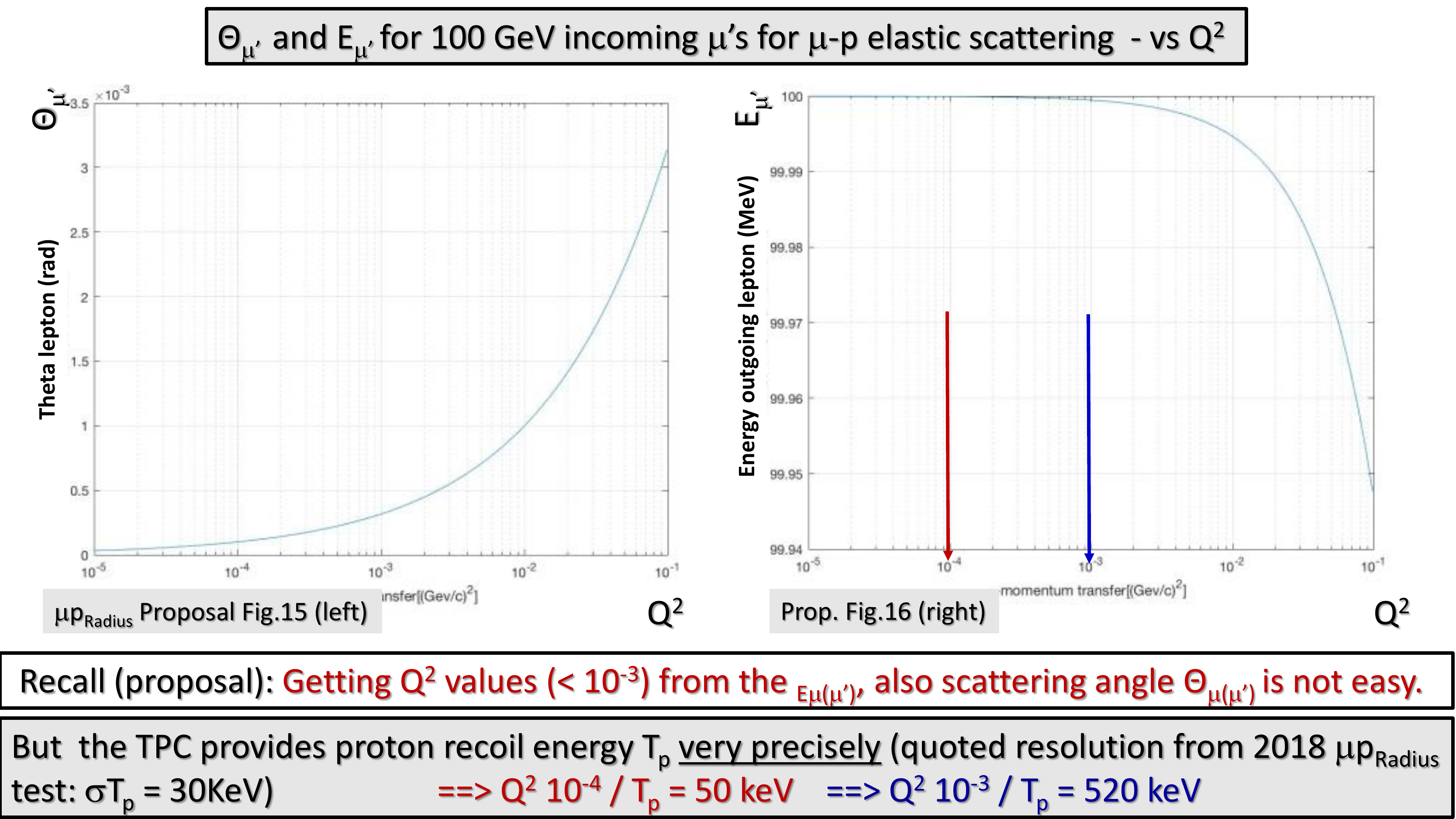}
\end{center}
\caption{Scattering angle $\theta_{\mu'}$ and energy $\rm E_{\mu'}$
  for the outgoing muon, assuming an incoming energy $\rm E_{\mu}$ of
  100 GeV.}
\label{fig:kinematics}
\end{figure}

Concerning the impact on the $\mu$-proton-radius measurement of the
expected degradation of the incoming $\mu$-beam energy (see above) it
is interesting to quote inputs from the proposal. In section 2.2.2 of
\revised{version 1 of the proposal} \cite{Friedrich:2286954} we read
``All kinematic quantities only depend on $Q^2$ and are almost
independent of the beam energy, except for the muon scattering angle
shown in Fig.~16a for three different possible values of the incoming
muon energy''. Also interesting are the proposal plots (see
figure~\ref{fig:kinematics}) which confirm that getting the low $Q^2$
values from the scattered muon angle $\theta_{\mu\mu'}$ also outgoing
$\rm E_{\mu'}$ versus incoming $\rm E_{\mu}$ is very difficult. For
low $Q^2$, its precise value is provided by the measurement of
$\rm T_{p}$ given by the TPC, see figure~\ref{fig:energies}~(bottom).

\clearpage

\section{Conclusions}
\label{sec:concl}

We conclude this report with a brief summary of the proposals and their physics highlights. \\

Experiments with gas targets at the LHC (\textbf{LHC-FT gas}) open up
a unique region of acceptance in hadron-hadron reactions.  Drell-Yan
as well as heavy flavour production in such a setup would
significantly improve the determination of PDFs, both in the proton
and in nuclei.  Especially in the region of large parton momentum
fraction $x$, such measurements would be highly competitive and timely
for the analysis of data taken at the LHC in collider mode.

A programme with polarised gas targets at the LHC would allow
measurement of asymmetries in Drell--Yan production that address core
issues in QCD spin physics, most notable the Sivers effect.  This
would be unique and highly complementary to existing and future
measurements in lepton-proton collisions, because the asymmetries in
question have a process dependence between $pp$ and $\ell p$ that is
predicted by theory.

A critical question for the realisation of such a programme is the impact of a fixed target on the LHC beams.  Furthermore, the target location with respect to the detector has a significant impact on the acceptance.  This directly influences the kinematic and hence physics reach of a setup. \\

A measurement of the magnetic moments of short lived heavy baryons with bent crystals at the LHC \textbf{(LHC-FT crystal)} would yield interesting information for heavy-quark physics.  This proposal is also of interest from a technological side, and it offers the possibility to be adapted to other goals such as measuring the magnetic moment of the $\tau$ lepton.  The proposed measurements are challenging, and further R\&D is necessary to establish their feasibility. \\

The measurement of $\mu p$ elastic scattering at high energy, proposed
by \textbf{COMPASS++}, would be unique and could significantly improve
our understanding of elastic lepton-proton scattering at low momentum
transfer.  Progress in this area would be highly welcome, given the
unsettled discrepancies in the determination of the electromagnetic
proton radius by different methods.  The per mil point-to-point
cross-section precision required for the proposed measurement is
challenging and will require further R\&D in instrumentation and data
acquisition.

The partonic structure of the pion is of special interest in QCD,
given the special role of the pion in chiral symmetry breaking.  A
measurement campaign of Drell--Yan production with conventional
$\pi^+$ and $\pi^-$ beams would significantly improve our knowledge of
pion PDFs, providing for the first time a handle on separating sea
quarks from valence quarks at momentum fractions $x$ above $0.1$.
Planned investigations of pion PDFs at JLab 12 and the EIC make use of
pions radiated from a target proton.  The interpretation of such
measurements is delicate because these pions are off shell, and
detailed pion Drell--Yan data would provide a most valuable baseline
for the field.

A broad range of unique possibilities would be opened up by RF-separated beams at the SPS, whose production was studied in the PBC Conventional Beams working group \cite{PBC-convbeam}.  Highlights among the experiments proposed by COMPASS++ are the detailed investigation of the kaon excitation spectrum, which remains poorly known, and the use of the Primakov process to gain information on the electromagnetic kaon polarisabilities.  Drell--Yan or prompt-photon production with RF-separated kaon beams would allow the investigation of kaon structure at the quark-gluon level. \\

The quantitative understanding of chiral dynamics in channels that involve strange quarks remains a challenge.  In addition to the kaon polarisabilities just mentioned, the elastic $\pi K$ scattering lengths are benchmark quantities for the comparison of nonperturbative calculations in QCD with experiment.  The study of mesonic atoms at the SPS proposed by \textbf{DIRAC++} could determine the difference $|a_{1/2} - a_{3/2}|$ with precision at par with current theory predictions, in addition to further improving our knowledge of the $\pi\pi$ scattering lengths. \\

The aim of the \textbf{MUonE} proposal is to extract the contribution of hadronic vacuum polarisation to $(g-2)_\mu$ with a precision competitive with the one currently obtained from $e^+ e^-$ annihilation into hadrons and from hadronic $\tau$ decays.  This would be an extremely valuable independent determination for the value $(g-2)_\mu$ in the Standard Model.  The uncertainties from pure lattice determinations are currently too large to be competitive.  A measurement by MUonE would be timely in view of imminent measurements at FNAL and J-PARC.  The envisaged accuracy puts highest demands on experimental and theoretical precision, and further work in both areas is required to establish the feasibility of this experiment. \\

With collision energies of $\sqrt{s_\textrm{NN}} \sim 5 - 17$\, GeV at
the {SPS}, the proposed experiments \textbf{NA60++} and
\textbf{NA61++} cover a very important part of the density regime in
the QCD phase diagram, that possibly contains the critical end
point. In particular, with its open charm measurements,
\textbf{NA61++} offers a unique possibility to access this regime, and
to study effects related to the onset of deconfinement. This also
allows
to discriminate between competing phenomenological explanations for the production of charm pairs. \textbf{NA60++} offers a precise determination of the initial fireball temperature, that would allow to pin down the first order regime for baryon densities above the critical end point. \\

Several of the fixed-target projects discussed here also aim at providing input to the interpretation of data from cosmic-ray measurements and accelerator-based neutrino experiments. Particle fluxes from nuclear collisions at various collision energies could be measured at \textbf{NA61++}, \textbf{COMPASS++}, and \textbf{LHC-FT gas}, covering a range in \(\sqrt{s}\) from 8.8--115 GeV, with a special focus on \(\bar{p}\) production in view of high-precision results on the \(\bar{p}\) flux from AMS-02. Charm production in the atmosphere constitutes an important background for neutrino fluxes and can be better understood with data on the charm content at large-\(x\) in the nucleus, which is part of the LHC-FT gas program. Long-baseline neutrino experiments require an excellent knowledge of the neutrino flux resulting from hadron-nucleus collision in the production target. Dedicated measurements using target replica are foreseen at \textbf{NA61++}.\\


In summary, the proposals discussed in the PBC QCD working group
reveal a number of physics opportunities at the CERN accelerator
complex, with the potential to significantly bring forward our
understanding of strong interaction physics in a broad range of areas,
from the quark-gluon structure of hadrons to low-energy dynamics and
to the collective phenomena responsible for the QCD phase transition.
Some of the measurements would greatly benefit other fields, from the
interpretation of precision measurements probing the limits of the
Standard Model to the propagation of cosmic rays.

For several proposals, further investigation is required to establish
whether they are able to reach their physics goals.  Further studies
are also required to determine the technical boundary conditions for
the optimal use of the M2 beam line of the SPS, which would be the
location of the experiments proposed by COMPASS++ and MUonE, and also
of a beam-dump experiment with muons proposed by NA64++ and discussed
in the PBC BSM working group \cite{PBC-BSM}.  It is hoped that the
extension of the PBC mandate until mid-2020 will help in making
progress on several of these open issues.

\acknowledgments

Individual authors of this report have received financial support from
ERC Ideas Consolidator Grant No.\ 771642 SELDOM (European Union);
CNRS (France) via: COPIN-IN2P3 agreement, IN2P3 project "TMD@NLO", Franco-Spanish PICS "Excitonium", Quarkonium4AFTER project of the Franco-Chinese LIA FCPPL; the P2I Department of Paris-Saclay University (France);
DFG Collaborative Research Centre SFB 1225 ``ISO-QUANT'' (Germany); BMBF grant 05P18VHFCA (Germany);
INFN (Italy);
GVA(Spain); Ikerbasque (Spain); MINECO (Spain);
RFBR/CNRS grant 18-52-15007 (Russia and France).



\phantomsection
\addcontentsline{toc}{section}{List of Figures}
\listoffigures
\phantomsection
\addcontentsline{toc}{section}{List of Tables}
\listoftables

\clearpage
\phantomsection
\addcontentsline{toc}{section}{References}

\bibliographystyle{JHEP}
\bibliography{qcd}

\end{document}